\documentclass[twocolumn,showpacs,preprintnumbers,amsmath,amssymb]{revtex4}
\usepackage{graphicx}% Include figure files
\usepackage{dcolumn}% Align table columns on decimal point
\usepackage{mathrsfs}
\usepackage{bm}% bold math

\usepackage{caption}
\usepackage{subcaption}
\usepackage{placeins}

\usepackage{xcolor}

\graphicspath{{./Figures/}}

\newcommand{\reals}{\mathbb{R}}

\newcommand{\bra}[1]{\left\langle #1 \right\rvert }
\newcommand{\ket}[1]{\left\lvert #1 \right\rangle }
\newcommand{\inner}[2]{\left\langle #1 \right\rvert \left.\hspace{-2.5pt} #2 \right\rangle}

\newcommand{\avg}[1]{\left\langle #1 \right\rangle}

\newcommand{\re}[1]{\text{Re}\left[ #1 \right]}
\newcommand{\imag}[1]{\text{Im}\left[ #1 \right]}

\newcommand{\intd}{\thinspace\mathrm{d}}

\begin{document}

\preprint{APS/123-QED}

\title{Probing localization properties of many-body Hamiltonians via an imaginary vector potential}% Force line breaks with \\

\author{Liam O'Brien}
\email{lobrien@caltech.edu}
%\affiliation{%
%Institute for Quantum Information and Matter, California Institute of Technology, Pasadena, CA 91125
%}%
\affiliation{%
Department of Physics and Institute for Quantum Information and Matter, California Institute of Technology, Pasadena, CA 91125
}%
 \author{Gil Refael}%
 \email{refael@caltech.edu}
\affiliation{%
Department of Physics and Institute for Quantum Information and Matter, California Institute of Technology, Pasadena, CA 91125
}%
%\affiliation{%
%Department of Physics, California Institute of Technology, Pasadena, CA 91125
%}%

\date{\today}% It is always \today, today,

\begin{abstract}
Identifying and measuring the ``localization length'' in many-body systems in the vicinity of a many-body localization transition is difficult. Following Hatano and Nelson, a recent work [S. Heu\ss{}en, C. D. White, and G. Refael, Phys. Rev. B \textbf{103}, 064201 (2021)] introduced an ``imaginary vector potential'' to a disordered ring of interacting fermions, in order to define a many-body localization length (corresponding, in the non-interacting case, to the end-to-end Green's function of the hermitian system). We extend these results, by connecting this localization length to the length scale appearing in the avalanche model of delocalization. We use this connection to derive the distribution of the localization length at the MBL transition, finding good agreement with our numerical observations. Our results demonstrate how a localization length defined as such probes the localization of the underlying ring, without the need to explicitly construct the l-bits. 
\end{abstract}

\pacs{Valid PACS appear here}% PACS, the Physics and Astronomy
\maketitle

\section{Introduction}
\label{intro}

Conventional statistical mechanics assumes that isolated systems reach thermal equilibrium - effectively acting as their own heat bath - when left to evolve under their own dynamics. The process by which this happens for quantum systems is highly non-trivial \cite{Deutsch-QSM,Srednicki-Thermalization,Rigol-Thermalization,Rigol-Srednicki-ETH-alts}, and has been the subject of intense research in recent decades - see \cite{Mori-Ikeda-Kamishi-Ueda-Review,Ueda-Review,Nandkishore-Huse-Review} for recent reviews. Through this body of work, certain classes of systems have been discovered which appear to violate this assumption, and do not thermalize under their own dynamics. One such class that has received much theoretical attention are lattice systems with quenched disorder. For non-interacting systems, Anderson localization occurrs, and the single-particle eigenstates are exponentially localized around a specific lattice site, with a decay length that is a function of the disorder and the energy \cite{Anderson}. Interacting systems display many-body localization (MBL), which is characterized by zero DC transport, area-law entanglement entropy, Poissonian level statistics, logarithmic growth of entanglement entropy, long-time memory of initial conditions, and more \cite{Nandkishore-Huse-Review}. MBL has emerged as a platform for novel quantum order \cite{Huse-Nandkishore-Oganesyan-Pal-Sondhi-Localization-order,Pekker-Refael-Altman-Demler-Oganesyan-Hilbert-glass,Chandran-Khemani-Laumann-Sondhi-Localization-SPT,Vassuer-Potter-Parameswaran-Quantum-critical-hot,Potter-Vishwanath-Protection-topo-preprint,Bahri-Vosk-Altman-Vishwanath-Localization-topo,Vasseur-Friedman-Parameswaran-PHS-MBL-topo,Kells-Moran-Meidan-p-wave,Iadecola-Schecter-Inverse-Freezing,Friedman-Vasseur-Potter-Parameswaran-Localization-order-non-Abelian,Kuno-MBL-protected-XXZ} (see \cite{Parameswaran-Vasseur-MBL-symm-topo-review} for a review), with implications for quantum information \cite{Santos-Dykman-Shapiro-Izrailev-Localized-qubits,Altshuler-Krovi-Rolan-Adiabatic-quantum-opt,Laumann-Moessner-Scardicchio-Sondhi-Quantum-annealing,Yao-Laumann-Vishwanath-MBL-state-transfer,Khemani-Nandkishore-Sondhi-Nonlocal-response,Goihl-Walk-Eisert-Tarantino-SPT-memory}. 

The existence of a many-body localized regime has been demonstrated numerically in small systems \cite{Pal-Huse,Bauer-Nayak,Znidaric-Prosen-Prelovsek,Kjall-Bardarson-Pollmann,Luitz-MBL-trans} and proven analytically \cite{Imbrie} for one-dimensional spin chains under certain assumptions, but many open questions still remain. In particular, the nature (universality class) of the transition between the thermal/ergodic phase and the MBL phase still remains to be understood \cite{Nandkishore-Huse-Review}. Some recent works even call into question whether the transition exists at all in the thermodynamic limit \cite{Suntajs-Bonca-Prosen-Vidmar-Quantum-Chaos,Sels-Polkovnikov-Dynamical-Obstruction}; subsequent works have argued the observed effects underlying those conclusions are the product of finite-size effects \cite{Localization-from-Chaos} and many-body resonances  \cite{Gopalakrishnan-Muller-Khemani-Knap-Demler-Huse-low-frequency-cond,Villalonga-Clark-Resonances-1,Villalonga-Clark-Resonances-2,Garratt-Roy-Chalker-resonances,Crowley-Chandran-RM,Morningstar-Colmenarez-Khemani-Luitz-Huse-Av-Resonances,Long-Crowley-Khemani-Chanrdan-prethermal-MBL,Garratt-Roy-resonances-observables,Ha-Morningstar-Huse-res-av}. Much remains to be done in probing the nature and stability of the localization in systems that are believed to be MBL.

The most natural way to quantify localization is through the localization length. In the non-interacting case, this is straightforward: it is the decay length of the single-particle eigenstates. In an interacting system, the desired ``localization length'' is less obvious, since there is no notion of single particle orbitals. One of the most useful ways to quantify the localization, then, is through the use of localized conserved quantities.  In addition to the properties mentioned above, the MBL phase comes equipped with a complete set of (quasi-)local integrals of motions (LIOMs) \cite{Huse-Nandkishore-Oganesyan,Ros-Muller-Scardicchio} - the so-called ``l-bits''. Roughly speaking, the l-bits are conserved quantities that are obtained by ``dressing'' the physical degrees of freedom (``p-bits'') with unitaries that are exponentially localized in real space. The decay of the l-bits thus offers a way of calculating a localization length in interacting systems. 

A number of methods have been put forth to address the issue of constructing the l-bits \cite{Serbyn-Papic-Abanin-LIOM,Chandran-LIOM,Rademaker-Ortuno,Pekker-Clark-Oganesyan-Refael-Wegner-Wilson}, but the procedure is computationally taxing and physically ambiguous (since the mapping of the set of l-bit configurations onto the computational basis states is not unique). This ambiguity means that the localization lengths measured depend on how the l-bits are constructed. In principle, it is possible to uniquely specify an assignment of l-bits that is ``most localized'' in the original basis \cite{Huse-Nandkishore-Oganesyan}, but there is no algorithm known rigorously to construct such an assignment (though algorithms based on Wegner-Wilson flow \cite{Wegner,Kehrein-Book} work fairly well in practice \cite{Pekker-Clark-Oganesyan-Refael-Wegner-Wilson}). In spite of these difficulties, a number of numerical studies \cite{Peng-Li-Yan-l-bits-loc-len, Kulshreshtha-Pal-Wahl-Simon-l-bits-loc-len, Varma-Raj-Gopalakrishnan-Oganesyan-Pekker-l-bits-loc-len, Kelly-Nandkishore-Marino-l-bits-environment,Thomson-Schiro-time-evo,Thomson-Schiro-powerlaw,Thomson-Magano-Schiro-Floquet-LIOMS,Thomson-Schiro-quasiperiodic-LIOMS,Quito-Titum-Pekker-Refael-Wegner-Flow} have extracted decay lengths from the l-bits and analyzed properties of the finite-size MBL phase and MBL-thermal crossover. 

An alternate method for finding a localization length is to introduce an imaginary vector potential, or tilt, that makes hopping preferential in one direction \cite{NHMBL}. By varying the strength of the vector potential/tilt, one finds that various eigenvalues will develop non-zero imaginary parts (as the system is no longer Hermitian). The point at which a given eigenvalue develops a non-zero imaginary part in this modified model defines a length scale $\xi$ that, in non-interacting systems, directly measures the localization length (defined in terms of an end-to-end Green's function) of the analogous (single-particle) eigenstate of the underlying model without the vector potential. This procedure does not require constructing the l-bits, and Ref \cite{NHMBL} showed the corresponding $\xi$ can be used to identify the critical parameters of the MBL-thermal crossover in interacting systems. 

Here, we connect the length scale $\xi$ to the \textit{avalanche model} of delocalization. Much of the current literature describes the asymptotic MBL to thermal transition in terms of such ``avalanches'', whereby rare (locally) thermal regions are able to induce a cascade of thermalization that overpowers the localization in the rest of the system \cite{DeRoeck-Huveneers-Stability-Av,Luitz-Av, Thiery-Av, Thiery-Av-Long, Johri-Env}. The instability of localized systems to avalanches is characterized by a length scale $\lambda$, which captures the decay of matrix elements coupling the l-bits and rare thermal regions. It is natural, then, to expect that $\xi$ and $\lambda$ are related to each other, as they both act as a measure for the localization-delocalization transition; we make this relation explicit.

Moreover, $\xi$ has a distribution (with respect to quenched disorder). We calculate this distribution numerically at the finite-size MBL-thermal crossover, and examine whether the relation between $\xi$ and $\lambda$ holds for the distributions as well. Indeed, we find it does, so that the distributions of $\xi$ at the MBL crossover contain information about the decay of matrix elements at the corresponding transition.

Our paper is organized as follows: In section \ref{model}, we present and discuss the model we will study and some of its salient features. In section \ref{avalanche} we show $\xi$ can be predicted by proliferation of non-hermitian avalanches, and we connect $\xi$ to the decay length $\lambda$ appearing in the avalanche picture of the delocalization transition. Finally, in section \ref{distributions}, we use this connection to derive an analytic form of the distribution for $\xi$, and compare with the numerically observed histograms.

\section{Non-hermitian MBL Model}
\label{model}

\subsection{Hamiltonian}
\begin{figure}
\includegraphics[scale=.33]{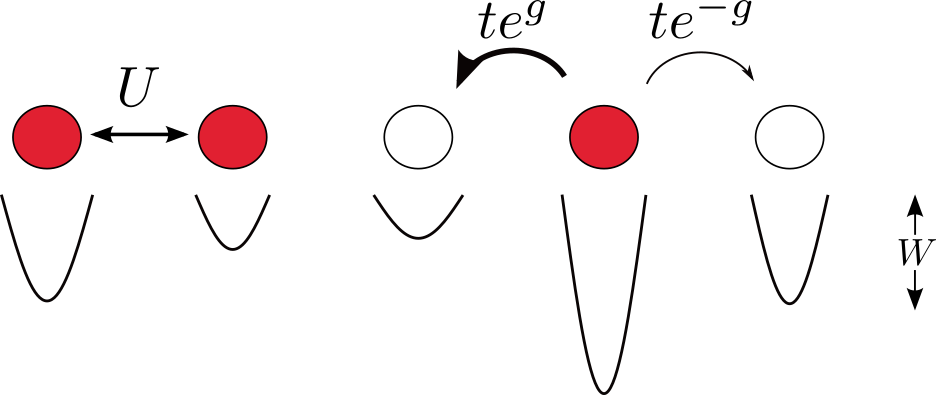}
\caption{A schematic of the model Hamiltonian \eqref{H}. Fermionic particles (filled red circles) can hop (preferentially to the left) with strength $te^{\pm g}$, interact with their nearest neighbor with strength $U$, and are also subject to a random potential whose strength is determined by the disorder width $W$.}

\label{schematic}
\end{figure}

We consider the following Hamiltonian for spinless fermions on a one-dimensional lattice with sites $i = 1,\ldots, L$
\begin{equation}
\resizebox{.425\textwidth}{!}{$\displaystyle H = t\sum_{i=1}^L[e^g c_i^{\dag}c_{i+1} + e^{-g}c_{i+1}^{\dag}c_i] + \sum_{i=1}^L h_i n_i + U\sum_{i=1}^L n_i n_{i+1}$},
\label{H}
\end{equation}
where $c_i^{\dag}/c_i$ are fermionic creation/annhiliation operators, $n_i = c_i^{\dag}c_i$ number operators, $t \in \reals$ is the hopping amplitude, $g \geq 0$ is the ``imaginary vector potential'' or ``tilt'', $U > 0$ is a parameter denoting the strength of interactions, and the $h_i$'s are random variables drawn independently and identically from a distribution characterized by a ``disorder strength'' $W$. This Hamiltonian describes particles hopping on a disordered lattice, subject to nearest neighbor repulsion, with an imaginary vector potential (of magnitude $g$) making left hopping preferential -  see Fig. \ref{schematic} for a schematic of this model.

The Hamiltonian in \eqref{H} has been studied as an effective model to describe driven open systems \cite{Panda-Banerjee-NESS}, and can also be mapped onto the statistical mechanics of depinning flux lines from columnar defects in two-dimensional type II superconductors  via the path integral formalism \cite{Hatano-Nelson-Flux,Hatano-Nelson-Eigenfunctions,Hatano-Nelson-Vortex}. In such a mapping, $g$ characterizes an external magnetic field that is tilted with respect to the random columnar defects - hence the choice of names for $g$.

Since $g$ is a real number, we generically expect some eigenvalues to be complex for $g\neq 0$. The values of $g$ at which certain eigenvalues develop a non-zero imaginary part are examples of so-called \textit{exceptional points} \cite{Kato}, which have been a topic of research since the 1990s \cite{Heiss-Sanninno-Level-Crossing,Heiss-Sannino-Transition,Heiss-Repulsion,Heiss-Global-Local,Heiss-Harney-Chirality,Heiss-Exc-Points,Heiss-Chirality-3,Heiss-Physics-Exc-Points,Luitz-Piazza-Exc-Points}. Of particular interest has been the connection between the exceptional points and localization/delocalization of eigenstates in models similar to \eqref{H} \cite{Hatano-Nelson-Flux,Hatano-Nelson-Eigenfunctions,Hatano-Nelson-Vortex,Hamazaki,Feinberg-Zee,Feinberg-Zee-Spectral-curve,Brezin-Zee}. We seek to exploit this connection, as we lay out in the next section.

\subsection{Exceptional points of $H$ and their connection to localization/delocalization}
\begin{figure}
\begin{subfigure}{.4\textwidth}
\includegraphics[scale=.18]{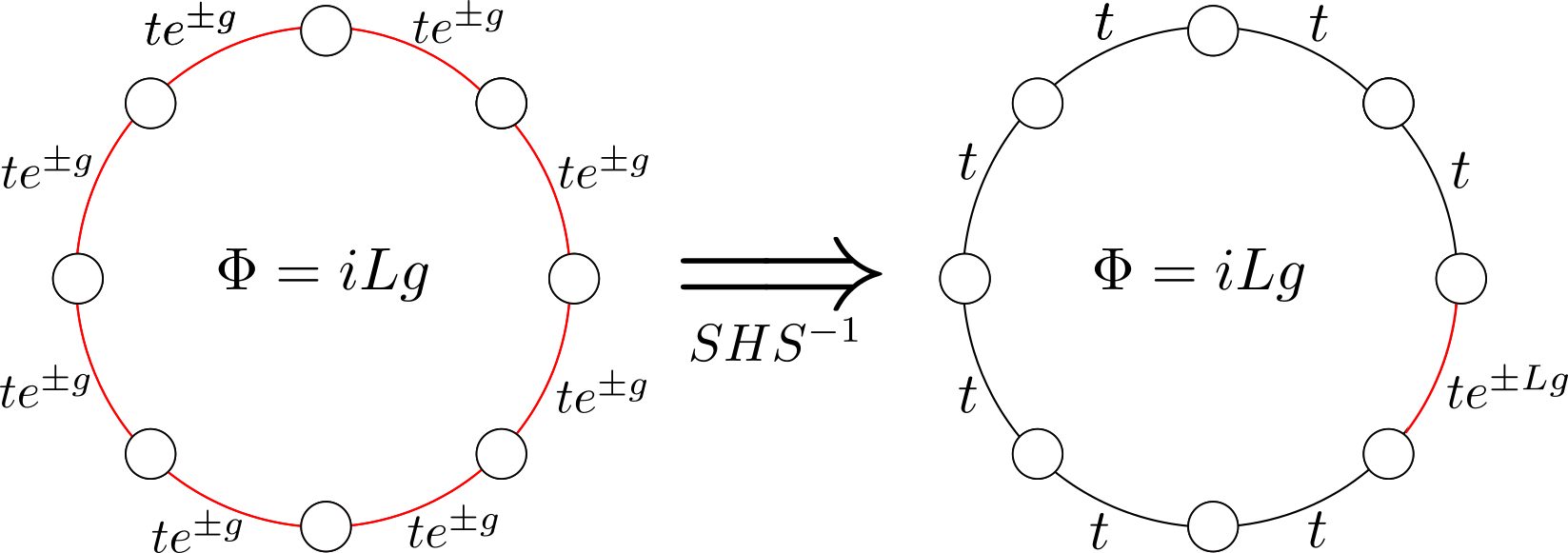}
\caption{}
\label{flux}
\end{subfigure}
\hfill\\
\begin{subfigure}{.4\textwidth}
\includegraphics[scale=.3]{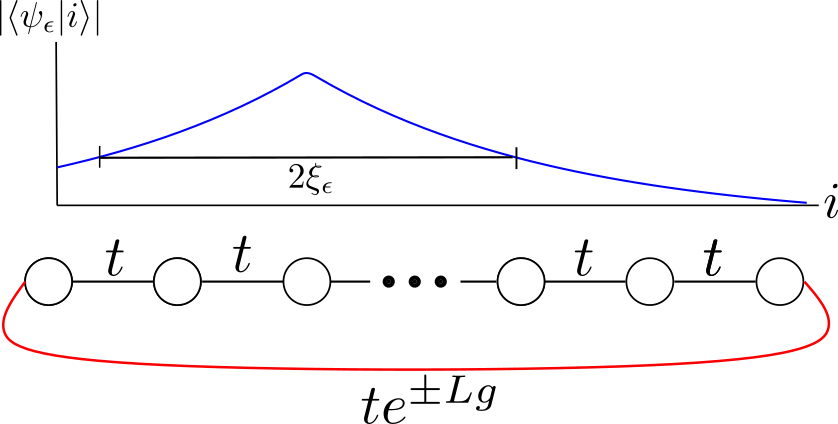}
\caption{}
\label{flux-loc-length}
\end{subfigure}
\caption{ a) Illustration of the imaginary flux $\Phi = iLg$ bound through our system (with periodic boundary conditions), as well as the action of the gauge transformation \eqref{S_gauge}. b) Illustration of the competition between the decay of the wavefunction $\ket{\psi_{\epsilon}}$ and the imaginary flux $iLg$. Heuristically, we expect the exceptional point $g_c(\epsilon)$ to occur where the product of the end-to-end tunneling amplitude $\inner{1}{\psi_{\epsilon}}\hspace{-2pt}\inner{\psi_{\epsilon}}{L}$ and hopping strength $|t|e^{Lg}$ is comparable to the level spacing. }
\label{flux-and-decay}
\end{figure}

To expand on the connection between exceptional points and delocalization in the model \eqref{H}, it is instructive to consider the following ``gauge transformation'' \cite{NHMBL}
\begin{equation}
S = \exp\left(\sum_{j=1}^L jg n_j\right).
\label{S_gauge} 
\end{equation}
Applying this transformation to our Hamiltonian \eqref{H} with open boundary conditions ($c_{L+1} = 0$) eliminates $g$ entirely. The Hamiltonian is similar to a Hermitian operator, and thus has a real spectrum, for any $g$. Put another way, we can ``gauge away'' the imaginary vector potential. Conversely, if we have periodic boundary conditions ($c_{L+1} = c_1$) the presence of $g$ is not removed, but rather is shifted entirely to the bond between sites $L$ and $1$ (or any other two sites, by an appropriate shift of indices in \eqref{S_gauge}). In this case, the non-hermiticity cannot be removed; there is a fixed ``flux'' $iLg$ through the ring. See Fig. \ref{flux} for an illustration.

Now, let us consider the single-particle sector of the Hamiltonian \eqref{H} with open boundary conditions and $g=0$; we know that the states will be localized for any non-zero disorder strength $W$. Let us consider an eigenstate $\ket{\psi_{\epsilon}}$ at energy density $\epsilon$ \footnote{For a given eigenvalue $E$, the energy density is defined as the excitation energy of $E$ normalized by the bandwidth: $$\epsilon = \frac{E - E_{\min}}{E_{\max} - E_{\min}}$$ Here $E_{\max}$, $E_{\min}$ are maximum and minimum energies in the spectrum, respectively.} in this open chain. This eigenstate will be localized at some lattice site - call it $j$. Based on the discussion of gauge freedom above, let us introduce an imaginary flux by adding a term of the form $e^{Lg}c_{L}^{\dag}c_1 + e^{-Lg}c_1^{\dag}c_L$ to our system. Borrowing intuition from perturbation theory, we expect the non-hermiticity to become important when
\begin{equation}
\bra{\psi_{\epsilon}}|t|e^{Lg}c_{L}^{\dag}c_{1}\ket{\psi_{\epsilon}} \sim \delta,
\label{g-wavefn-decay}
\end{equation}
where $\delta$ is the level spacing in the chain. See Fig. \ref{flux-loc-length} for an illustration.

Using the localized state ansatz $\ket{\psi_{\epsilon}} \sim \sum_k e^{-|k-j|/\xi_{\epsilon}}\ket{k}$ in \eqref{g-wavefn-decay} above, one finds the exceptional point $g_c(\epsilon)$ for this eigenstate to be $g_c(\epsilon) \sim 1/\xi_{\epsilon}$ to leading order in $L$. In other words, the tilt competes directly with the localization of the underlying hermitian Hamiltonian. This fact has been studied in detail by Hatano and Nelson for the non-interacting case \cite{Hatano-Nelson-Flux,Hatano-Nelson-Eigenfunctions} and Hamazaki et al \cite{Hamazaki} for the interacting case. Hamzaki et al focused on the localization properties of the non-hermitian Hamiltonian \eqref{H} at nonzero $g$, as opposed to using the real-complex transition to probe the localization properties of the underlying $g=0$ Hamiltonian (which is what the relation $g_c \sim 1/\xi$ encodes).

The above argument, while suggestive, can be made precise. Specifically, let $\ket{\psi_n}$ denote the $n^{\text{th}}$ single-particle eigenstate of the $g=0$ Hamiltonian \eqref{H} with open boundary conditions, and let $g_{c,\, n}$ be the exceptional point for the $n^{\text{th}}$ eigenstate in the closed chain with tilt (ordered by real part of the eigenvalue). Then one can show \cite{NHMBL}
\begin{equation}
g_{c,\,n} = \frac{1}{\xi_n} + \mathcal{O}\left(\frac{1}{L}\right),
\label{g_c}
\end{equation}
where $1/\xi_n = -\frac{1}{L}\log(\bra{\psi_n}c_1^{\dag}c_L\ket{\psi_n})$ is the ``eigenstate localization length'' (equivalently, the logarithm of the end-to-end Green's function in the vicinity of $E_n$). While derived only for the single-particle sector of \eqref{H}, Ref \cite{NHMBL} showed that defining a length scale via \eqref{g_c} in interacting systems has the expected behavior of a localization length - namely, a finite-size scaling collapse of $\xi$ identifies critical parameters that agree with other numerical studies of finite-size MBL systems, and $\xi$ appears to diverge at the asymptotic transition. Thus, by searching for the exceptional points of the Hamiltonian \eqref{H} as we vary $g$, we can directly probe the localization properties of the underlying $g=0$ Hamiltonian.

\section{Connection to Avalanche Model}
\label{avalanche}

In this section, we seek to connect the localization length $\xi$ to the length scale $\lambda$ appearing in the avalanche model of delocalization. We do so in several steps.

\subsection{Hermitian Avalanche Model}
\label{herm-av}

We begin with a brief review of the avalanche model of the localization-delocalization transition \cite{DeRoeck-Huveneers-Stability-Av,Luitz-Av, Thiery-Av, Thiery-Av-Long, Johri-Env}. In this picture, localized chains are coupled weakly to a thermal bath (or thermal sub-region of the chain), which can thermalize the sites bordering it. The new, larger, ``effective'' thermal bath comprised of the original bath and its thermalized neighbors can now thermalize the subsequent sites bordering it, and this can continue, triggering an ``avalanche'' that will either halt at some point - leaving the chain (partially) localized - or thermalize the whole chain.

To be more concrete, we consider the following toy model that captures the essential physics of such avalanches \cite{DeRoeck-Huveneers-Stability-Av,Luitz-Av}. The system is a 1d chain of length $L+L_b$, with sites $-L_b+1,\ldots,0$ being a thermal ``bath'', and sites $1,\ldots, L$ being many-body localized l-bits. The Hamiltonian is
\begin{equation}
H = H_{\text{bath}} + \mathcal{G}_0\sum_{j=1}^{L} e^{-j/\lambda} [b_{0}^{\dag}\tilde{c}_{j} + \tilde{c}_{j}^{\dag}b_{0}] + \sum_{j=1}^{L} u_j \tilde{n}_j  ,%$}
\label{H_av}
\end{equation}
where $H_{\text{bath}}$ is the bath Hamiltonian (acting only on the bath sites), $b_i/b_i^{\dagger}$ are bath creation/annihilation operators, $\tilde{c}_i/\tilde{c}_i^{\dagger}/\tilde{n}_i$ are l-bit creation/annihilation/number operators, and $\{u_i\}_{i=1}^L$ are the single-particle energies associated with each l-bit. Additionally, $e^{-1/\lambda} \in (0,1]$ is the base of exponential decay of the matrix elements coupling the bath to the localized part of the chain, with $\lambda$ being the decay length of said  matrix elements. This model neglects interactions between the l-bits, which are argued to induce only higher order corrections \cite{Luitz-Av}.

In the avalanche picture of delocalization, the eigenstate thermalization hypothesis (ETH) holds for a given l-bit if and only if it sucessfuly hybridizes with the bath according to Fermi's golden rule \cite{Potirniche-Banerjee-Altman}. Symbolically, this criteria amounts to the condition $\mathcal{T} \gg \delta$, where $\delta$ is the level spacing in the bath, and $\mathcal{T}$ is a matrix element of the hopping $b_0^{\dagger}\tilde{c}_i$ between eigenstates of the unperturbed Hamiltonian $H_{\text{bath}} + \sum_{i=1}^L u_i \tilde{n}_i$. Assuming ETH for the bath, a typical matrix element between two energy eigenstates has the form $\mathcal{T} \sim \kappa\sqrt{\rho(\epsilon,\omega)\delta}$, where the spectral function $\rho$ is a smooth positive function, $\epsilon$ and $\omega$ are the average and difference of the bath energies of the eigenstates, and $\kappa = \mathcal{G}_0 e^{-j/\lambda}$ is direct coupling between the bath and l-bit $j$ \cite{DeRoeck-Huveneers-Stability-Av}. 

The l-bit most strongly coupled to the bath will be at site 1, and this l-bit hybridizes with the bath if $\mathcal{G}_0 e^{-1/\lambda} \sqrt{\rho(\epsilon,\omega)/\delta} \gg 1$. Should this l-bit hybridize, it will effectively become part of the bath; this (approximately) halves the level spacing $\delta \mapsto \delta/2$ and (approximately) leaves the spectral function unchanged: $\rho'(\epsilon,\omega) \approx \rho(\epsilon,\omega)$. One can now pose the hybridization criteria to the l-bit at site 2, with the level spacing of the ``effective'' bath half that of the original bath level spacing. Should the l-bit at site 2 hybridize with the bath, the level spacing will be further reduced by a factor of 2. This process continues at each site a distance $r$ from the bath so long as the hybridization condition
\begin{equation}
\frac{\mathcal{T}(r)}{\delta(r)} \gg 1 \quad\Longrightarrow\quad \mathcal{G}_0 e^{-r/\lambda} 2^{r/2}\sqrt{\frac{\rho(\epsilon,\omega)}{2\delta_0}}  \gg 1
\label{av_cond}
\end{equation}
holds. Here, we've taken $\delta(r) \approx \delta_0 2^{-r+1}$, with $\delta_0$ the level spacing of the original bath. 

For the entire chain to thermalize, \eqref{av_cond} should hold for all $r$. Since the left hand side of \eqref{av_cond} is monotonic, full thermalization can be determined by the behavior at the end of the chain ($r= L$). Taking logarithms, full thermalization of the chain thus amounts to
\begin{equation}
-\frac{1}{\lambda} + \frac{\log 2}{2} + \mathcal{O}\left(\frac{1}{L}\right) \gtrapprox 0 .
\label{av_cond_log}
\end{equation}
In the thermodynamic limit, we can see this condition will always be satisfied for $\lambda > 2/\log2$. We thus see the length scale $\lambda$ controls the delocalization transition, which occurs at a critical decay length $\lambda_c = 2/\log2$.

\subsection{Non-hermitian delocalization \& avalanches}
\label{avalanche-toy}

In numerical studies of non-hermitian Hamiltonians with tilt $g$ whose $g=0$ counterpart is many-body localized, it has been observed that a non-hermitian MBL regime exists for small $g$, and is eventually destroyed for large enough $g$ \cite{Hamazaki,Panda-Banerjee-NESS}. We conjecture that this non-hermitian delocalization occurrs by the same avalanche mechanism as in the hermitian case. That is, we propose that the imaginary vector potential helps couple ergodic grains to the rest of the chain in such a way as to successively delocalize neighboring sites of the chain. 

Note that Ref \cite{Hamazaki} also argues that the real-complex transition roughly coincides with the non-hermitian delocalization transition in their numerics, and that the coincidence is exact in the thermodynamic limit. Under our conjecture, the exceptional points $g_c$ thus measure the location of this avalanche-based transition (i.e., $g_c \sim \lambda_c - \lambda$).

The conjecture that a non-hermitian avalanche drives the non-hermitian delocalization transition is a highly non-trivial statement, so we first verify it on a simple toy model. To that end, we consider the following non-hermitian version of the avalanche Hamiltonian:
\begin{widetext}
\begin{equation}
H = H_{\text{bath}}+ \sum_{j=1}^{L} u_j \tilde{n}_j + \mathcal{G}_0\sum_{j=1}^{L} e^{-j/\lambda}\left[e^{jg} b_{0}^{\dag}\tilde{c}_{j} + e^{-jg} \tilde{c}_{j}^{\dag}b_{0} \right]  +\, \mathcal{G}_0\sum_{j=1}^{L} e^{-j/\lambda} \left[e^{jg} \tilde{c}_{L-j+1}^{\dag}b_{-L_b+1} + e^{-jg} b_{-L_b+1}^{\dag}\tilde{c}_{L-j+1}\right].
\label{H_av_NH}
\end{equation}
\end{widetext}
This is the Hamiltonian \eqref{H_av}, with the inclusion of a non-hermitian tilt. Note also that we include hopping from both ends of the bath, since the effect of $g$ can be gauged away for open boundary conditions.

\begin{figure}
\begin{center}
\begin{subfigure}{.22\textwidth}
\includegraphics[scale=.22]{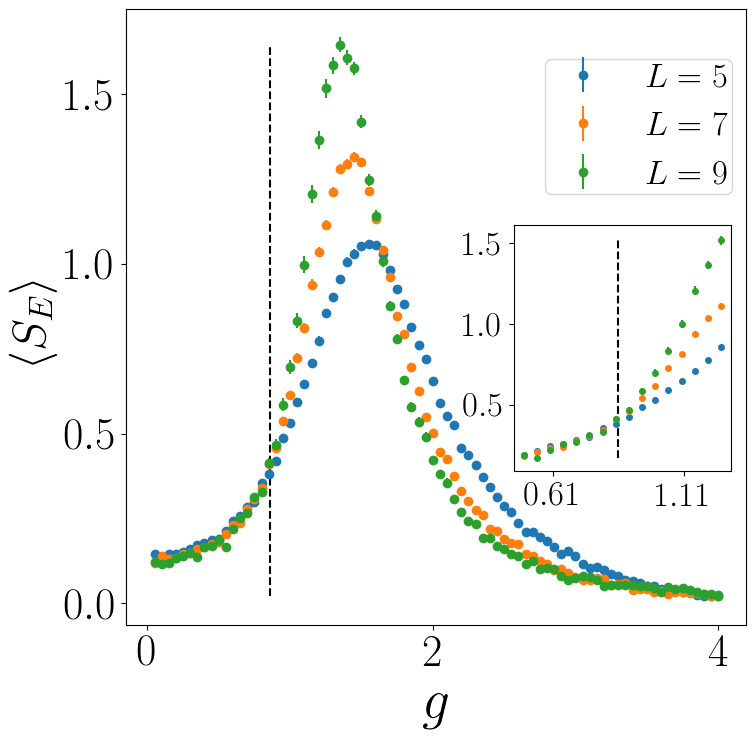}
\caption{}
\label{NH_av_numerics_S_E}
\end{subfigure}
\hfill
\begin{subfigure}{.22\textwidth}
\includegraphics[scale=.22]{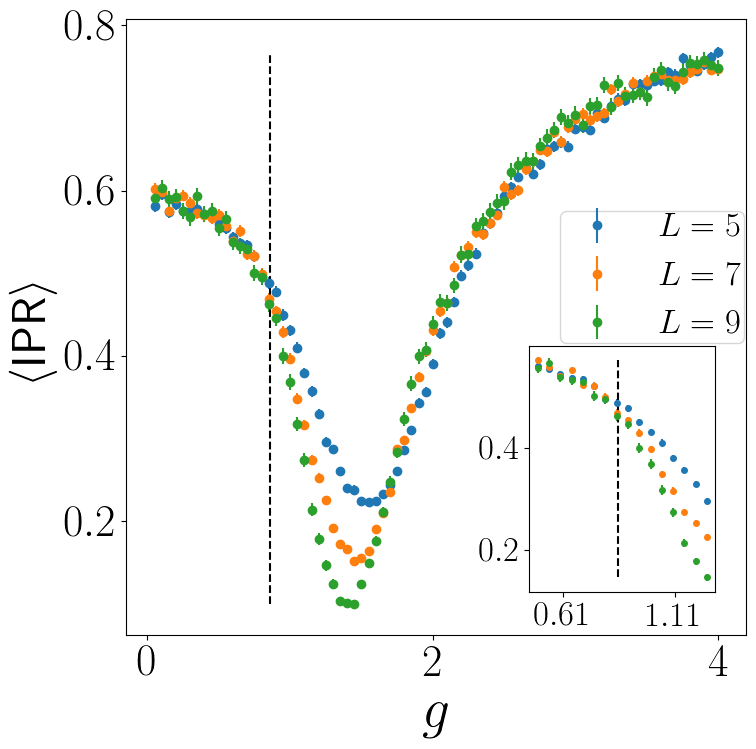}
\caption{}
\label{NH_av_numerics_IPR}
\end{subfigure}\\
\begin{subfigure}{.24\textwidth}
\includegraphics[scale=.22]{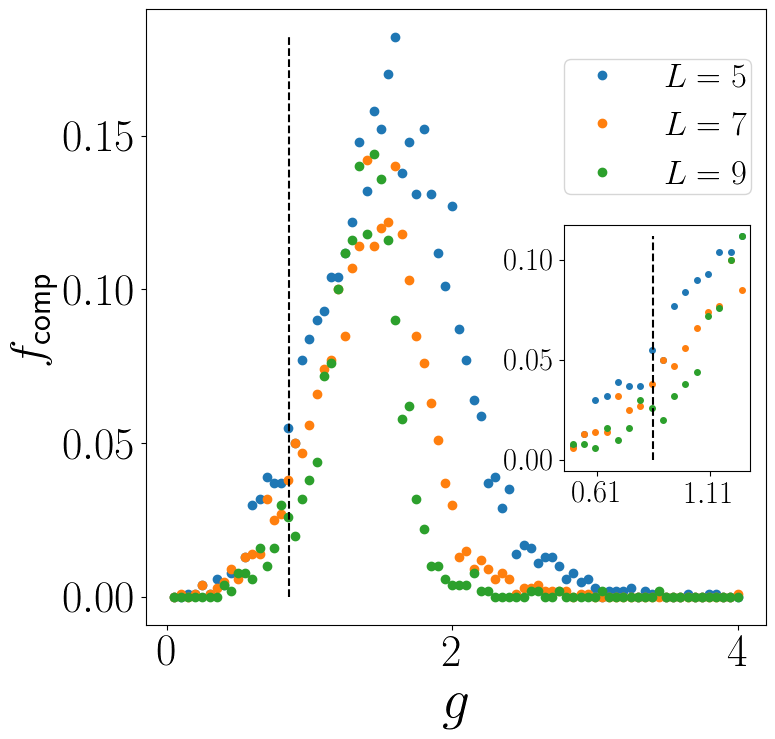}
\caption{}
\label{NH_av_numerics_compfrac}
\end{subfigure}
\end{center}
\caption{Disorder-averaged localization metrics of the eigenstates of \eqref{H_av_NH} for $e^{-1/\lambda} = 0.3$ and various $L$, as a function of $g$. Going clockwise: a) half-chain entanglement entropy $\langle S_E\rangle$, b) Inverse participation ratio $\langle IPR \rangle$, and c) Fraction $f_{\text{comp}}$ of disorder realizations whose central eigenvalue ($\epsilon = 0.5$) is non-real (note that $\avg{\cdot}$ denotes a disorder average). Error bars indicate the standard error of the mean. From a) and b), we see a crossover from localized behavior to delocalized behavior,  and from c), we see a similar real-complex crossover. Insets show a ``zoomed in'' look at the crossover region. The crossovers occurr roughly at $g = 1/\lambda - 1/\lambda_c$ (black dashed lines), the value one would expect from a non-hermitian avalanche criterion, supporting the idea that non-hermitian avalanches are responsible for these crossovers. All quantities were computed for the eigenstate at the center of the spectrum ($\epsilon = 0.5$) in the half-filling sector with $\mathcal{G}_0 = 1$, $L_b = 3$, and $H_{\text{bath}}$ an $8\times 8$ GOE matrix. The l-bit energies were drawn from a uniform distribution $u_i \sim \text{Uni}[-w,w]$ with $w=10$, and we averaged all quantities over 1000 ($L=5,7$) and 500 ($L=9$) disorder realizations. The entanglement entropies were computed using right eigenvectors only. }
%Disorder-averaged localization metrics of the eigenstates of the non-hermitian avalanche Hamiltonian \eqref{H_av_NH} for $e^{-1/\lambda} = 0.3$ and various $L$, as a function of $g$. Going clockwise: a) half-chain entanglement entropy $\langle S_E\rangle$, b) Inverse participation ratio $\langle IPR \rangle$, and c) Fraction $f_{\text{comp}}$ of disorder realizations whose central eigenvalue ($\epsilon = 0.5$) is non-real (note that $\avg{\cdot}$ denotes a disorder average). Error bars on $\langle S_E\rangle$ and $\langle IPR \rangle$ indicate the standard error of the mean. From a) and b), we see a crossover from localized behavior to delocalized behavior,  and from c), we see a similar real-complex crossover. Insets show a ``zoomed in'' look at the crossover region. These crossovers occurr roughly at $g = 1/\lambda - 1/\lambda_c$ (black dashed lines), the value one would expect from a non-hermitian avalanche criterion, supporting the idea that non-hermitian avalanches are responsible for these crossovers. All quantities were computed for the eigenstate at the center of the spectrum ($\epsilon = 0.5$) in the half-filling sector with $\mathcal{G}_0 = 1$, $L_b = 3$, and $H_{\text{bath}}$ an $8\times 8$ GOE matrix. The l-bit energies were drawn from a uniform distribution $u_i \sim \text{Uni}[-w,w]$ with $w=10$, and we averaged all quantities over 1000 ($L=5,7$) and 500 ($L=9$) disorder realizations. The entanglement entropies were computed using right eigenvectors only. }
\label{NH_av_numerics}
\end{figure}

Figure \ref{NH_av_numerics} shows numerical results from diagonalizing the Hamiltonian \eqref{H_av_NH} for several system sizes, in the half-filling sector, for $e^{-1/\lambda} = 0.3$. We can see a crossover in both the half-chain entanglement entropy and the inverse participation ratio from constant to increasing/decaying with system size (respectively), indicating a localization to delocalization crossover. These crossovers also occurr roughly at the value of $g$ we would expect from modifying the avalanche criteria \eqref{av_cond} with $e^{-r/\lambda} \mapsto e^{-r/\lambda + rg}$, i.e. $g = 1/\lambda - 1/\lambda_c$ (with $\lambda_c = 2/\log 2$ \footnote{Note that we use the value of $\lambda_c$ for the \textit{open} chain here, since we expect from the non-interacting case that the exceptional points of the ring should probe the localization of the open chain formed by removing one of the bonds.}). We also see similar crossovers for varying values of $e^{-1/\lambda}$ - see appendix \ref{appendix-NH} for additional examples. These results support the idea that a non-hermitian avalanche mechanism is driving delocalization in this model.

Figure \ref{NH_av_numerics_compfrac} also shows the fraction $f_{\text{comp}}$ of disorder realizations that have a non-real eigenvalue at energy density $\epsilon = 0.5$. Though there is no sharp transition or crossover, the window over which a non-zero fraction of disorder realizations have a complex eigenvalue overlaps with the crossover window observed in Figs. \ref{NH_av_numerics_S_E} and \ref{NH_av_numerics_IPR}. This suggests that the real-complex transition may not only coincide with the many-body localization-delocalization transition (as previously suggested in the literature), but that it is consistent with a non-hermitian avalanche mechanism.

\subsection{Mapping chain hamiltonians to avalanche hamiltonians}
\label{mapping-to-avalanche}
Having evidence that a non-hermitian avalanche drives the delocalization and real-complex transition in a simple toy model, we now would like to verify that the exceptional points $g_c(\epsilon)$ in more generic Hamiltonians such as \eqref{H} are described by a similar mechanism. To do so, we need a way to explicitly connect such Hamiltonians to ``avalanche'' Hamiltonians resembling \eqref{H_av_NH}. 

Let us consider a generic system comprised of a thermal bath/grain and a chain (which we take to be MBL). The Hamitonian is of the form
\begin{equation}
H = H_{\text{bath}} + H_{\text{chain}} + H_{\text{bc}},
\label{H_with_bath}
\end{equation}
where $H_{\text{bath}}$ and $H_{\text{chain}}$ act only on the bath and chain (respectively), and $H_{\text{bc}}$ couples the bath and chain subsystems. To bring this into the desired ``avalanche'' form, we need to re-express $H_{\text{chain}}$ and $H_{\text{bc}}$ in terms of $l$-bits. In principle, this can be done straightforwardly by diagonalizing $H_{\text{chain}}$ and performing a change of basis on $H_{\text{bc}}$. When using a generic diagonalization routine, however, it is not obvious how to extract the l-bit occupation numbers for each eigenstate. As the avalanche model relies on a cascade effect from thermalizing successive l-bits, we need to employ a diagonalization routine that allows us to access this crucial information.

\begin{figure}
\centering
\begin{subfigure}{.22\textwidth}
\includegraphics[scale=.165]{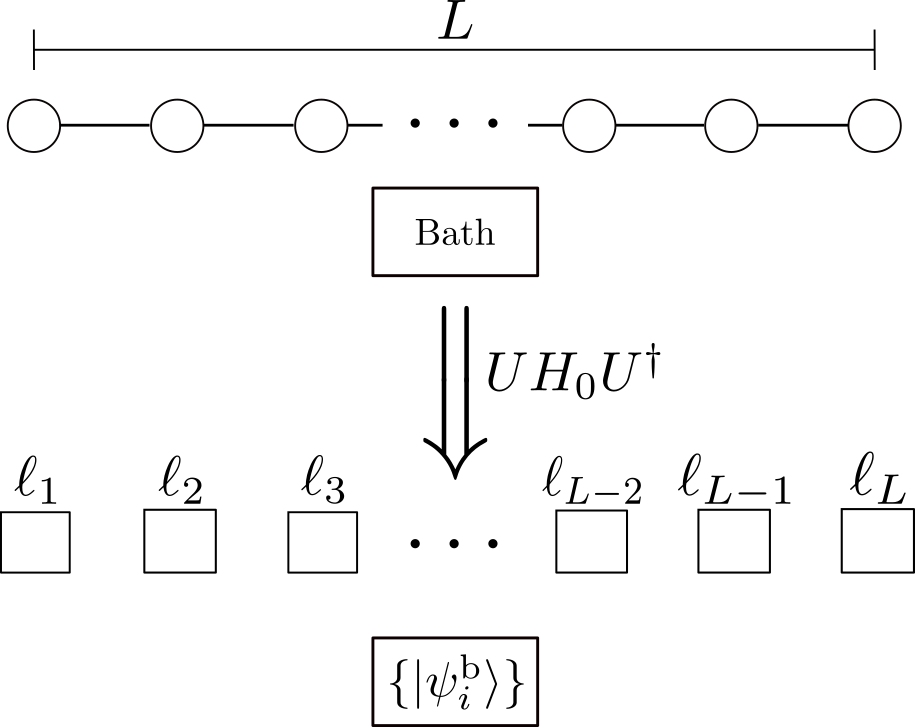}
\caption{}
\label{mapping-separate}
\end{subfigure}
\hfill
\begin{subfigure}{.22\textwidth}
\includegraphics[scale=.165]{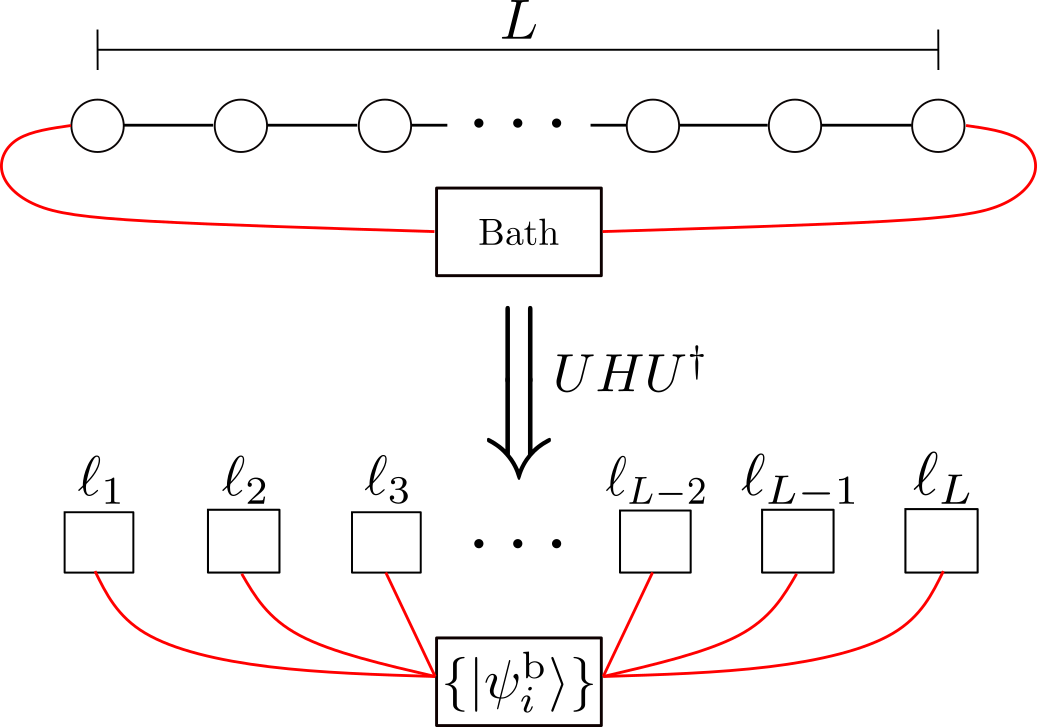}
\caption{}
\label{mapping-combined}
\end{subfigure}
\caption{A schematic illustration of how we map chains with a bath to avalanche-like Hamiltonians. a) We first diagonalize the \textit{decoupled} chain and bath system (above the arrow), by iteratively applying displacement transformations to eliminate the largest off-diagonal hopping terms (solid lines). This leaves the system in diagonal form (below arrow), with $\ket{\psi^{\text{b}}_i}$ the bath eigenstates and squares representing the l-bits $\ell_i$. b) We then apply the same transformations found in (a) to the \textit{full} Hamiltonian $H$, including the bonds between the bath and chain (red lines, above arrow). This yields avalanche-like hopping between the bath eigenstates and l-bits (red lines, below arrow). } 
\label{mapping}
\end{figure}

To achieve this, we employ a mapping that combines the ``displacement transformations'' of Rademaker and Ortu\~{n}o \cite{Rademaker-Ortuno} with the principles of Wegner-Wilson Flow \cite{Pekker-Clark-Oganesyan-Refael-Wegner-Wilson,Quito-Titum-Pekker-Refael-Wegner-Flow}. Specifically, we construct and apply displacement transformations to eliminate individual off-diagonal terms in the Hamiltonian $H_0 = H_{\text{bath}} + H_{\text{chain}}$. We iterate this procedure, repeatedly eliminating the largest remaining off-diagonal term (as in the flow equation approach) until all such terms in $H_0$ are below some tolerance - see Appendix \ref{appendix-Mapping} for details. At the end, this procedure yields similarity transformations $U_{\text{bath}}$ and $U_{\text{chain}}$ that diagonalize $H_{\text{bath}}$ and $H_{\text{chain}}$, respectively. We then apply these transformations to $H_{\text{bc}}$, which generates an avalanche-like series of terms involving hopping between bath eigenstates and the l-bits. Figure \ref{mapping} illustrates this procedure schematically.

There are several important points about this procedure that we wish to emphasize. The first is that the transformation $U_{\text{chain}}$ obtained from the Wegner-Wilson-flow like approach will well approximate the ``near-optimal'' mapping of basis states to l-bit configurations alluded to in Section \ref{intro} \cite{Pekker-Clark-Oganesyan-Refael-Wegner-Wilson}. Thus, we can read off the l-bit occupation numbers from the physical site occupation numbers in the original computational basis. The second is that by completely diagonalizing the decoupled bath/chain system $H_0$ - rather than diagonalizing only $H_{\text{chain}}$ - we have access to the spectrum of the decoupled system. We can therefore label each l-bit/eigenstate hopping by its energy density in the uncoupled system, allowing for direct comparison with $g_c$ at the same energy density. The final key point is that, due to the gauge freedom, we can take $H_0$ to be hermitian without loss of generality. More specifically, we can gauge all of the flux onto the coupling Hamiltonian $H_{\text{bc}}$ (i.e., the red bonds in Fig \ref{mapping-combined}), and perform our mapping on $H_0$ with $g=0$. The transformations $U_{\text{bath}}$ and $U_{\text{chain}}$ will be unitary, and significantly easier to construct. The influence of the imaginary flux is also now simply a multiplicative weight for the various terms generated from $H_{\text{bc}}$ by the change of basis (rather than being part of the transformations $U_{\text{bath}}$ and $U_{\text{chain}}$, as it would be otherwise). We will leverage these last two points in the next two sections.

%As a final remark, we note this mapping provides a useful indicator of the MBL to ergodic transition. Namely, we observe that the ratio of the hopping amplitudes between consecutive l-bits (which is itself a random variable due to the quenched disorder) has a distribution which is peaked at zero and one in the MBL and ergodic phases (respectively), and is uniformly distributed in the crossover regime - see Appendix \ref{appendix-Amplitudes} for discussion and examples.

\subsection{Applying the mapping}
\label{mapping-results}
Let us now apply the mapping described in the previous section to a concrete model describing our system of interest. To that end, we take $H_{\text{bath}}$ to be a $2^{L_\text{b}} \times 2^{L_\text{b}}$ GOE matrix (where $L_{\text{b}}$ is the bath size) and
\begin{subequations}
\begin{equation}
\begin{split}
 H_{\text{chain}} =& \displaystyle \sum_{i=1}^{L-1}\left[c_{i}^{\dag}c_{i+1} + c_{i+1}^{\dag}c_i\right] \\ 
 &+ \sum_{i=1}^L w_i n_i + U\sum_{i=1}^{L-1}n_in_{i+1} ,
\end{split}
\label{H_chain}
\end{equation}
\begin{equation}
 \begin{split}
H_{\text{bc}} =& b_0^{\dag}c_1 + c_1^{\dag}b_0 + c_L^{\dag}b_{-L_b+1} + b_{-L_b+1}^{\dag}c_L ,
\end{split}
\label{H_bc}
\end{equation}
\label{H_av_2}
\end{subequations}
where $b_i$, $c_i$ are bath/chain fermion operators (respectively), $n_i$ are chain number operators,  $U > 0$ is the interaction strength, the $w_i$ are independently and identically drawn from a distribution characterized by strength $W$, and $L$ is the length of the chain. Note that we have taken the bare hopping strength $t$ (c.f. the Hamiltonian \eqref{H}) to be unity, so that all quantities are in units of $t$. Additionally, per the discussion of the previous section, we apply the mapping to a hermitian Hamiltonian without the vector potential $g$; we can restore the influence of $g$ by placing a flux $i(L+1)g$ onto the bonds in and out of the bath.

\begin{figure}
\centering
\begin{subfigure}{.45\textwidth}
\includegraphics[scale=0.25]{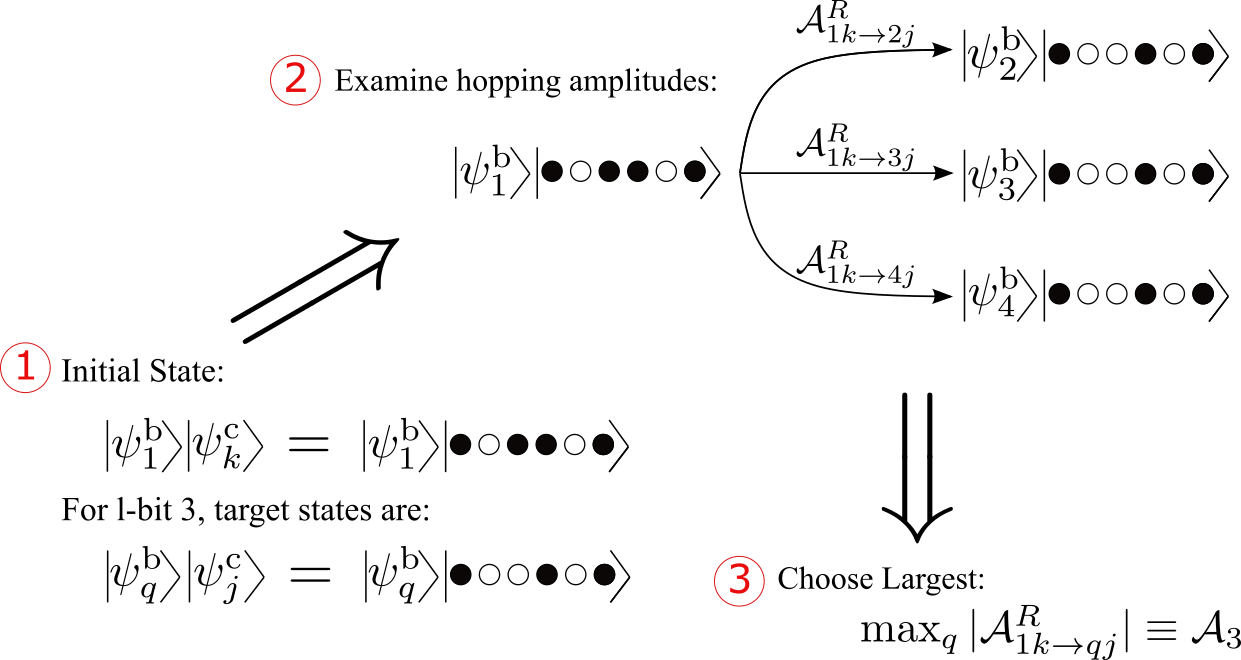}
\caption{}
\label{amplitude_choosing}
\end{subfigure}
\\
\begin{subfigure}{.45\textwidth}
\includegraphics[scale=0.2]{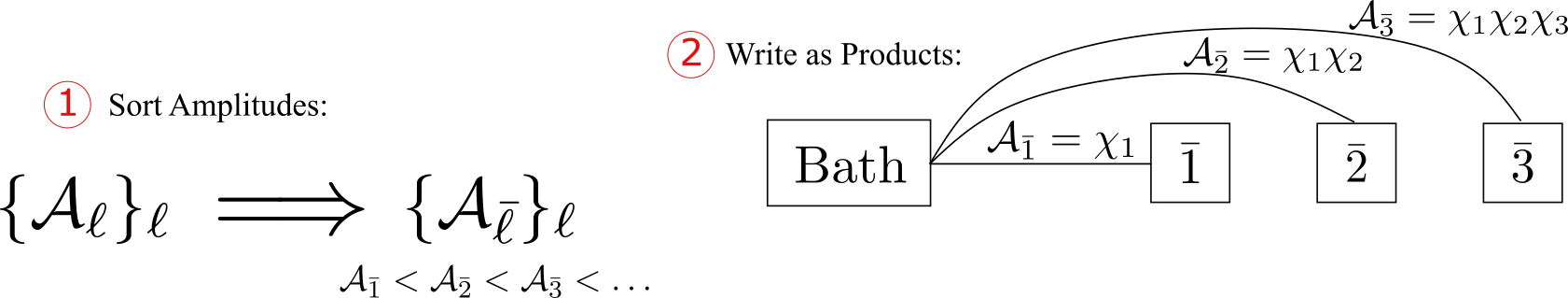}
\caption{}
\label{chi_definition}
\end{subfigure}

\caption{Schematic examples illustrating how we extract the $A_{\ell}$ and $\chi_i$'s in a small chain. a) An example of how to choose the amplitude $\mathcal{A}_{\ell}$ for $\ell = 3$. Starting from an initial state $\ket{\psi^{\text{b}}_1}\hspace{-2pt}\ket{\psi^{\text{c}}_k}$ with $\ket{\psi^{\text{c}}_k}$ having l-bit 3 occupied, we identify the set of target states (Step 1). These are the states of the form $\ket{\psi^{\text{b}}_q}\hspace{-2pt}\ket{\psi^{\text{c}}_j}$, where $\ket{\psi^{\text{b}}_q}$ is any bath eigenstate and $\ket{\psi^{\text{c}}_j}$ has the same l-bit configuration as $\ket{\psi^{\text{c}}_k}$ except for l-bit 3, which is now unoccupied. We then examine the matrix elements connecting the initial state to all possible target states (Step 2). We choose the amplitude with the largest magnitude, and assign its magnitude to be $\mathcal{A}_{3}$ (Step 3). We can then repeat this process for different initial states and l-bits $\ell$. In our computations, the initial states are chosen by energy density $\epsilon$, so we label the resulting amplitudes by energy density as $\mathcal{A}_{\ell}(\epsilon)$. b) An example of how we define the $\chi_i$'s. We first sort the $\mathcal{A}_{\ell}$'s (for a given initial state) in descending order (Step 1). We then write each (sorted) amplitude as a product $\mathcal{A}_{\bar{\ell}} = \prod_{i=1}^{\ell} \chi_i$ such that $\mathcal{A}_{\overline{\ell+1} }= \chi_{\ell+1}\mathcal{A}_{\bar{\ell}}$ (Step 2). Note that we write the l-bit indices with bars to emphasize the fact that the amplitudes have been sorted - we drop such notation in the main text for brevity.}
\end{figure}

For each disorder realization, after applying the mapping, we need to extract the decay of matrix elements coupling each l-bit to the bath. To do so, we first separate the bath-chain coupling $H_{\text{bc}}$ into couplings from the left and right end of the bath: $H_{\text{bc}}^L = c_L^{\dag}b_{-L_b+1} + b_{-L_b+1}^{\dag}c_L$ and $H_{\text{bc}}^R = b_0^{\dag}c_1 + c_1^{\dag}b_0$, respectively. In terms of the \textit{uncoupled} eigenstates $\ket{\psi^{\text{b}}_i}\hspace{-2pt}\ket{\psi^{\text{c}}_j}$ (where $\ket{\psi^{\text{b}}_i}$, $\ket{\psi^{\text{c}}_j}$ are eigenstates of $H_{\text{bath}}$ and $H_{\text{chain}}$, respectively), we obtain matrix elements: 
$$\mathcal{A}^{s}_{kl \rightarrow ij} = \bra{\psi^{\text{b}}_i}\hspace{-2pt}\bra{\psi^{\text{c}}_j}H_{\text{bc}}^s \ket{\psi^{\text{b}}_k}\hspace{-2pt}\ket{\psi^{\text{c}}_l}$$%\bra{\Psi_{ij}}H_{\text{bc}}^s\ket{\Psi_{kl}} .$$
We wish to extract specific matrix elements for each l-bit $\ell$ and energy density $\epsilon$. To that end, we choose $k = k_{\epsilon}$ and $l = l_{\epsilon}$ above such that $\ket{\psi^{\text{b}}_{k_{\epsilon}}}\hspace{-2pt}\ket{\psi^{\text{c}}_{l_{\epsilon}}}$ is the eigenstate closest to energy density $\epsilon$ in the uncoupled system. Then, for each l-bit $\ell$, we define the amplitude $\mathcal{A}_{\ell}(\epsilon)$ via
\begin{equation}
\scalebox{.84}{$\displaystyle
\mathcal{A}_{\ell}(\epsilon) = \begin{cases} \max_i \big{\lvert}\bra{\psi^{\text{b}}_i}\left(\bra{\psi^{\text{c}}_{l_{\epsilon}}}\tilde{c}^{\dag}_{\ell}\right)H^R_{\text{bc}}\ket{\psi^{\text{b}}_{k_{\epsilon}}}\hspace{-2pt}\ket{\psi^{\text{c}}_{l_{\epsilon}}}\big{\rvert} & \ell \text{ occupied} \\ \vspace{2pt}
\max_i \big{\lvert}\bra{\psi^{\text{b}}_i}\left(\bra{\psi^{\text{c}}_{l_{\epsilon}}}\tilde{c}_{\ell}\right)H^L_{\text{bc}}\ket{\psi^{\text{b}}_{k_{\epsilon}}}\hspace{-2pt}\ket{\psi^{\text{c}}_{l_{\epsilon}}}\big{\rvert} & \ell \text{ unoccupied} \end{cases} .$}
\label{A_ell}
\end{equation}
%i.e., we find the largest (by magnitude) hopping connecting $\ket{\Psi_{k_{\epsilon}l_{\epsilon}}}$ to a state with l-bit $\ell$ unoccupied and all other l-bit occupations the same. For each \textit{uncoccupied} l-bit $\ell$, we define analogously:
%\begin{equation}
%\mathcal{A}_{\ell}(\epsilon) = \max_i \bigg{\lvert}\bra{\phi_i}\left(\bra{\psi_{l_{\epsilon}}}\tilde{c}_{\ell}\right)H^L_{\text{bc}}\ket{\Psi_{k_{\epsilon}l_{\epsilon}}}\bigg{\rvert}
%\label{A_ell_unocc}
%\end{equation}
%i.e., we find the largest (by magnitude) hopping connecting $\ket{\Psi_{k_{\epsilon}l_{\epsilon}}}$ to a state with l-bit $\ell$ occupied, and all other l-bit occupations the same.  Note that the choice of $H_{\text{bc}}^R$ in \eqref{A_ell_occ} and $H^L_{\text{bc}}$ in \eqref{A_ell_unocc} reflects the fact that the non-hermitian tilt $g$ enhances left hopping, and suppresses right hopping. 
In words, we choose the largest (by amplitude) matrix element conecting  $\ket{\psi^{\text{b}}_{k_{\epsilon}}}\hspace{-2pt}\ket{\psi^{\text{c}}_{l_{\epsilon}}}$ to a state with the occupation of l-bit $\ell$ flipped and all other l-bit occupations the same - see Fig. \ref{amplitude_choosing} for a schematic example. Note that the choice of $H^R_{\text{bc}}$ versus $H^{L}_{\text{bc}}$ reflects the fact that hopping left (right) is enhanced (suppressed) when we eventually restore the imaginary vector potential $g$; hence, we choose the matrix element that will be enhanced by a factor of $e^{(L+1)g}$ in each case above.

In the event that it is impossible to find such a hopping (if, for example, l-bit $\ell$ is unoccupied but the bath is empty), then we examine the eigenstate with energy density next closest to $\epsilon$.

\begin{figure}
\centering
\begin{subfigure}{.23\textwidth}
\includegraphics[scale=.275]{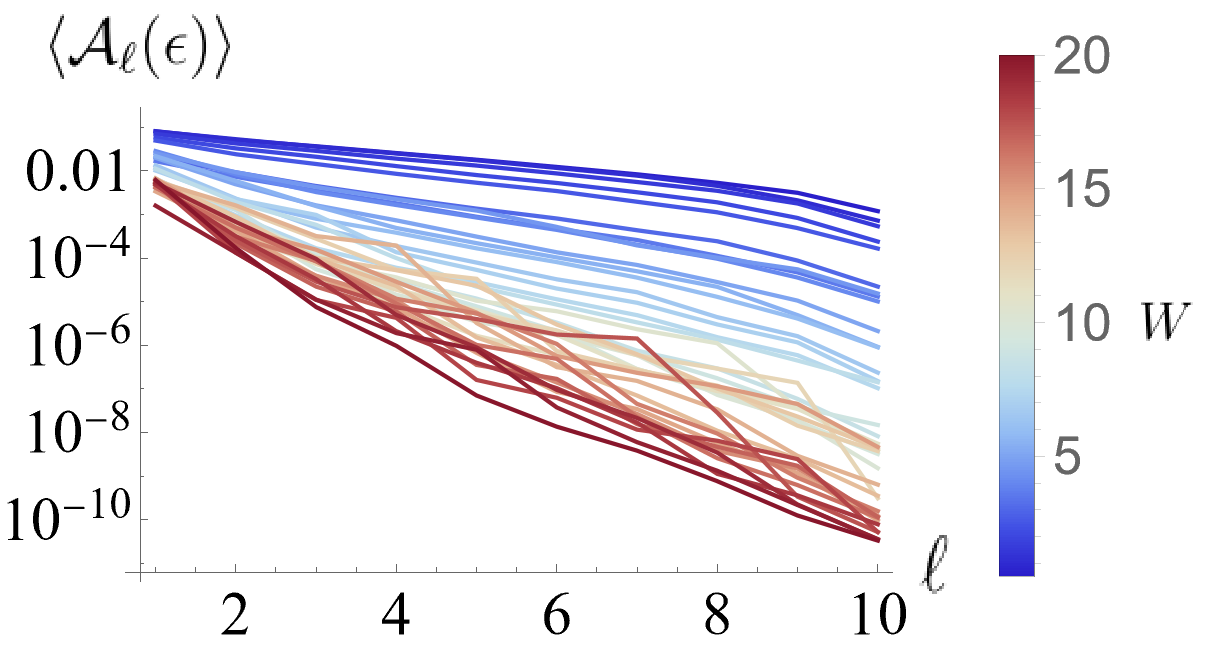}
\caption{}
\label{amp-means-U=1}
\end{subfigure}
\hfill
\begin{subfigure}{.23\textwidth}
\includegraphics[scale=.275]{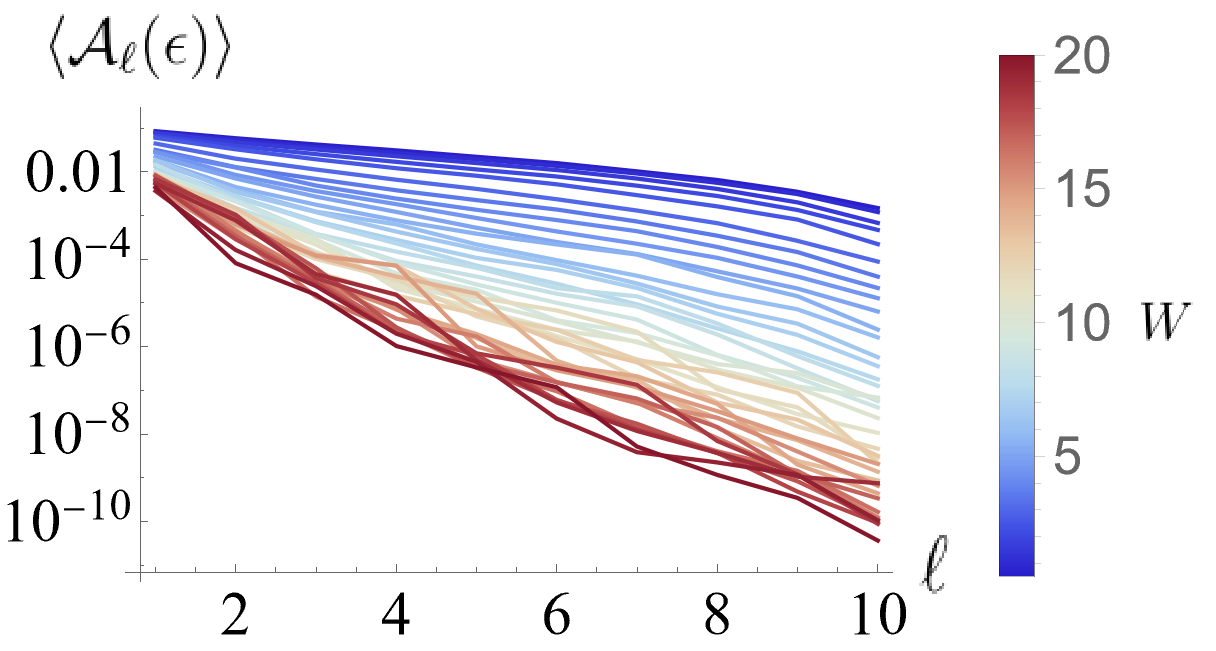}
\caption{}
\label{amp-means-U=2}
\end{subfigure}
\\
\begin{subfigure}{.23\textwidth}
\includegraphics[scale=.275]{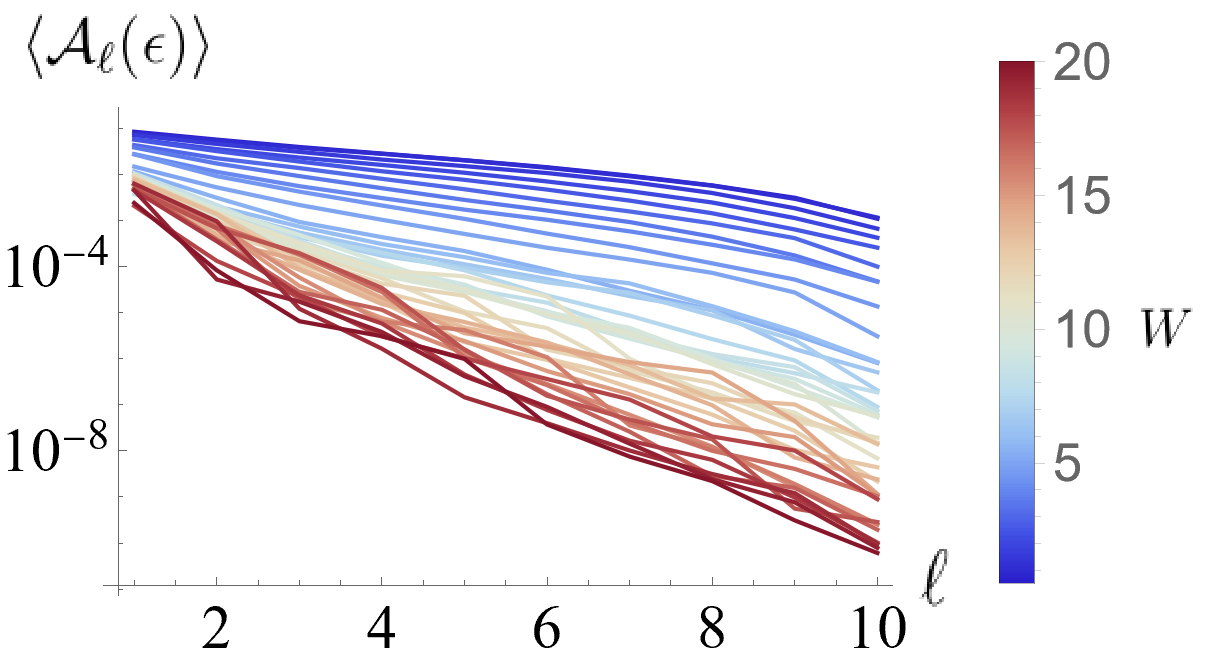}
\caption{}
\label{amp-means-U=3}
\end{subfigure}
\hfill
\begin{subfigure}{.23\textwidth}
\includegraphics[scale=.275]{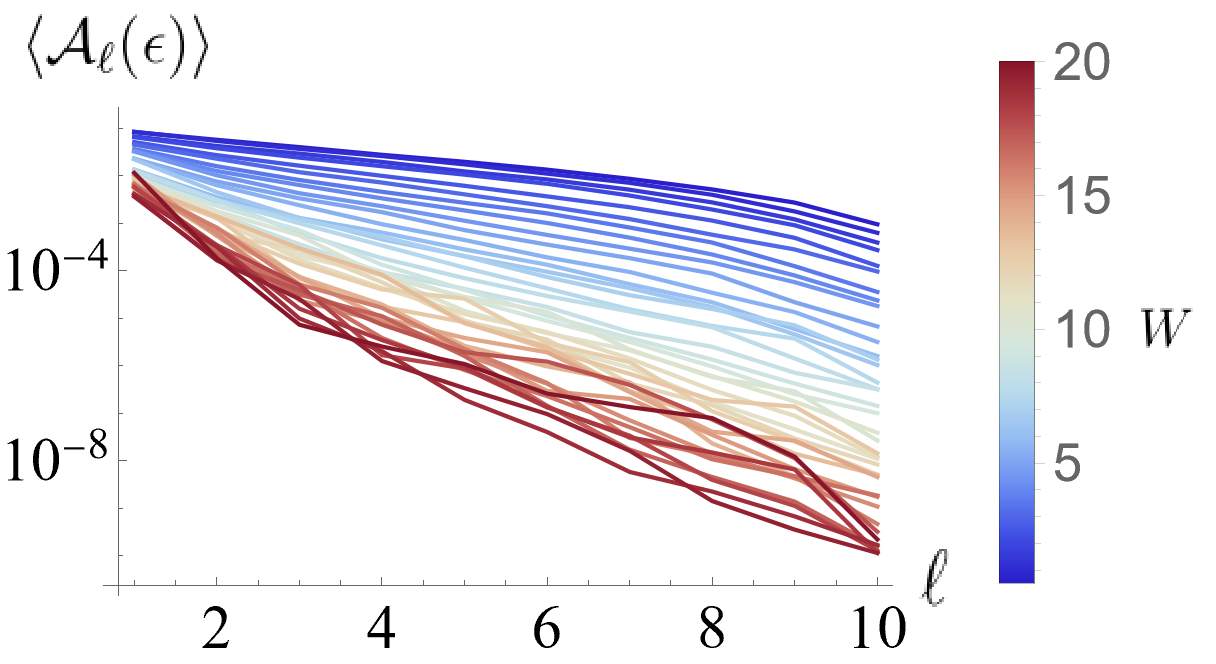}
\caption{}
\label{amp-means-U=4}
\end{subfigure}
\\

\begin{subfigure}{.23\textwidth}
\includegraphics[scale=.275]{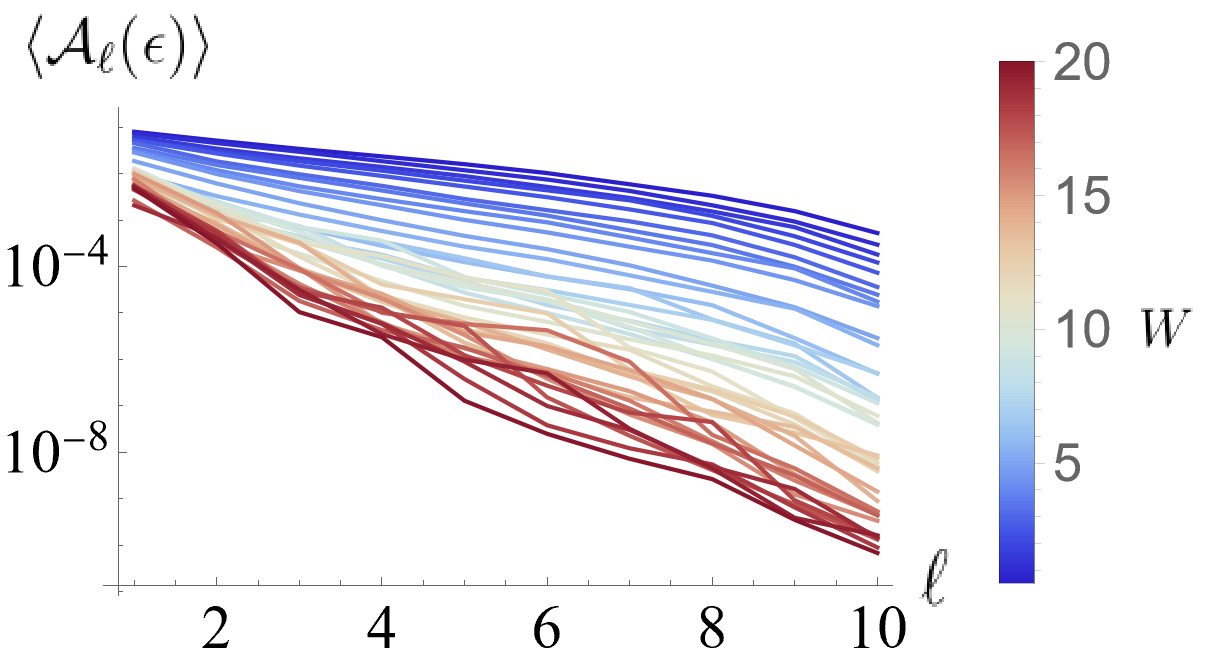}
\caption{}
\label{amp-means-U=5}
\end{subfigure}
\hfill\\

\caption{ Decay of $\avg{\mathcal{A}_{\ell}(\epsilon)}$ (with $\avg{\cdot}$ denoting disorder average) versus $\ell$ at the center of the spectrum ($\epsilon = 0.5$), for interaction strengths a) $U=1$, b) $U=2$, c) $U = 3$, d) $U=4$, e) $U=5$. Within each panel, each line traces out the decay of $\avg{\mathcal{A}_{\ell}(\epsilon)}$ for a given disorder strength $W$, with the color of the line indicating the value of $W$. We see that the average amplitudes appear to decay exponentially for all disorder strengths, with the rate of decay increasing with the disorder strength. All quantities are computed with $L = 11$, $L_b = 3$, in the $N = 3$ sector, and the averages are over 500 disorder realizations.}
\label{amp-means}
\end{figure}

In the context of an avalanche mechanism, we are interested in studying how the couplings $\mathcal{A}_{\ell}(\epsilon)$ decay for each energy density. To that end, we assume without loss of generality that the $\mathcal{A}_{\ell}(\epsilon)$ are sorted in $\ell$ in descending order. The amplitudes $\mathcal{A}_{\ell}(\epsilon)$, as well as their decay  in $\ell$, will generically be random (as they depend on the random disorder) - this is in stark contrast with the \textit{deterministic} decay of the canonical avalanche model discussed in Section \ref{herm-av}. We do, however, observe that upon averaging over disorder realizations, the amplitudes decay exponentially in $\ell$ - see Fig. \ref{amp-means} for examples of how this is borne out in chains of size $L=11$, $L_{\text{b}} = 3$, in the $N=3$ occupation sector \footnote{The mapping described in section \ref{mapping-to-avalanche} is computationally very expensive to run, so we use a smaller filling factor here in order to reach larger system sizes while keeping the Hilbert space dimension roughly fixed. We have verified that we attain qualitatively similar results at half-filling with smaller system sizes.}. Consequently, the canonical avalanche model of Section \ref{herm-av} emerges from our results upon disorder averaging. 

\begin{figure}
\centering
\begin{subfigure}{.23\textwidth}
\includegraphics[scale=.275]{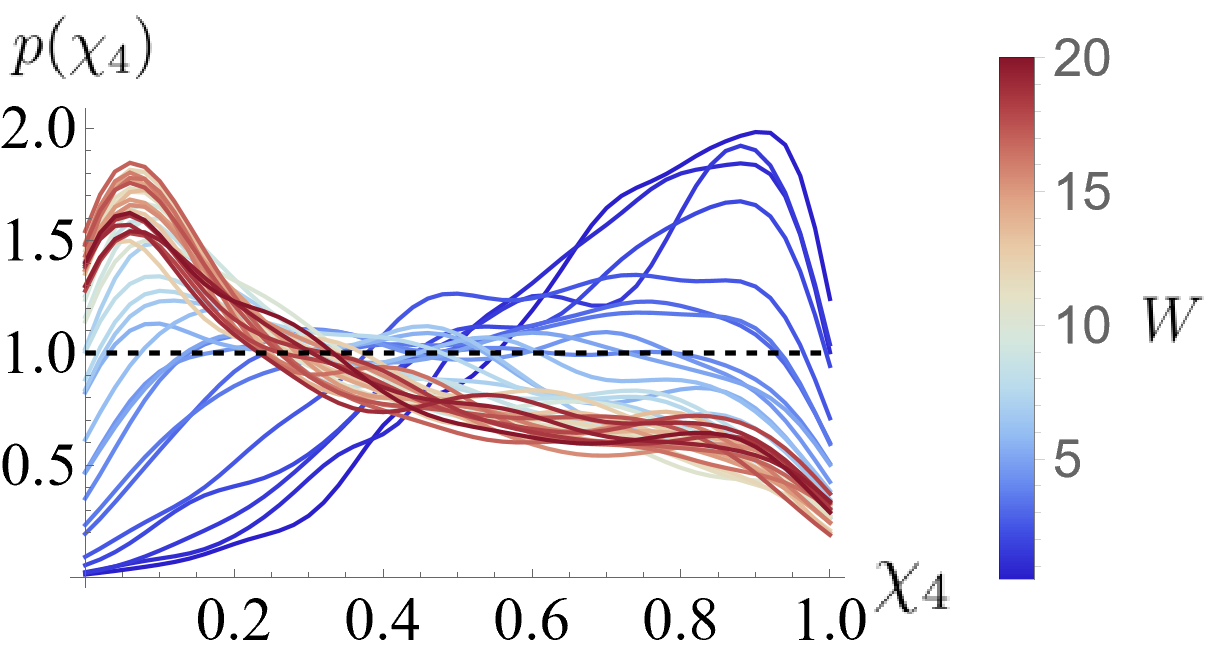}
\caption{}
\label{manyWs-U=1}
\end{subfigure}
\hfill
\begin{subfigure}{.23\textwidth}
\includegraphics[scale=.275]{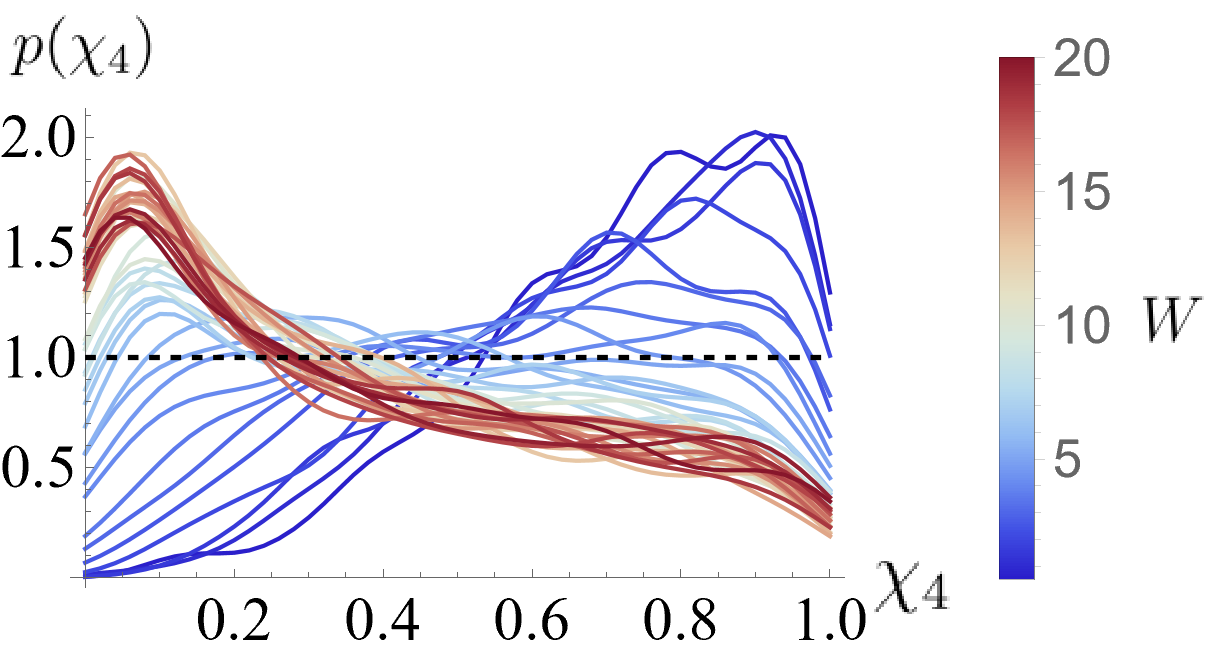}
\caption{}
\label{manyWs-U=2}
\end{subfigure}
\\
\begin{subfigure}{.23\textwidth}
\includegraphics[scale=.275]{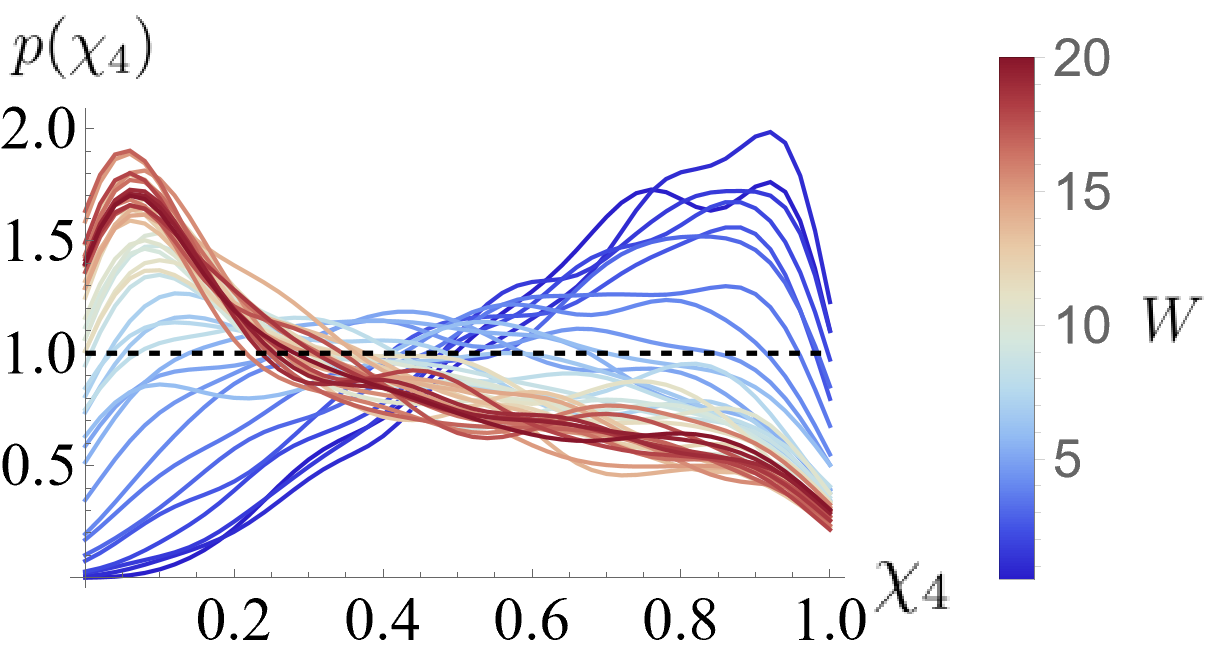}
\caption{}
\label{manyWs-U=3}
\end{subfigure}
\hfill
\begin{subfigure}{.23\textwidth}
\includegraphics[scale=.275]{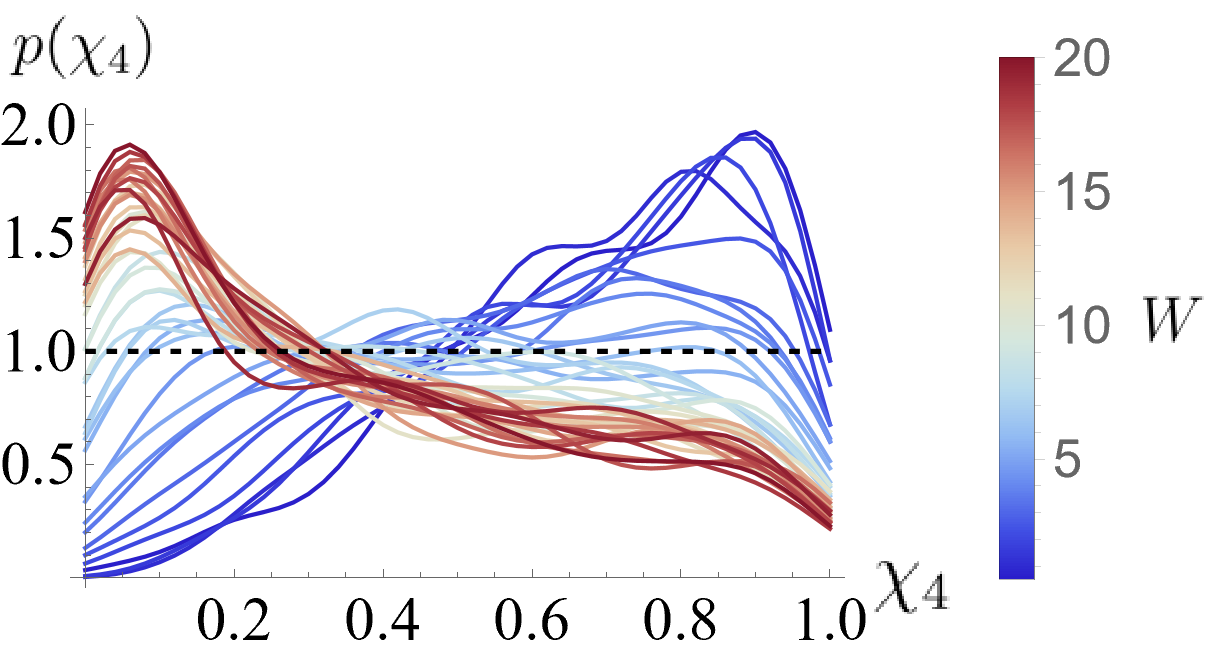}
\caption{}
\label{manyWs-U=4}
\end{subfigure}
\\

\begin{subfigure}{.23\textwidth}
\includegraphics[scale=.275]{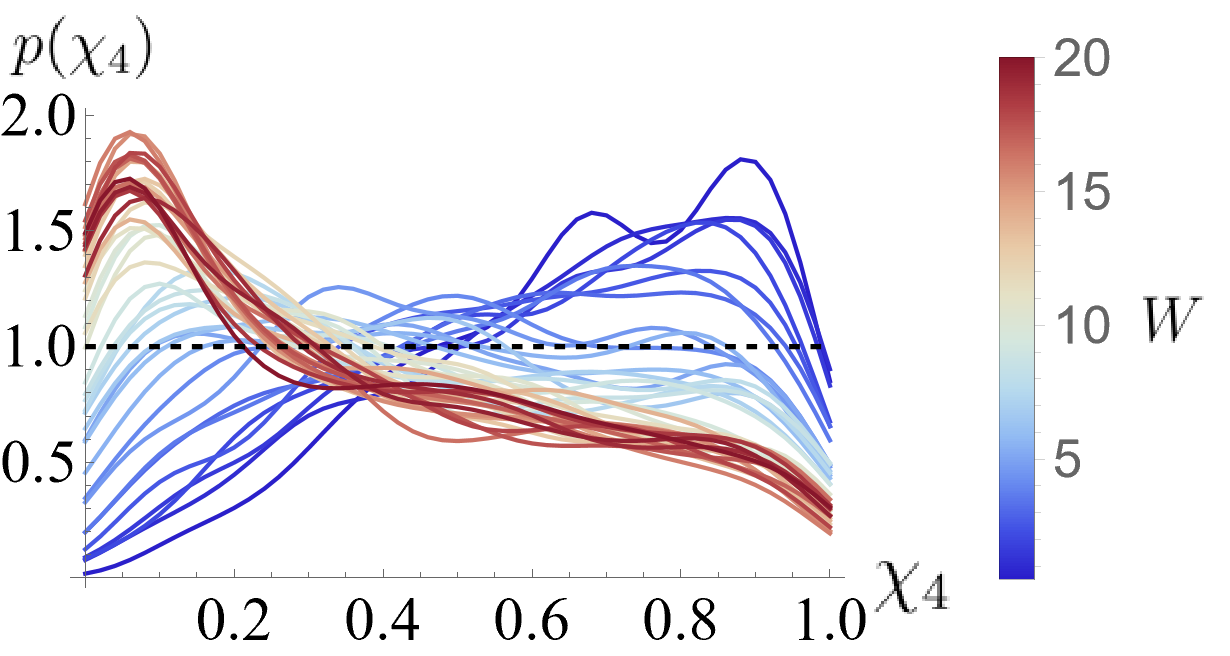}
\caption{}
\label{manyWs-U=5}
\end{subfigure}
\hfill\\

\caption{ Distributions of $\chi_4(\epsilon)$ (with respect to disorder realizations) at $\epsilon = 0.5$ for a variety of $W$, for a) $U=1$, b) $U=2$, c) $U = 3$, d) $U=4$, e) $U=5$. Each line is the observed probability distribution $p(\chi_4)$ (smoothed using a Gaussian kernel density estimator) for a given disorder strength, with the color of the line indicating the disorder strength. The dashed line traces out a uniform distribution for comparison. We see a qualitative difference between the distributions as we tune from the ergodic phase (dark blue) to deep in the MBL phase (dark red); the distributions cross over from peaked at $1$ to peaked near zero, with an intermediate critical regime of being approximately uniform. All quantities are computed with $L = 11$, $L_b = 3$, in the $N = 3$ sector, and the distributions are over 500 disorder realizations.}
\label{amps-manyWs}
\end{figure}

\begin{figure}
\begin{subfigure}{.23\textwidth}
\includegraphics[scale=.3]{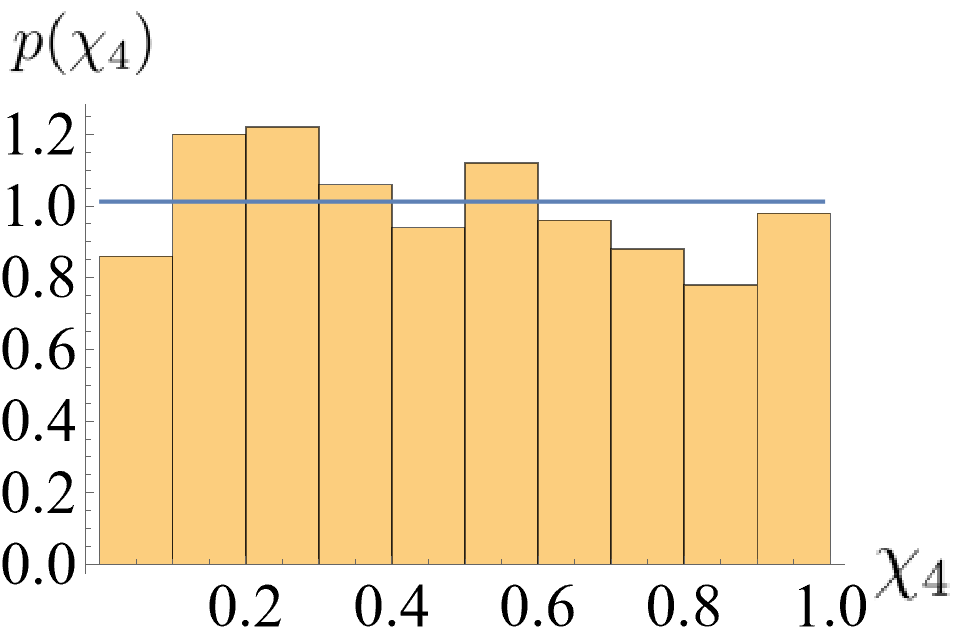}
\caption{}
\label{site-4}
\end{subfigure}
\hfill
\begin{subfigure}{.23\textwidth}
\includegraphics[scale=.3]{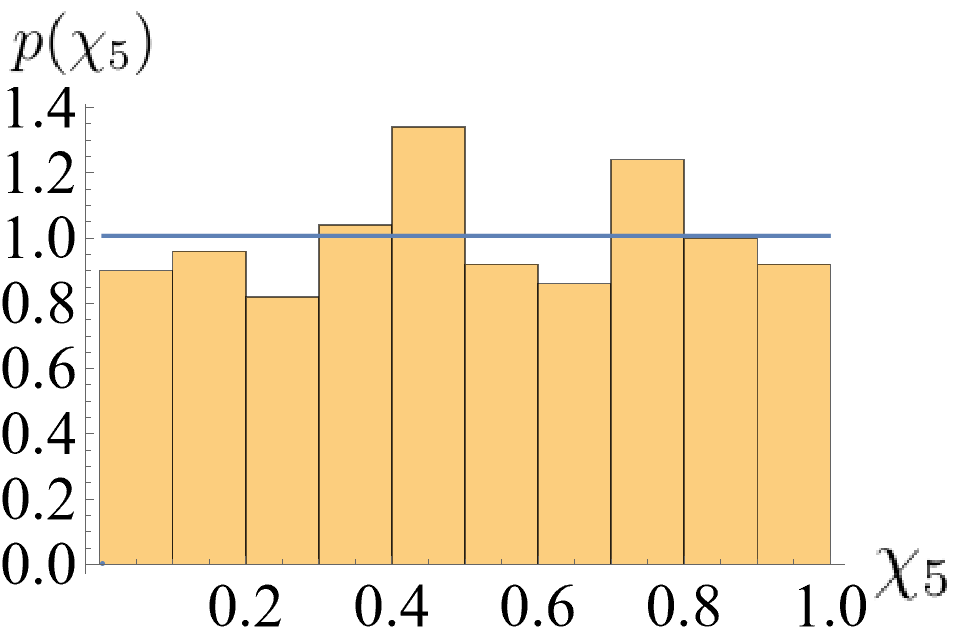}
\caption{}
\label{site-5}
\end{subfigure}
\\
\begin{subfigure}{.23\textwidth}
\includegraphics[scale=.3]{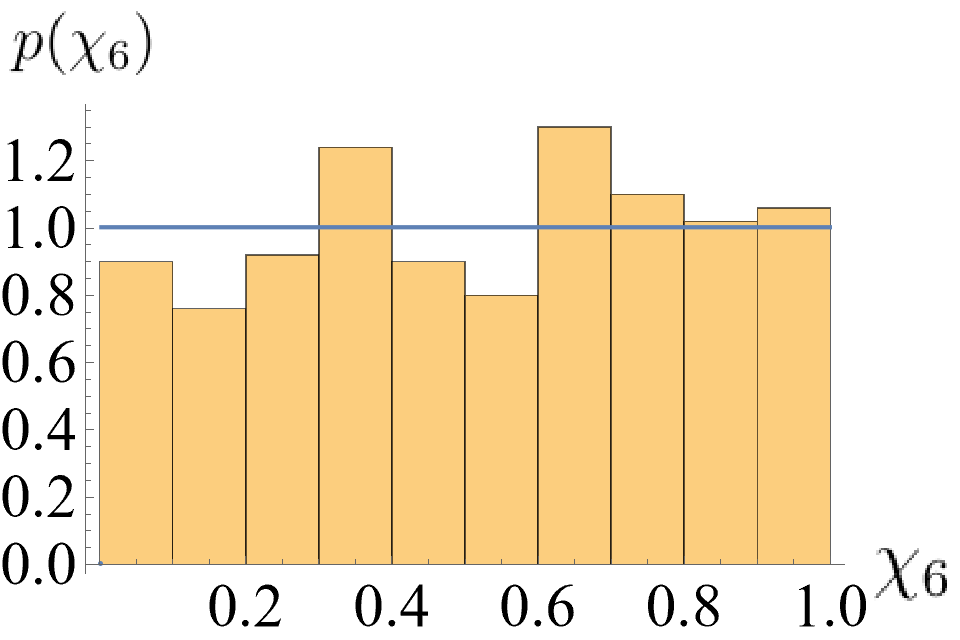}
\caption{}
\label{site-6}
\end{subfigure}
\hfill
\begin{subfigure}{.23\textwidth}
\includegraphics[scale=.3]{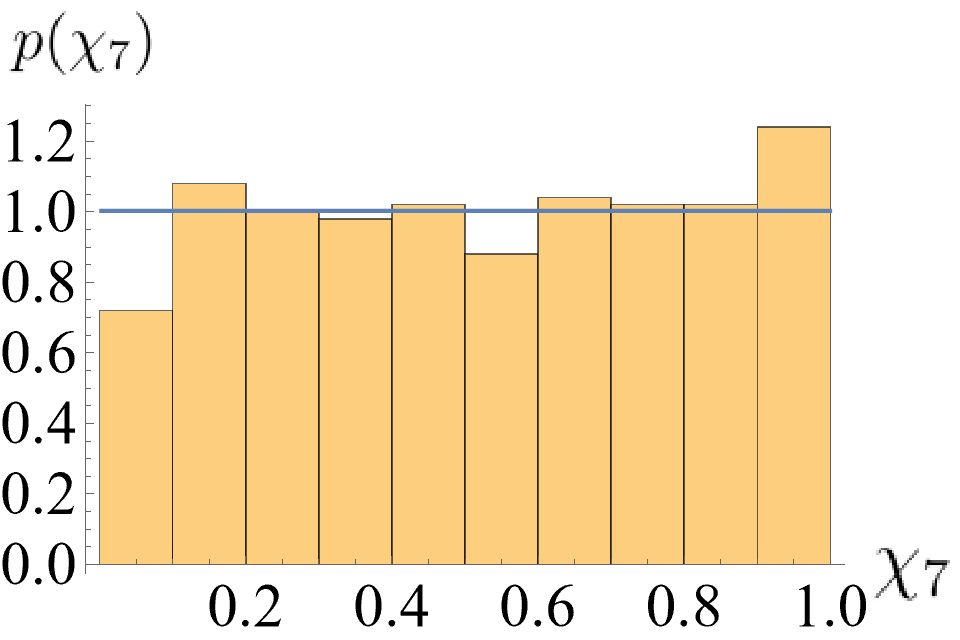}
\caption{}
\label{site-7}
\end{subfigure}

\caption{Histogram of amplitude ratios $\chi_{\ell}(\epsilon) = \mathcal{A}_{\ell}(\epsilon)/\mathcal{A}_{\ell-1}(\epsilon)$ for energy density $\epsilon = 0.5$ and interaction/disorder strength $U=1$, $W=4.8$. Shown are the histograms for a) $\ell = 4$, b) $\ell = 5$, c) $\ell = 6$, d) $\ell = 7$. Blue lines overlaid are the ``best fit'' distributions minimizing the Akaike information (see text). The best fit is a uniform distribution in all cases, suggesting $\chi_i \sim \text{Uni}[0,1]$ at the finite-size crossover. All quantities are computed with $L = 11$, $L_b = 3$, in the $N = 3$ sector, and the distributions are over 500 disorder realizations.} 
\label{amps-crit}
\end{figure}

This mapping allows us to go beyond studying the disorder-averaged amplitudes - in particular, we have access to the distributions of $\mathcal{A}_{\ell}(\epsilon)$ and their decay in $\ell$. To that end, we define the quantities
\begin{equation}
\chi_{\ell}(\epsilon) = \frac{\mathcal{A}_{\ell}(\epsilon)}{\mathcal{A}_{\ell-1}(\epsilon)} ,
\label{chi_ell}
\end{equation}
where we take $\mathcal{A}_0(\epsilon) = 1$. By construction, we have $\mathcal{A}_{\ell}(\epsilon) = \prod_{i=1}^{\ell}\chi_i(\epsilon)$, with $0 \leq \chi_i(\epsilon) \leq 1$ for each $\ell$ (as the $\mathcal{A}_{\ell}(\epsilon)$ are sorted to be monotonically decreasing in $\ell$). See Fig. \ref{chi_definition} for a schematic illustration of the $\chi_i$'s.  Such a decomposition amounts to replacing the \textit{deterministic} decay $e^{-1/\lambda}$ in \eqref{H_av_NH} by the \textit{random}, energy dependent $\chi_{\ell}(\epsilon)$. The distributions of these $\chi_{\ell}(\epsilon)$ show very interesting behavior as we bring the system from MBL to ergodic. Fig.  \ref{amps-manyWs} shows the observed probability densities of $\chi_{\ell}(\epsilon)$ (with $\ell = 4$) as we tune the disorder strength from deep in the MBL regime to ergodic. In the MBL and thermal regimes, the distributions are peaked near 0 and 1, respectively. This is consistent with the intuition that the amplitudes should decay slowly in the thermal regime and quickly in the MBL regime (which we observe to be the case on average in Fig. \ref{amp-means}). Significantly, there is a window between the two extreme regimes where the distributions are approximately uniform. To make this statement more precise, we ``fit'' the numerically observed probability distributions in the vicinity of the MBL-thermal crossover, by minimizing the Akaike information over a large class of known continuous probability distributions. The Akaike information is defined as
\begin{equation}
I_{\text{A}} = 2k - \log \hat{L} ,
\label{AIC}
\end{equation}
where $k$ is the number of parameters in the model, and $\hat{L}$ is the likelihood function. We choose to minimize the Akaike information, as it acts as an \textit{unbiased} estimator for the expected Kullback-Leibler divergence (whereas the Bayesian information is a \textit{biased} estimator), a standard measure for the information difference between two probability density functions \cite{Burnham-Anderson-AIC-BIC}. Fig \ref{amps-crit} shows the numerically observed distributions of $\chi_{\ell}(\epsilon)$ for several $\ell$ at the center of the spectrum for $U=1$ and $W = 4.8$ (the critical point for the chain Hamiltonian \eqref{H_chain} - see Appendix \ref{appendix-FSS}), along with the results of carrying out this fit. In all cases, the best-fit distribution is a uniform distribution on $[0,1]$. Though we only show results for the center of the spectrum for $U=1$, we find similar distributions at the transition throughout the spectrum, as well for $U=2,3,4,5$. This observation that the amplitudes $\chi_{\ell}(\epsilon)$ are uniformly distributed at the crossover will be key in the subsequent sections when we derive expressions for the distribution of $g_c$ at the critical point.

As a final note, the behavior of the distributions observed in Figures \ref{amps-manyWs} and \ref{amps-crit} is generic throughout the spectrum for $ 2 \leq \ell \lessapprox 10$ (though the peak of the distributions at 0 in the MBL phase is less pronounced for $\ell \gtrapprox 6$). For the largest and smallest values of $\ell$, we do see (not shown) deviation from the behavior in Fig. \ref{amps-manyWs}; we expect this to be caused chiefly by finite-size and finite numerical precision effects. Going forward, we assume that the qualitative behavior observed in Figures \ref{amps-manyWs} and \ref{amps-crit} describes the asymptotic ($L \to \infty$) behavior of the system, and we attempt to characterize our finite-size numerics by this expected behavior.

\subsection{Connecting $g_c$ to $\mathcal{A}_{\ell}$}
\label{g_c-and-avalanche}

In section \ref{avalanche-toy}, we found evidence that the non-hermitian many-body delocalization transition is consistent with a non-hermitian avalanche mechanism, and that $g_c \approx 1/\lambda - 1/\lambda_c$ measured the location of this transition, for the toy Hamiltonian \eqref{H_av_NH}. With the mapping of the previous two sections, we are now in position to test if an analogous relation holds for more realistic Hamiltonians (like the Hamiltonian \eqref{H}).

To that end, we take the Hamiltonian specified by \eqref{H_av_2}, add an imaginary flux $i(L+1)g$, and compute $g_c(\epsilon)$ for a variety of energy densities (see Appendix \ref{appendix-numerical-procedure} for details on computing $g_c$). We do so for the same disorder realizations as shown in the previous section, so that we have \textit{both} $g_c(\epsilon)$ and $\mathcal{A}_{\ell}(\epsilon)$ for the same disorder realizations.

To connect $g_c(\epsilon)$ to the $\mathcal{A}_{\ell}(\epsilon)$, let us consider an avalanche proliferating in this chain. As remarked in the last section, the \textit{deterministic} decay of hopping amplitudes $e^{-1/\lambda}$ in the toy model \eqref{H_av_NH} must be replaced by the \textit{random variables} $\mathcal{A}_{\ell}(\epsilon)$. Because of this, the ratio of matrix element to bath level spacing (in analogy with \eqref{av_cond}) is not necessarily monotonic, and we must examine this ratio for every l-bit to determine delocalization. We also incorporate the imaginary vector potential by enhancing each bond by a factor of $e^{(L+1)g}$ (since, due to the gauge freedom, we can put all influence of $g$ onto the bonds connecting the bath and chain). Thus, the analogous condition to \eqref{av_cond} (at energy density $\epsilon$) for \textit{all} l-bits should read
\begin{equation}
g \gtrapprox -\frac{1}{L+1}\min_{1\leq \ell \leq L} \left[ \log\mathcal{A}_{\ell}(\epsilon) + (\ell-1)\frac{\log 2}{2}\right] .
\label{g_av_cond}
\end{equation} 
In words, the flux $i(L+1)g$ must make up for the maximal ``difference'' between the (random) decay of the hoppings and the level spacing.

\begin{figure}
\begin{subfigure}{.22\textwidth}
\includegraphics[scale=.18]{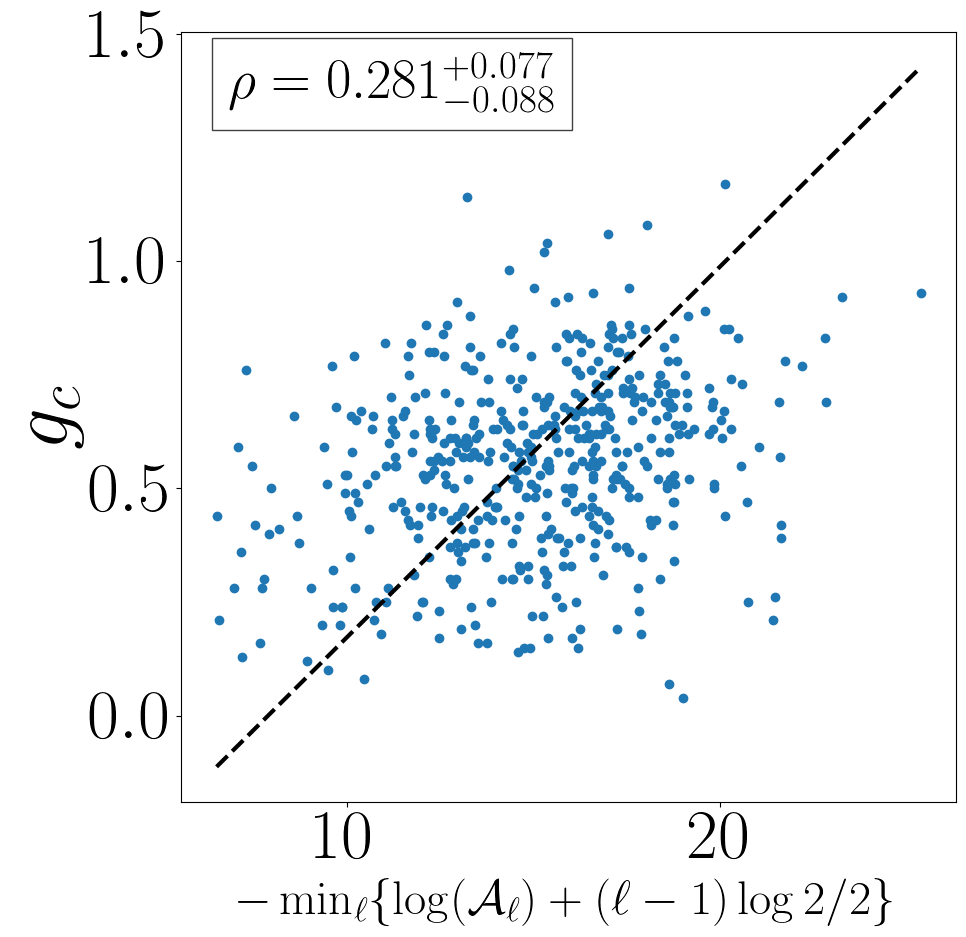}
\caption{}
\label{realizations_gs_amps_U=1}
\end{subfigure}
\hfill
\begin{subfigure}{.22\textwidth}
\includegraphics[scale=.18]{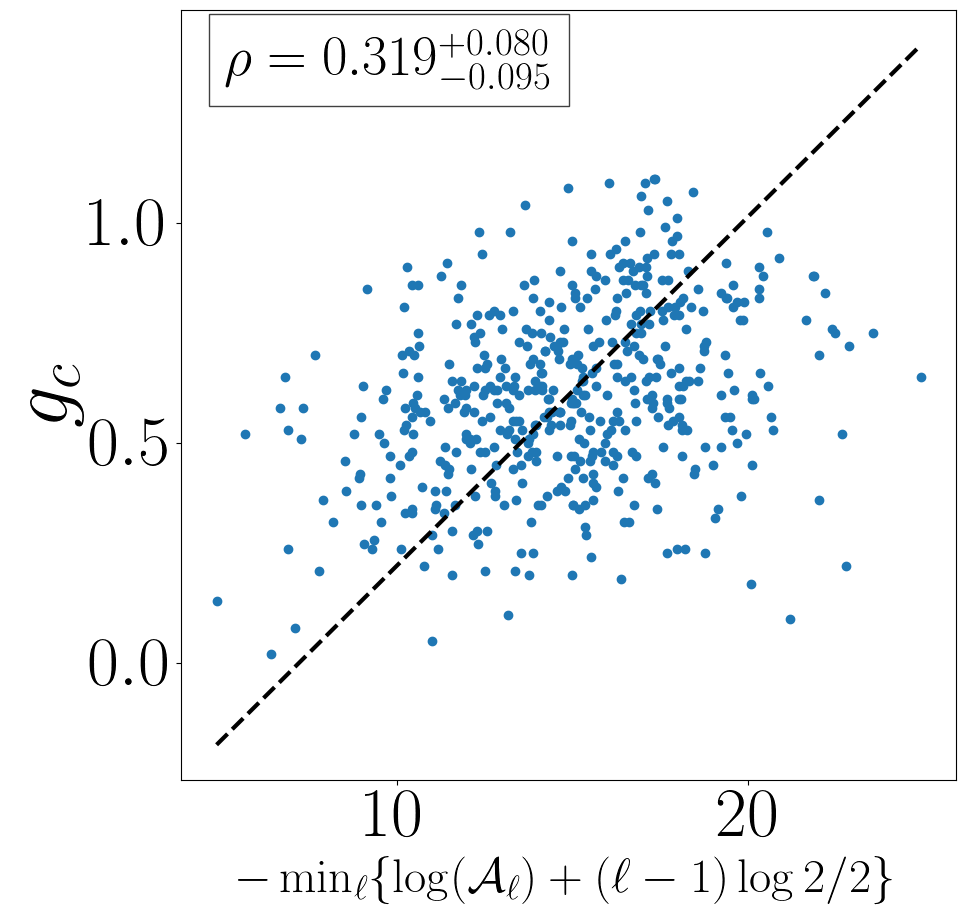}
\caption{}
\label{realizations_gs_amps_U=2}
\end{subfigure}
\\
\begin{subfigure}{.22\textwidth}
\includegraphics[scale=.18]{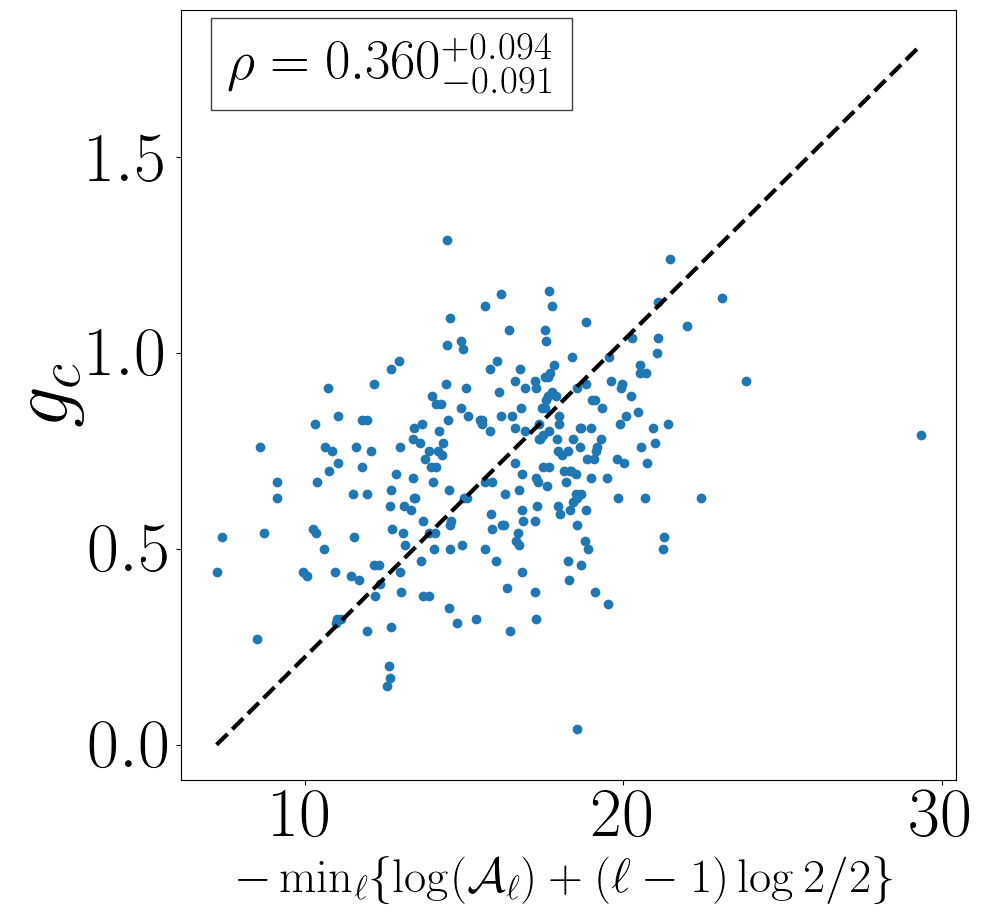}
\caption{}
\label{realizations_gs_amps_U=3}
\end{subfigure}
\hfill
\begin{subfigure}{.22\textwidth}
\includegraphics[scale=.18]{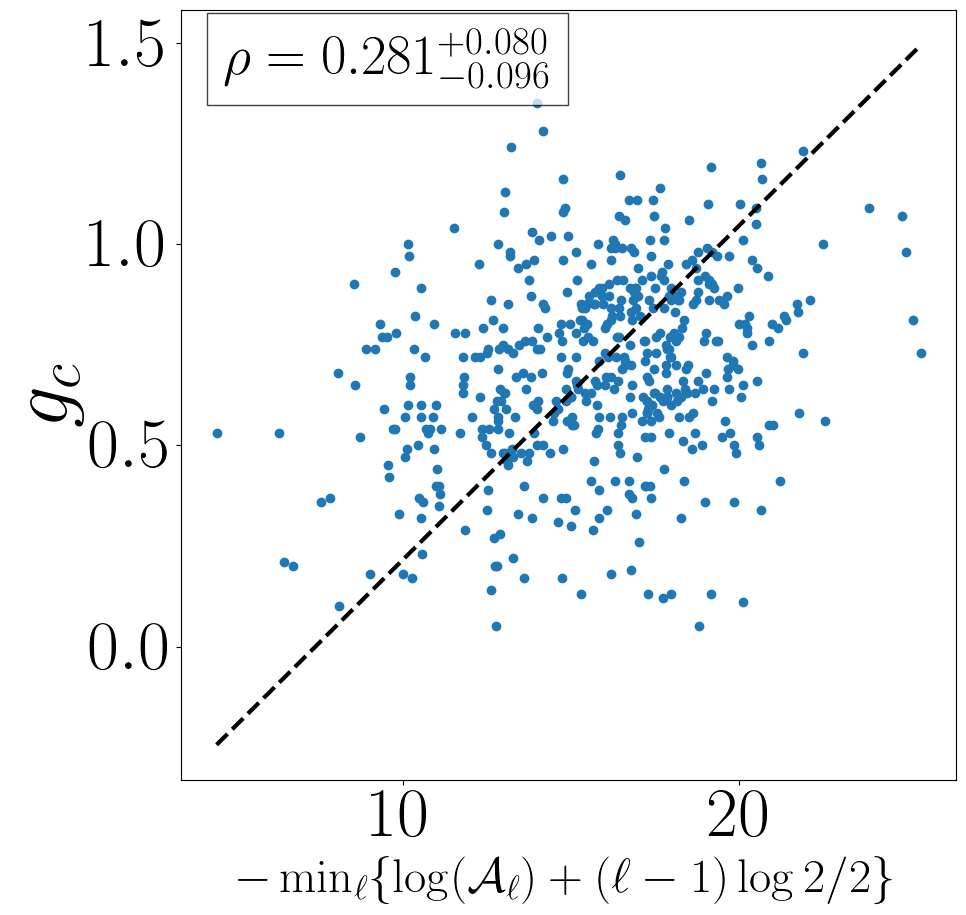}
\caption{}
\label{realizations_gs_amps_U=4}
\end{subfigure}
\\
\begin{subfigure}{.22\textwidth}
\includegraphics[scale=.18]{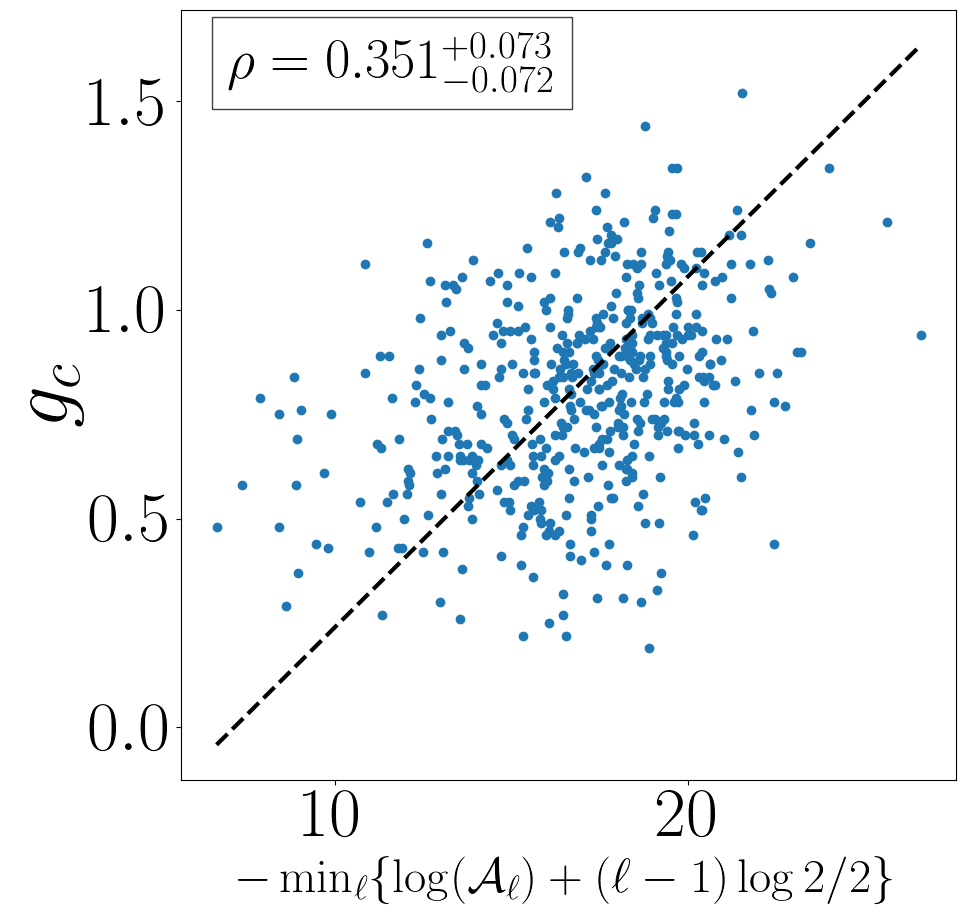}
\caption{}
\label{realizations_gs_amps_U=5}
\end{subfigure}

\caption{Example scatter plots of exceptional points $g_c(\epsilon)$ and ``avalanche parameter'' $\mathscr{A}(\epsilon)$ (computed from the $\mathcal{A}_{\ell}(\epsilon)$ found in Section \ref{mapping-results}), slightly above the critical point. Each point represents a single disorder realization (for a fixed disorder strength), with the ordinate and abscissa $g_c$ and $\mathscr{A}$ at the center of the spectrum ($\epsilon = 0.5$), respectively. The various panels show interaction/disorder strengths a) $U=1, W=5.5$, b) $U=2, W=6.0$, c) $U=3, W=7.0$, d) $U=4, W=7.0$, e) $U=5, W=8.0$. There is no obvious visual relationship between $g_c(\epsilon)$ and $\mathscr{A}(\epsilon)$, though the Pearson Correlation $\rho(X,Y) = \text{cov}(X,Y)/\sigma_X\sigma_Y$ (where $\text{cov}$ is the covariance) indicates weak positive correlation ($\rho \sim 0.3$). We mark $\rho$ in the upper left of each plot, with error estimates from the 95\% bootstrap confidence interval. For reference, we also overlay the disorder averaged fits of Fig. \ref{gs_amps} (dashed lines). All data was taken from the $N=3$ sector of a chain with $L=11$, $L_b=3$, and computed for 500 disorder realizations.}
\label{gs_amps_noavg}
\end{figure}

\begin{figure}
\begin{subfigure}{.22\textwidth}
\includegraphics[scale=.175]{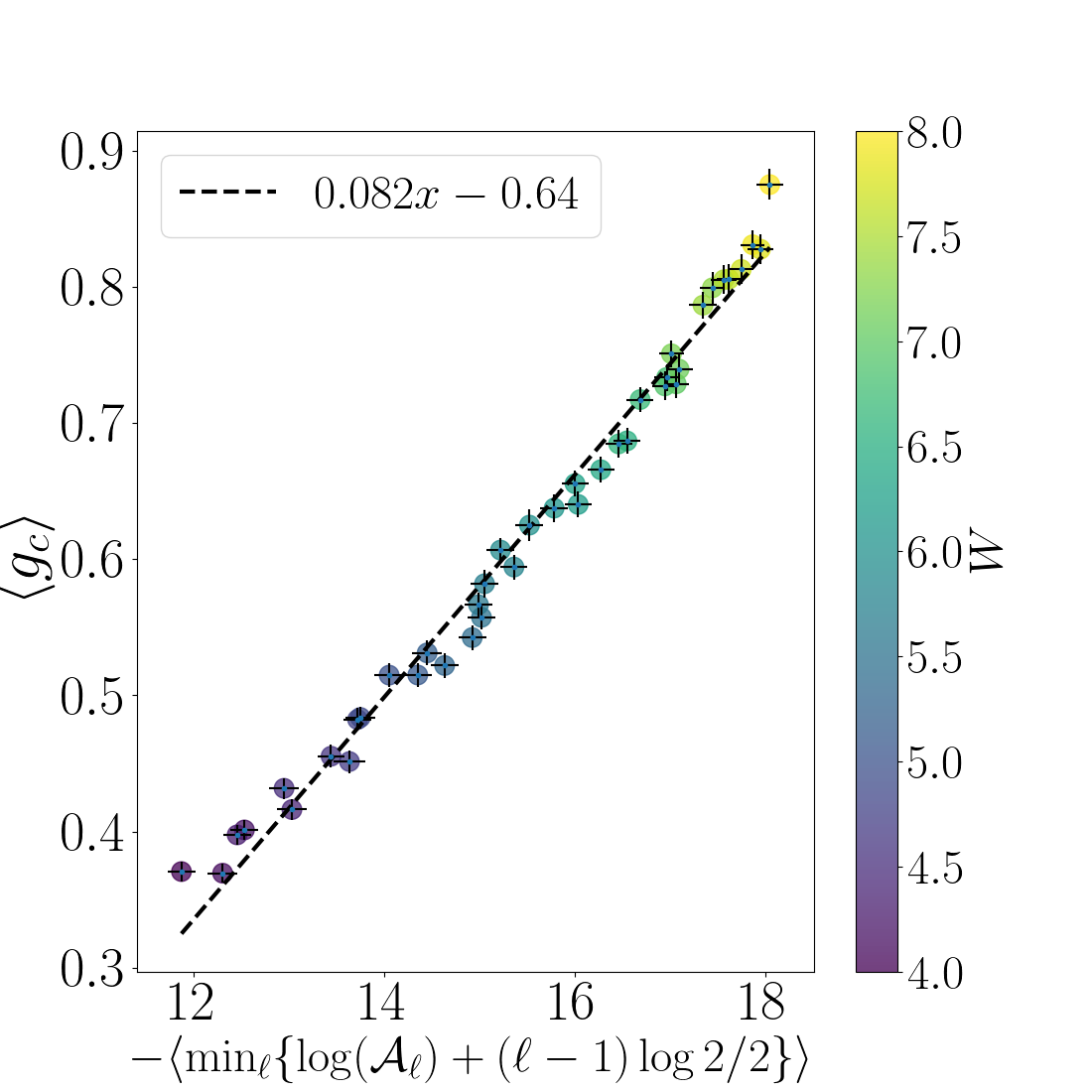}
\caption{}
\label{gs_amps_U=1}
\end{subfigure}
\hfill
\begin{subfigure}{.22\textwidth}
\includegraphics[scale=.175]{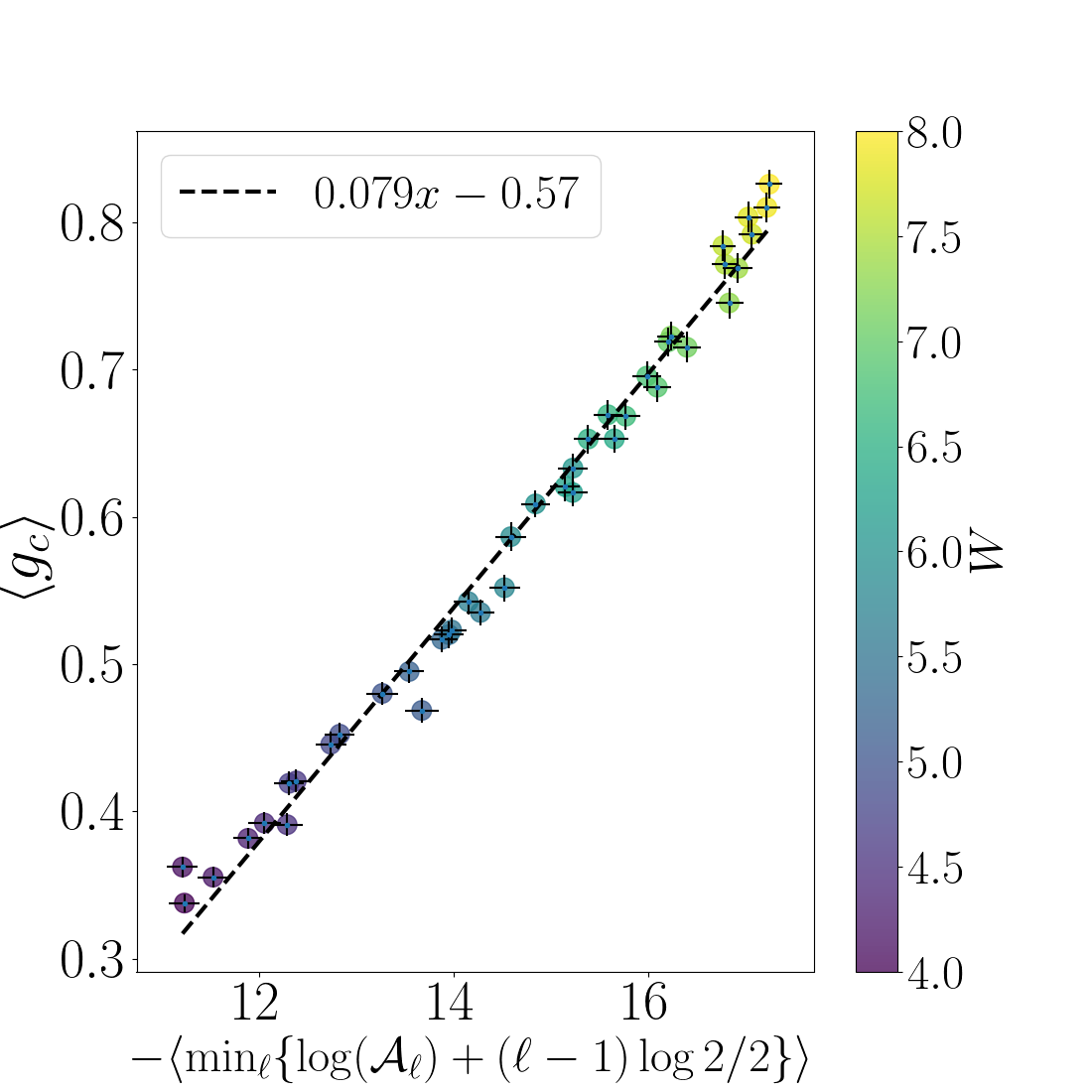}
\caption{}
\label{gs_amps_U=2}
\end{subfigure}
\\
\begin{subfigure}{.22\textwidth}
\includegraphics[scale=.175]{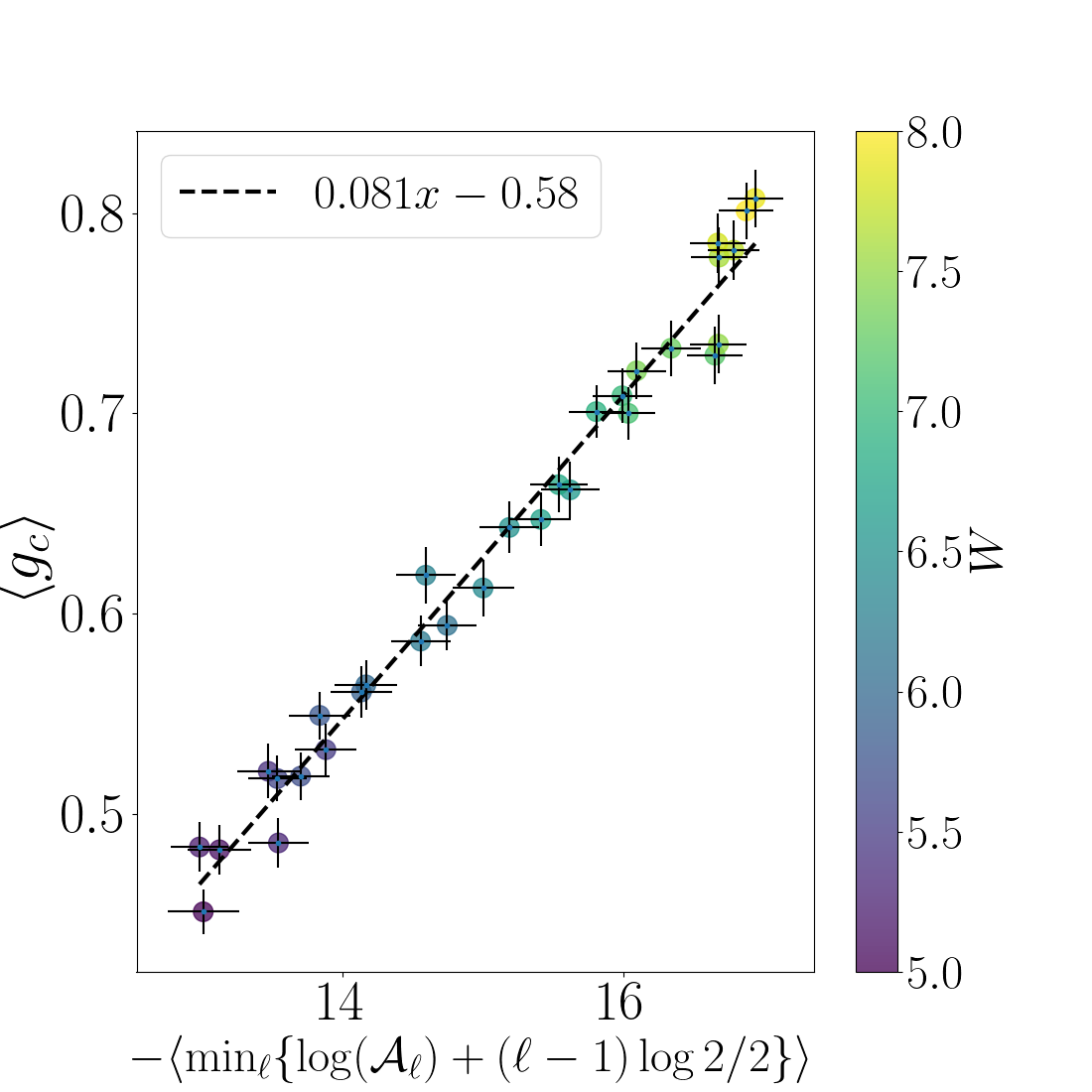}
\caption{}
\label{gs_amps_U=3}
\end{subfigure}
\hfill
\begin{subfigure}{.22\textwidth}
\includegraphics[scale=.175]{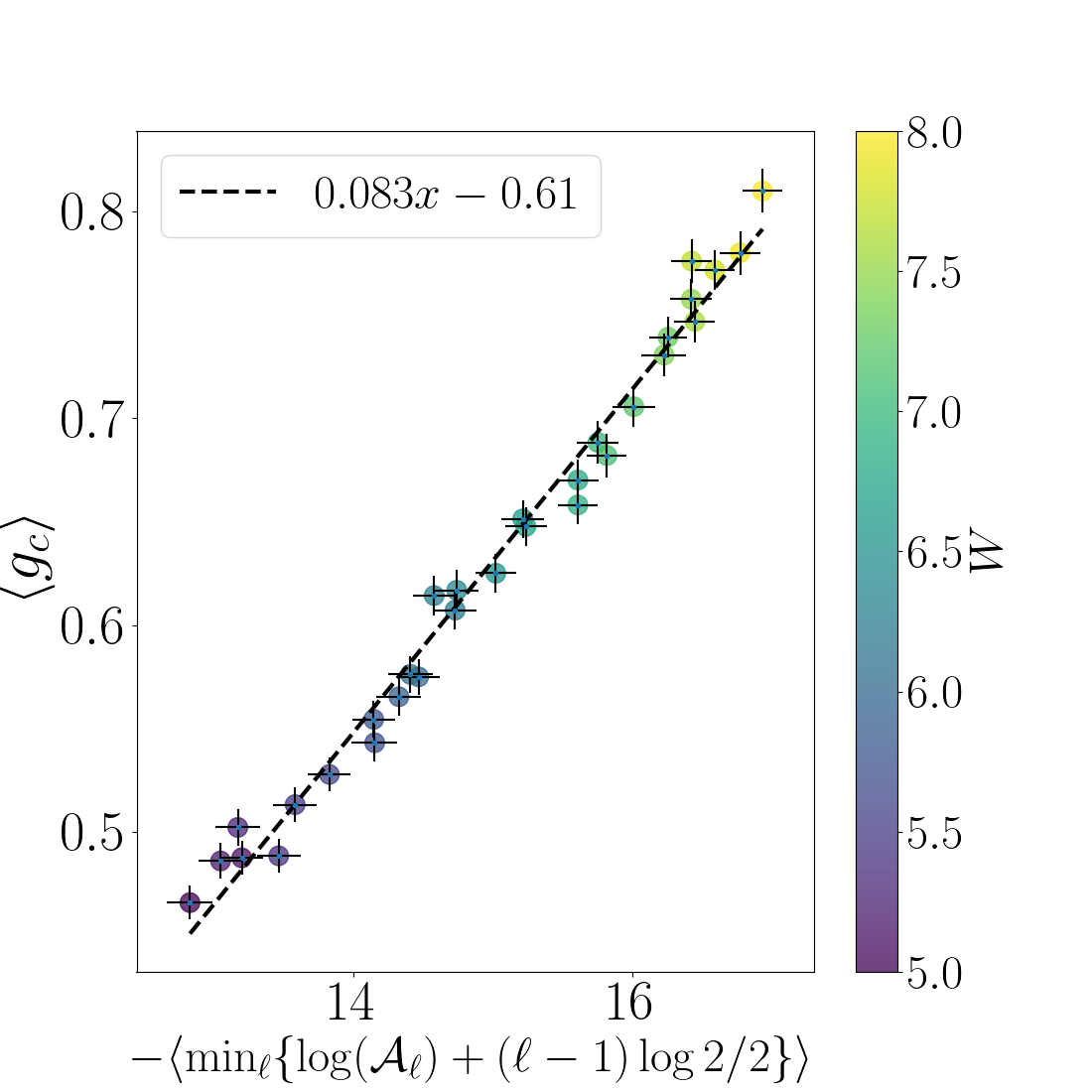}
\caption{}
\label{gs_amps_U=4}
\end{subfigure}
\\
\begin{subfigure}{.22\textwidth}
\includegraphics[scale=.175]{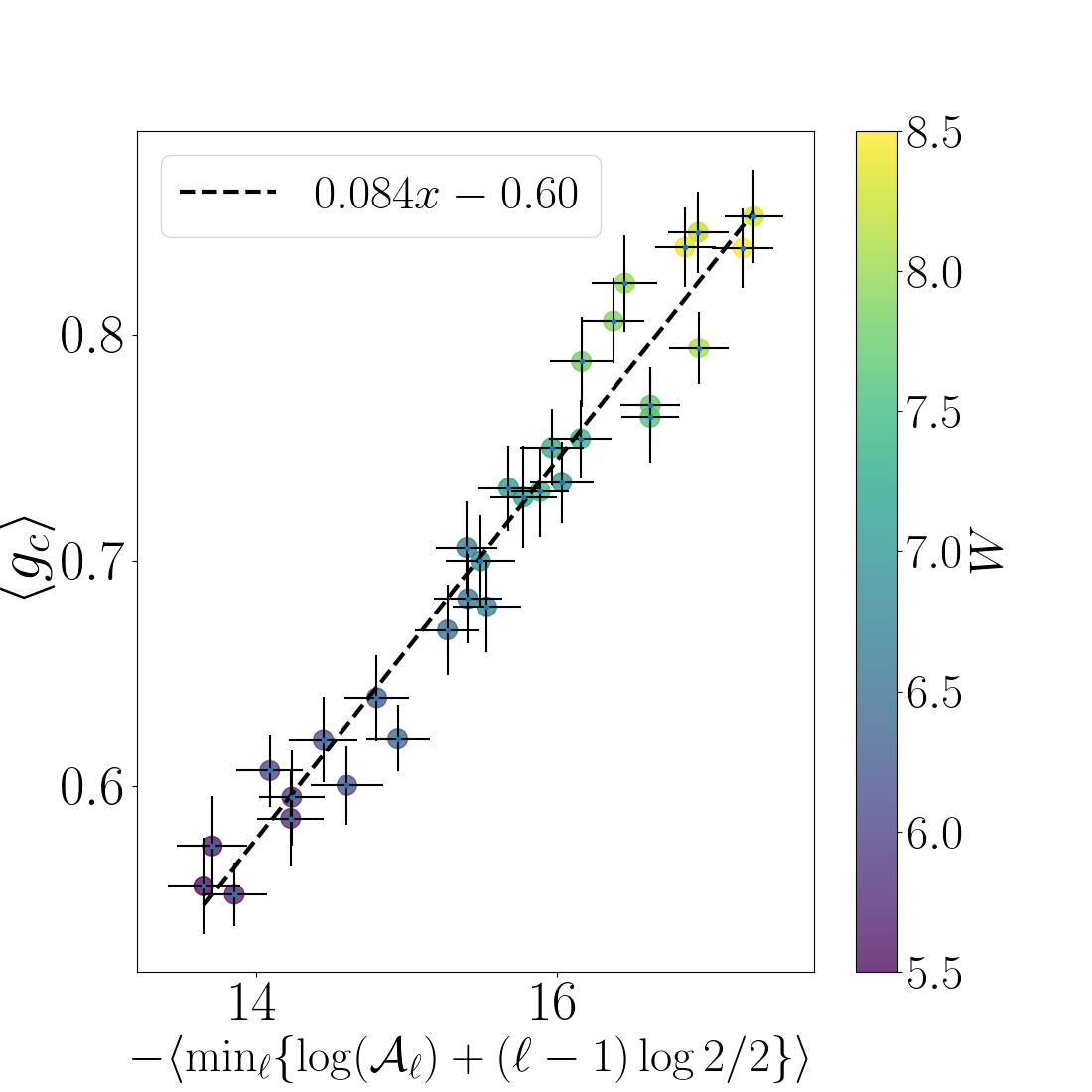}
\caption{}
\label{gs_amps_U=5}
\end{subfigure}

\caption{Scatter plots of disorder-averaged exceptional points $g_c(\epsilon)$ and ``avalanche parameter'' $\mathscr{A}(\epsilon)$(computed from the $\mathcal{A}_{\ell}(\epsilon)$ found in Section \ref{mapping-results}) for a variety of disorder strengths. Each point in the scatter plot corresponds to a specific disorder strength $W$ (the coloration indicating the numerical value of $W$), with the ordinate and abscissa the average (over disorder realizations) of $g_c$ and $\mathscr{A}$ at the center of the spectrum ($\epsilon = 0.5$), respectively. The various panels show interaction strengths a) $U=1$, b) $U=2$, c) $U=3$, d) $U=4$, e) $U=5$. Dashed lines indicate a linear fit performed via orthogonal distance regression. We see good agreement between the fits and the data points, providing evidence that the real-complex/localization-delocalization transition is driven by non-hermitian avalanches. All data was taken from the $N=3$ sector of a chain with $L=11$, $L_b=3$. Averages were performed over 500 disorder realizations.}
\label{gs_amps}
\end{figure}

Based on the results of Section \ref{avalanche-toy}, we expect that the exceptional points $g_c(\epsilon)$ saturate this bound - that is, $g_c$ is \textit{exactly} enough to cover all of these ``differences''. We test this criterion numerically by computing 
\begin{equation}
\mathscr{A}(\epsilon) := -\min_{1\leq \ell \leq L} \left[ \log\mathcal{A}_{\ell}(\epsilon) + (\ell-1)\frac{\log 2}{2}\right]
\label{A_min}
\end{equation}
and $g_c(\epsilon)$ for each disorder realization, and comparing the two. Visually, there is no clear cut trend, though the Pearson correlation coefficient suggests a weak linear relationship - see Fig. \ref{gs_amps_noavg} for some example scatter plots. We observe a much clearer trend by averaging $g_c(\epsilon)$  and $\mathscr{A}(\epsilon)$ over disorder realizations for multiple disorder strengths $W$ - see Fig. \ref{gs_amps} for examples at the center of the spectrum. We are able to fit a line to the scatter plot of $\avg{g_c(\epsilon)}(W)$ vs $\avg{\mathscr{A}(\epsilon)}\hspace{-3pt}(W)$ (where $\avg{\cdot}$ denotes a disorder average) - these fits are shown in Fig. \ref{gs_amps} (and overlaid in Fig. \ref{gs_amps_noavg}), and Table \ref{linear_fit} shows the resulting fit parameters (with uncertainties \footnote{Uncertainties on the fit parameters were found via bootstrapping: for each disorder strength, we sample with replacement from the disorder realizations, average $g_c(\epsilon)$ and $\mathscr{A}(\epsilon)$, and perform a linear fit. The uncertainties on the slope and intercept are the the standard deviation of each quantity over 1000 such resamplings.}). We find that the disorder-averaged data is well-fit by a line, and the slope agrees within error bars (at least at the center of the spectrum) with the expected value of $1/12 = .0825$ (predicted by \eqref{g_av_cond} with $L=11$). Curiously, the intercepts of these fits are non-zero. At present, we do not have an explanation for this, but we note in passing that the intercepts are roughly consistent with $\log 2 \approx 0.693$, which would correspond to a shift of about $\log 2$ per non-bath site to $\mathscr{A}(\epsilon)$. In any case, there is unmistakably a linear relationship between $\avg{g_c}$ and $\avg{\mathscr{A}(\epsilon)}$, with (roughly) the correct slope of $1/(L+1)$. This results support our conjecture in Section \ref{avalanche-toy}, and the idea that the location of $g_c$ is given by saturating the avalanche criterion \eqref{g_av_cond}. This in turn means that the localization length $\xi = 1/g_c$ captures the difference between the hopping decay and the level spacing. 

In taking \eqref{A_min} as the ``avalanche condition'' determining $g_c$, we are assuming that delocalization occurrs when \textit{all} l-bits are delocalized by hybridizing with the bath. It is a worthwhile question to ask whether this is strictly necessary, and if delocalization of the system (indicated here by the eigenvalues acquiring a non-zero imaginary part) only requires \textit{a fraction} of the l-bits to hybridize with the bath. Requiring, say, only $L'$ out of $L$ l-bits to hybridize would entail replacing the minimum in \eqref{A_min} by the function that selects the $(L')$'th largest of the collection of $L$ values. In the extreme case of $L' = 1$, the minimum in \eqref{A_min} should be replaced by a maximum; we find (not shown) that scatter plots analogous to Fig. \ref{gs_amps_noavg} are very noisy and uncorrelated, and disorder-averaged fits analogous to those in Fig. \ref{gs_amps} are very poor. Consequently, this extreme case is unlikely to capture the physics (unlike the minimum, corresponding to $L'=L$).

It is not clear whether there even \textit{should} be a universal value of $L'$ for all disorder realizations. If $L'$ varies from realization to realization, this could explain the non-zero intercepts in the linear fits of Fig. \ref{gs_amps}; the intercept could capture the (disorder-averaged) variation between the minimum and $(L')$'th largest value of $\mathcal{A}_{\ell}(\epsilon) + (\ell-1)\log 2/2$ across disorder realizations. To attempt to address this question, we repeat the analysis presented above with an average over l-bits, instead of a minimum - see Appendix \ref{appendix-avg}. There, we find comparable results to those we have presented in this section, and the linear fits generically have smaller intercepts. As averaging over l-bits does not fix a particular universal value of $L'$, this suggests that $L'$ may indeed not have a universal value across disorder realizations. In spite of this - as we shall see in the next section - the avalanche condition \eqref{A_min} (i.e., taking $L'=L$) appears to be the most successful in capturing the physics of $g_c$ in the original Hamiltonian \eqref{H} without a bath, so we primarily focus on this case moving forward.

\begin{table*}
\centering
\begin{tabular}{|c|c|c|c|c|c|c|c|c|c|c|}
\hline 
 & \multicolumn{2}{|c|}{ $U=1$} & \multicolumn{2}{|c|}{ $U=2$}   & \multicolumn{2}{|c|}{$U=3$} & \multicolumn{2}{|c|}{ $U=4$}  & \multicolumn{2}{|c|}{$U=5$} \\ 
 \hline
 $\epsilon$ & Slope & Intercept & Slope & Intercept & Slope & Intercept & Slope & Intercept & Slope & Intercept \\
 \hline 
 $0.2$ & $.098 \pm .002$ & $-.895 \pm .033$ & $.086 \pm .002$ & $-.651 \pm .031$ & $.091 \pm .004 $ & $-.737 \pm  .075$ & $.085 \pm .003$ & $-.625 \pm .044$ & $.093 \pm .004$ & $-.694 \pm .068$\\
 \hline
$0.3$ & $.086 \pm .001$ & $-.711 \pm .023$ & $.080 \pm .001$ & $-.586 \pm .020$ & $.087 \pm .003$ & $-.691 \pm .052$ & $.083 \pm .002$ & $-.619 \pm .033$ & $.093 \pm .003$ & $-.751 \pm .048$\\
 \hline
$0.4$ & $.084 \pm .001$ & $-.683 \pm .019 $ & $.079 \pm .001$ & $-.573 \pm .018 $ & $.086 \pm .003$ & $-.664 \pm .044$ & $.083 \pm .002$ & $-.627 \pm .032$ & $.094 \pm .005$ & $-.769 \pm .072$\\
 \hline
$0.5$ & $.082 \pm .001$ & $-.643 \pm  .018$ & $.079 \pm .001$ & $-.574 \pm .017$ & $.081 \pm .002$ & $-.584 \pm .038$ & $.083 \pm .002$ & $-.613 \pm .031$ & $.084 \pm .004$ & $-.599 \pm .066$\\
 \hline
$0.6$ & $.082 \pm .001$ & $-.643 \pm .020$ & $.081 \pm .001$ & $-.581 \pm .018$ & $.089 \pm .003$ & $-.703 \pm .050$ & $.087 \pm .002$ & $-.653 \pm .037$ & $.100 \pm .006$ & $-.798 \pm .097$\\
 \hline
$0.7$ & $.088 \pm .002$ & $-.724 \pm .025$ & $.082 \pm .001$ & $-.586 \pm .021$ & $.091 \pm .004$ & $-.705 \pm .059$ & $.091 \pm .003$ & $-.663 \pm .046$ & $.106 \pm .004$ & $-.890 \pm .070$\\
 \hline
$0.8$ & $.092 \pm .002$ & $-.775 \pm .035$ & $.088 \pm .002$ & $-.640 \pm .030$ & $.091 \pm .006$ & $-.651 \pm .094$ & $.096 \pm .004$ & $-.684 \pm .073$ & $.103 \pm .006$  & $-.748 \pm .105$\\
 \hline
\end{tabular}
\caption{Best fit parameters for a linear fit of $\avg{g_c(\epsilon)}$ vs $\avg{\mathscr{A}(\epsilon)}$ (see the main text and Fig. \ref{gs_amps} for details regarding the fits), across the spectrum for various interaction strengths. The linear fits are performed using orthogonal distance regression, using the standard errors of $\avg{g_c(\epsilon)}$ and $\avg{\mathscr{A}(\epsilon)}$ as uncertainties. We estimate the uncertainties on the fit parameters as the standard deviations on the fit parameters across 1000 bootstrap resamplings. At weaker interaction strengths and near the center of the spectrum, we see good agreement with the expected slope of 1/12, while the parameters deviate from this expectation for stronger interactions and towards the band edges. All data was taken from the $N=3$ sector of a chain with $L=11$, $L_b=3$. Averages were performed over 500 disorder realizations.}
\label{linear_fit}
\end{table*}

\subsection{Single vs multi l-bit flips}
\label{single-vs-multi-flips}

Before moving on, we address a key aspect of the definition \eqref{A_ell} of the amplitudes $\mathcal{A}_{\ell}$: the fact they are derived only from considering \textit{single} l-bit flips. 

At first glance, such a definition may seem too restrictive; namely, that by considering only single l-bit flips, we are missing the dominant processes by which the system relaxes (as observed, e.g., in \cite{Ha-Morningstar-Huse-res-av}). We find, however (not shown), that extending the definition of $\mathcal{A}_{\ell}$ to include \textit{all} hoppings that flip l-bit $\ell$ yields quantitively worse results - in the sense that scatterplots akin to Fig. \ref{gs_amps_noavg} show near-zero Pearson correlation and linear fits akin to Fig. \ref{gs_amps} yield the wrong slope. Our numerics are thus more consistent with an avalanche mechanism that propagates via single l-bit flips only.

This is not to say that our results suggest that multi l-bit flips are not important to the finite-size MBL regime or the avalanche mechanism - the growing consensus in the literature is that they very much are \cite{Gopalakrishnan-Muller-Khemani-Knap-Demler-Huse-low-frequency-cond,Villalonga-Clark-Resonances-1,Villalonga-Clark-Resonances-2,Garratt-Roy-Chalker-resonances,Crowley-Chandran-RM,Morningstar-Colmenarez-Khemani-Luitz-Huse-Av-Resonances,Long-Crowley-Khemani-Chanrdan-prethermal-MBL,Garratt-Roy-resonances-observables,Ha-Morningstar-Huse-res-av}. Rather, we suspect the primary reason behind this apparent discrepancy is the fact that our imaginary vector potential couples to charge. The real-complex transition we detect thus corresponds to delocalization of \textit{charge}, rather than \textit{entropy} or \textit{energy} (which other studies such as \cite{Ha-Morningstar-Huse-res-av} are sensitive to). Correspondingly, our imaginary vector potential may not be sensitive to charge-neutral processes that thermalize the system, which based on our numerics seem to coincide with processes that flip multiple l-bits at a time.

%\FloatBarrier

\section{Distributions at the MBL crossover}
\label{distributions}

We have sucessfully connected the localization length $\xi = 1/g_c$ to the length scale appearing in the avalanche model. We now seek to leverage this relationship to predict the distribution of $g_c$ (with respect to disorder) - the distribution of $\xi$ then follows as the inverse distribution, per the relationship $\xi = 1/g_c$. We focus here on the distribution at the MBL-thermal crossover, as we have an analytic form for the distribution of the amplitudes $\mathcal{A}_{\ell}(\epsilon)$ (based on the observations of Section \ref{mapping-results}).

Based on the results of Section \ref{g_c-and-avalanche}, we consider the following generalized relation between $g_c$ and the $\mathcal{A}_{\ell}$'s:
\begin{equation}
g_c(\epsilon) \approx \frac{1}{F}\max_{1\leq \ell \leq N_{\ell}} \left[ -\log\mathcal{A}_{\ell}(\epsilon) - (\ell-1)\alpha\right] + g_0 ,
\label{g_c_relation}
\end{equation}  
where $iFg$ is the total flux through the system, $N_{\ell}$ the number of l-bits in the system, and $\alpha = \log d / 2$, where $d$ is the on-site Hilbert space dimension (in fermionic models, $d=2$). We also allow for an offset $g_0$, based on the results of Section \ref{g_c-and-avalanche}. Note that we have written the relation in terms of a maximum here, to ease analytic computation of the distribution. 

Let us assume this relation for $g_c$ holds for each disorder realization (i.e., we neglect the noise observed in Fig. \ref{gs_amps_noavg}), and let $P_F^{N_{\ell}}(g_c \leq x\,|\, g_0)$ denote the cumulative density function for $g_c$ given by this relation. As observed in Section \ref{mapping-results}, we have $\mathcal{A}_{\ell}(\epsilon) = \prod_{i=1}^{\ell}\chi_i(\epsilon)$ with $\chi_i(\epsilon)$ uniformly distributed on $[0,1]$ at the MBL crossover. Using this, we arrive at the following form for the cdf:
\begin{equation}
\scalebox{.88}{$\displaystyle P_F^{N_{\ell}}(g_c \leq x \,|\, g_0) = 1 - e^{-\tilde{x}}\left[1 + \tilde{x} \sum_{j=1}^{N_{\ell}-1}\frac{e^{-j\alpha}}{j!}(\tilde{x} + j\alpha)^j\right] ,$}
\label{CDF_gen}
\end{equation} 
where $\tilde{x} = F(x-g_0)$ - see Appendix \ref{appendix-CDF} for details on the derivation of this result. Crucially, we note here that the derivation in Appendix \ref{appendix-CDF} assumes the $\chi_i$ are independently and identically distributed. This is not strictly true, as the distributions of $\chi_i$ do vary in $i$ (as remarked in Section \ref{mapping-results}), and all the $\chi_i(\epsilon)$ arise from the same disorder realization, and are thus almost certainly correlated. We neglect these two facts here. 

We now would like to see if the analytic distribution \eqref{CDF_gen} describes the distribution of $g_c$'s (and thereby $\xi$'s) at the MBL crossover in the original model, \eqref{H}. To that end, we compute the histogram of $g_c(\epsilon)$ at the (energy-resolved) critical disorder strength identified from finite-size scaling (see Appendix \ref{appendix-FSS}). We cannot compare the numerically observed distributions directly to \eqref{CDF_gen}, however, as the Hamiltonian \eqref{H} doesn't explicitly include a thermal bath/grain. The size of the bath is thus itself also a random variable; we account for this by modifying \eqref{CDF_gen} to a mixture distribution of the form
\begin{equation}
P_{\boldsymbol{\beta},g_0}(g_c \leq x) = \sum_{i=1}^{L-1} \beta_i P_F^{L-i}(g_c\leq x \,|\, g_0) ,
\label{CDF_mixture}
\end{equation} 
with $0 \leq \beta_i \leq 1$, $\sum_i \beta_i = 1$. In this distribution, $\beta_i$ is the probability the bath/thermal grain is of size $i$, and $P_F^{L-i}$ is the CDF of $g_c$ conditioned on the bath being size $i$. Such a distribution follows from the law of total probability applied to the CDF of $g_c$.

\begin{figure}
\begin{subfigure}{.22\textwidth}
\includegraphics[scale=.35]{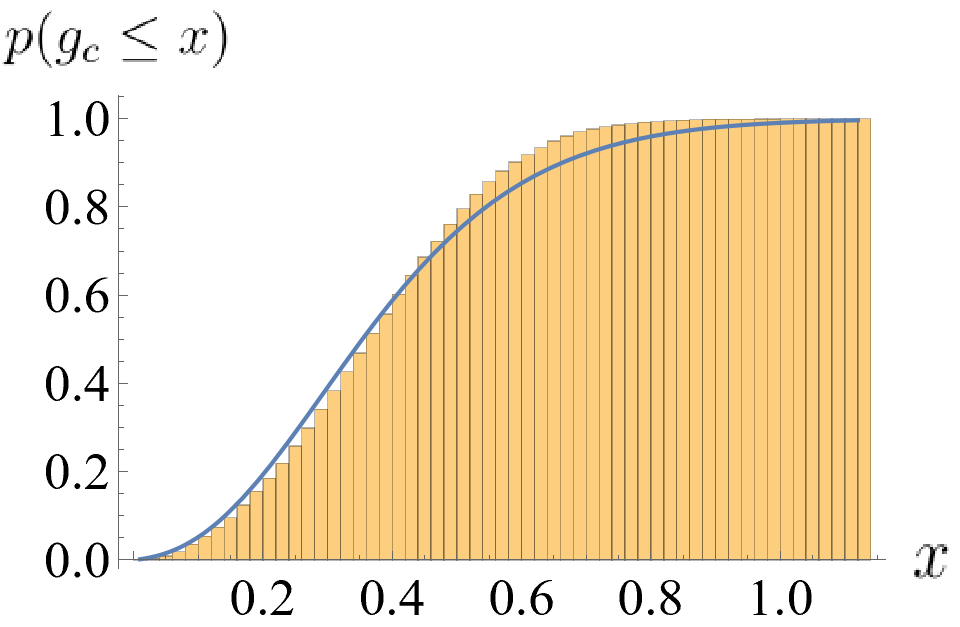}
\caption{}
\label{CDF_fits_U=1}
\end{subfigure}
\hfill
\begin{subfigure}{.22\textwidth}
\includegraphics[scale=.35]{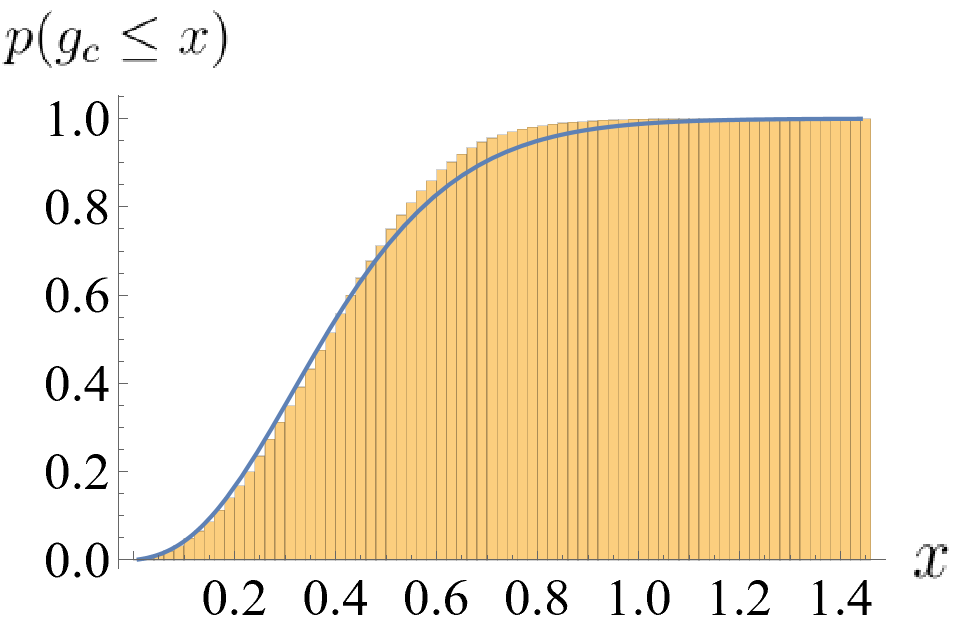}
\caption{}
\label{CDF_fits_U=2}
\end{subfigure}
\\
\begin{subfigure}{.22\textwidth}
\includegraphics[scale=.35]{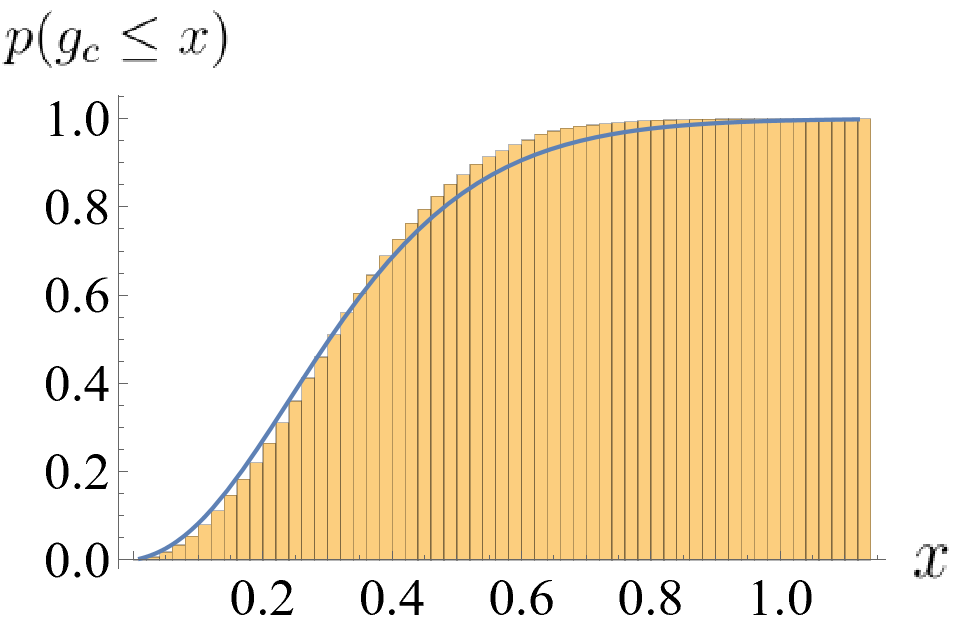}
\caption{}
\label{CDF_fits_U=3}
\end{subfigure}
\hfill
\begin{subfigure}{.22\textwidth}
\includegraphics[scale=.35]{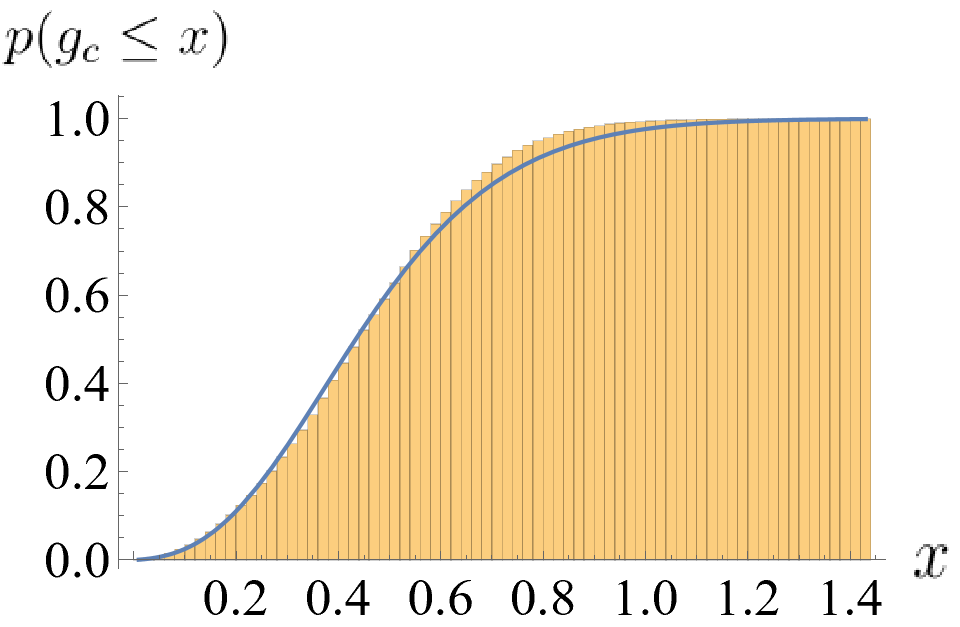}
\caption{}
\label{CDF_fits_U=4}
\end{subfigure}
\\
\begin{subfigure}{.22\textwidth}
\includegraphics[scale=.35]{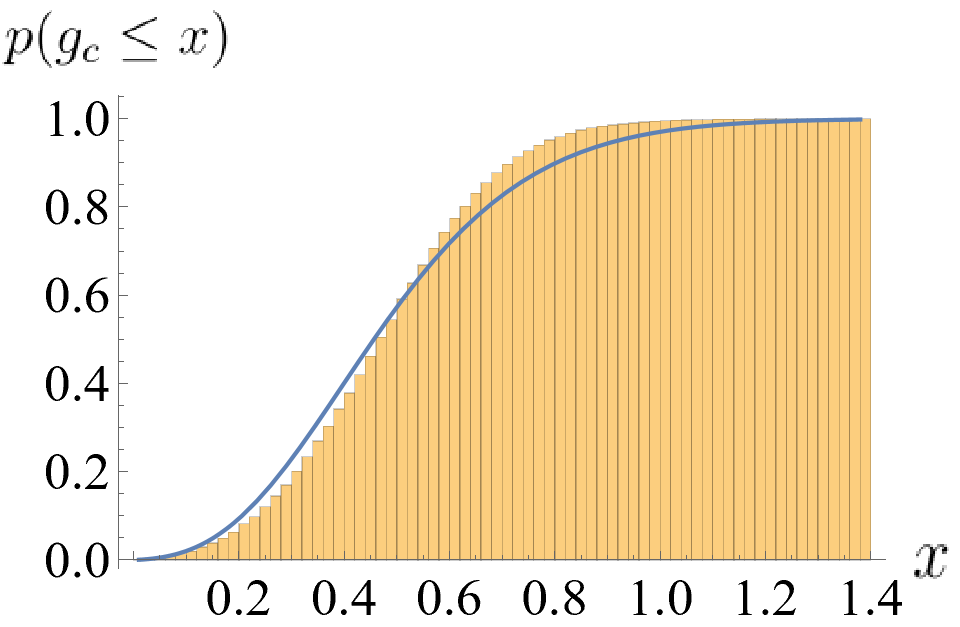}
\caption{}
\label{CDF_fits_U=5}
\end{subfigure}

\caption{Numerically observed cumulative density functions of $g_c$ for $\epsilon = 0.5$ at the critical point, for interaction strengths a) $U=1$ ($W=4.8$) , b) $U=2$ ($W = 5.4$), c) $U=3$ ($W = 5.9$), d) $U=4$ ($W = 6.1$), e) $U=5$ ($W = 6.7$). The blue solid lines overlaid are best fit mixture distributions of the form \eqref{CDF_mixture} (see main text for details). We see good agreement between fit and numerical observation, suggesting that our distribution derived from a non-hermitian avalanche criterion accurately describes the transition point. The parameters extracted from the fit are a) $\beta_5 = 0.366$, $\beta_6 = 0.634$ b) $\beta_5 = 0.769$, $\beta_6 = 0.231$, c) $\beta_6 = 0.435$, $\beta_7 = 0.565$, d) $\beta_4 = 0.827$, $\beta_5 = 0.173$ e) $\beta_3 = 0.243$, $\beta_4 = 0.757$, and $g_0 = 0$ in all cases. All quantities are computed in chains of size $L=12$ in the half-filling sector, and from 10000 disorder realizations. When computing the squared error \eqref{squared_error}, we sample $n = 100$ points uniformly spaced between the smallest and largest observed values of $g_c$.}
\label{CDF_fits}
\end{figure}

We fit a mixture distribution of the form \eqref{CDF_mixture} to the numerically observed CDF. We do so by computing a kernel density estimator $\hat{f}$ of the numerically observed CDF (using a Gaussian kernel), sampling this estimator at various points $\{x_i\}_{i=1}^n$, and numerically minimizing the squared error
\begin{equation}
\delta(\boldsymbol{\beta},g_0) = \sum_{i=1}^n\left(P_{\boldsymbol{\beta},g_0}(g_c \leq x_i) - \hat{f}(x_i)\right)^2 ,
\label{squared_error}
\end{equation} 
subject to the constraints $0 \leq \beta_i \leq 1$, $\sum_i \beta_i = 1$. The results of this fit at the center of the spectrum are shown in Fig \ref{CDF_fits}. We see good overall agreement between the observed CDF and the fitted mixture distribution, though the fitted distributions have longer tails than the observed distributions (as evidenced by the slower approach of the CDF to 1 - see also Appendix \ref{appendix-Distribution-Fits}). This suggests that the distributions of $g_c$ at the critical point are well-described by an avalanche-like picture of delocalization (which gives rise to the distribution \eqref{CDF_mixture}). The distribution of $\xi$, obtained by computing the inverse distribution of \eqref{CDF_mixture}, thus contains information about the distribution of the hopping amplitudes $\mathcal{A}_{\ell}$.   

Curiously, we find all the fits have $g_0 \approx 0$; the offset does not appear to play a role in the physics without the bath. In light of the discussion at the end of the previous section, the lack of offset would suggest that \textit{all} l-bits ($L' = L$) must hybridize when there is no explicit bath in the system. In Appendix \ref{appendix-avg}, we derive an analytic distribution of $g_c$ assuming an average avalanche condition and perform similar fits; the agreement between fitted and numerically observed distributions is not as good for the average avalanche condition (especially at larger interaction strengths). This further suggests that the physics in the system without a bath is captured by requiring \textit{all} l-bits to hybridize ($L' = L$). 

The mixture distribution defined by \eqref{CDF_mixture} contains $\mathcal{O}(L)$ free parameters (the $\beta_i$'s, with one of them fixed by $\sum_i \beta_i = 1$, and $g_0$). As a result, the agreement seen in Fig \ref{CDF_fits} appears not too surprising, despite the fact that this discrete set of parameters is paramterizing a continuous distribution relying on thousands of disorder realizations. However, one must keep in mind that the $\beta_i$'s represent how the system partitions itself into localized and ergodic subregions. It is \textit{this} information that parameterizes the distributions $P_{\boldsymbol{\beta},g_0}$, and this information is, in principle, calculable by other means (which could eliminate the $\beta_i$'s as fit parameters entirely). Indeed, in all the fits we show, only a subset of the fitted parameters (2 in the case of Fig. \ref{CDF_fits}, 2-4 for the fits shown in the Appendices) are non-zero, and the indices of the non-zero $\beta_i$'s seem to increase with increasing interaction strength $U$. This is consistent with the idea above that the $\beta_i$'s are not actually free, but rather determined by another unknown function: the distribution of sizes of thermal inclusions in the system. Determining this distribution, especially for small system sizes where one must account for resonances, is beyond the scope of our current work, so we instead treat the $\beta_i$'s as fit parameters.

As a final remark, we note that the fit quality of the mixture distribution \eqref{CDF_mixture} is even better towards the band edges - see Appendix \ref{appendix-Distribution-Fits} for examples and discussion.

\section{Conclusion \& Discussion}
\label{conclusion}

We have shown how to connect a localization length $\xi$, defined by introducing an imaginary vector potential to a disordered chain of interacting fermions, to the length scale $\lambda$ characterizing the decay of matrix elements in the avalanche model of delocalization. We have also derived an analytic form of how the $g_c$'s are distributed at the MBL crossover in finite systems, and shown that this form describes well the observed histograms. The distributions of $\xi$ at the crossover, which can be derived from those of $g_c$, thus contain information about the distribution of avalanche-like hoppings, giving insight into the avalanche mechanism at the finite-size crossover.

Determining $\xi$, while computationally difficult, has the advantage of not requiring one to explicitly construct the l-bits. That is, a given $H$ with a fixed disorder realization has a well-defined $g_c$, and thereby $\xi$, for each energy density $\epsilon$. Though our numerical verification of $g_c$ saturating the avalanche condition \eqref{g_av_cond} made use of a particular construction of the l-bits, the agreement of the numerically observed histograms to our fitted CDFs \eqref{CDF_mixture} suggests the form of avalanche criteria for $g_c$ does not depend on how the l-bits are constructed. This method thus allows one to obtain information about the localization of the underlying $g=0$ Hamiltonian whilst sidestepping issues with defining the l-bits.

The connection of $\xi$ to the localization properties of the underlying $g=0$ system relies on the proliferation of avalanches enhanced by the imaginary flux through the system. It is therefore an intereresting question to ask how a $\xi$ defined analogously for ``clean'' systems differs, as there are no rare regions driving thermalization in these systems. The recent wealth of work on the crucial role of many-body resonances in finite size MBL systems  \cite{Gopalakrishnan-Muller-Khemani-Knap-Demler-Huse-low-frequency-cond,Villalonga-Clark-Resonances-1,Villalonga-Clark-Resonances-2,Garratt-Roy-Chalker-resonances,Crowley-Chandran-RM,Morningstar-Colmenarez-Khemani-Luitz-Huse-Av-Resonances,Long-Crowley-Khemani-Chanrdan-prethermal-MBL,Garratt-Roy-resonances-observables,Ha-Morningstar-Huse-res-av} also raises the question of how $g_c$ is connected to these resonances; after all, avoided crossings are generically expected to give rise to exceptional points when the parameter space is expanded to the complex plane \cite{Heiss-Sanninno-Level-Crossing,Luitz-Piazza-Exc-Points}. In particular, it is worth asking how $g_c$ could be used to probe the statistics of many-body resonances in the finite-size MBL regime (and how such a connection affects the distributions we derived in Section \ref{distributions}). Additionally, in light of the discussion of Section \ref{single-vs-multi-flips}, it would be interesting to adapt this non-hermitian technique such that the real-complex transition corresponds to a non-zero \textit{information} current. This (presumably) would make this non-hermitian method sensitive to the destabilizing charge-neutral processes alluded to in Section \ref{single-vs-multi-flips}, and would be an interesting comparison with the results derived here. 

Finally, though we introduced the l-bits and the mapping of Section III C primarily as an intermediate step to connect $g_c$ to the avalanche model, the mapping and the numerical results in Section III D open several potential avenues of future work. Among other questions, the significance of the distributions of the $\chi_i$ - especially being uniformly distributed at the MBL-thermal crossover - is worthy of additional study. The implications of these distributions for the properties of the l-bits is particularly interesting. 

\begin{acknowledgments}
LO and GR thank David Huse for useful discussion regarding the distinction between single and multi l-bit flips when defining $\mathcal{A}_{\ell}$. LO also thanks Christopher David White for his guidance and advice during the early stages of this project, as well as Dan Borgnia for useful discussions. GR is grateful for support from the Simons Foundation as well as support from the NSF DMR grant number 1839271, and from the IQIM, an NSF Physics Frontiers Center. This work was performed in part at Aspen Center for Physics,
which is supported by National Science Foundation grant PHY-1607611.
\end{acknowledgments}

\appendix

\section{Numerical Procedure for Obtaining $g_c$}
\label{appendix-numerical-procedure}
Here, we briefly summarize our numerical procedure for obtaining the exceptional points $g_c$.

For a given instance of a Hamiltonian (i.e., a fixed disorder realization), the exceptional point $g_c(\epsilon)$ at energy density $\epsilon$ is defined as 
\begin{equation}
g_c(\epsilon) = \inf \left\{ g\, |\, g > 0,\, \text{Im}[E({\epsilon})] \neq 0\right\} ,
\label{g_c_def}
\end{equation}
where $E({\epsilon})$ is the eigenvalue at energy density $\epsilon$. The length scale $\xi(\epsilon)$ is then defined by $\xi(\epsilon) = 1/g_c(\epsilon)$ - from here on we refer to $\xi(\epsilon)$ as the localization length.

For finite-size systems, the energies are discrete, and we must consider the eigenvalue closest to energy density $\epsilon$, given by
\begin{equation}
E({\epsilon}) = \arg\min_{E \in \mathcal{S}\{H\}}\left\lvert \frac{\re{E} - \re{E_{\min}}}{\re{E_{\max}} - \re{E_{\min}}} - \epsilon \right\rvert ,
\label{E_eps}
\end{equation}
where $\mathcal{S}\{H\}$ is the spectrum of $H$, and $E_{\max}$, $E_{\min}$ are the eigenvalues with the maximal and minimal real parts (respectively).

To find $g_c(\epsilon)$ numerically for a fixed disorder realization, we repeatedly increment $g$ by a fixed step $\delta g$ (which we take to be .01), exactly diagonalizing the Hamiltonian at each value of $g$, until $|\imag{E(\epsilon)}|$ is larger than some tolerance (which we take to be $10^{-8}$). Note this numerical approach differs from previous studies that examine all eigenvalues within some energy window to determine $g_c$. We choose to sample only one state per disorder realization to sidestep subtleties involving correlations of eigenvectors from the same disorder realization \cite{Rodriguez-Vasquez-Louella-Slevin-Romer-Multifractality}. We have verified that this approach gives the same disorder-averaged results as averaging over an energy window. % - see Appendix \ref{appendix-FSS} for finite-size scaling results that reproduce the results of Ref \cite{NHMBL} for the Hamiltonian \eqref{H}. 

\section{Identifying the MBL crossover via finite-size scaling}
\label{appendix-FSS}

In this Appendix, we show finite-size scaling collapse on the disorder-averaged localization lengths to identify the location of the (finite-size) MBL-thermal crossover for the Hamiltonian \eqref{H}. We find results in good agreement with those of \cite{NHMBL}.

\begin{figure}
\centering
\begin{subfigure}{.22\textwidth}
\includegraphics[scale=.225]{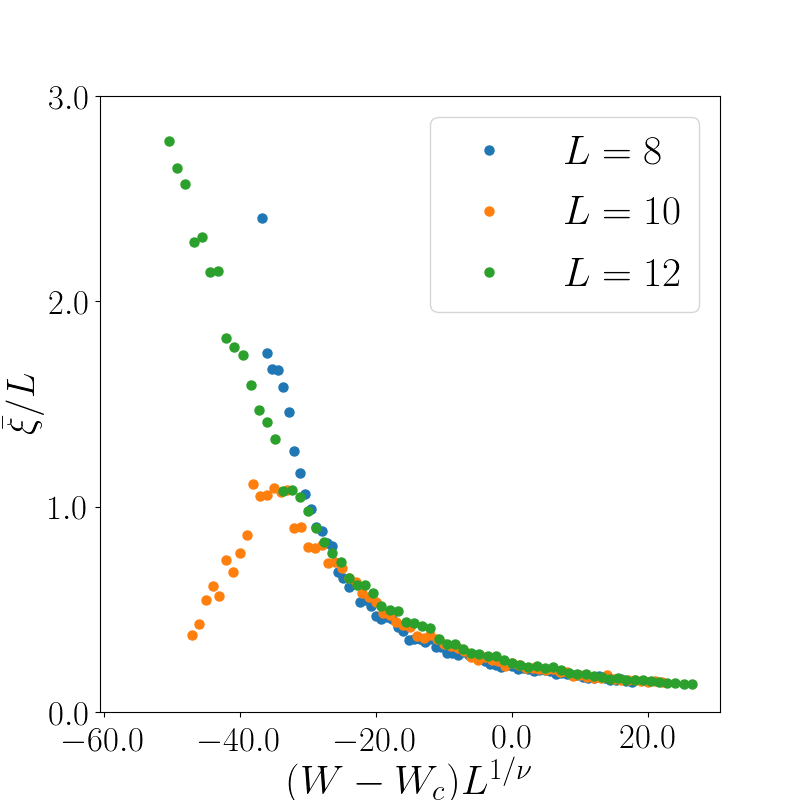}
\caption{}
\label{U=1}
\end{subfigure}
\hfill
\begin{subfigure}{.22\textwidth}
\includegraphics[scale=.225]{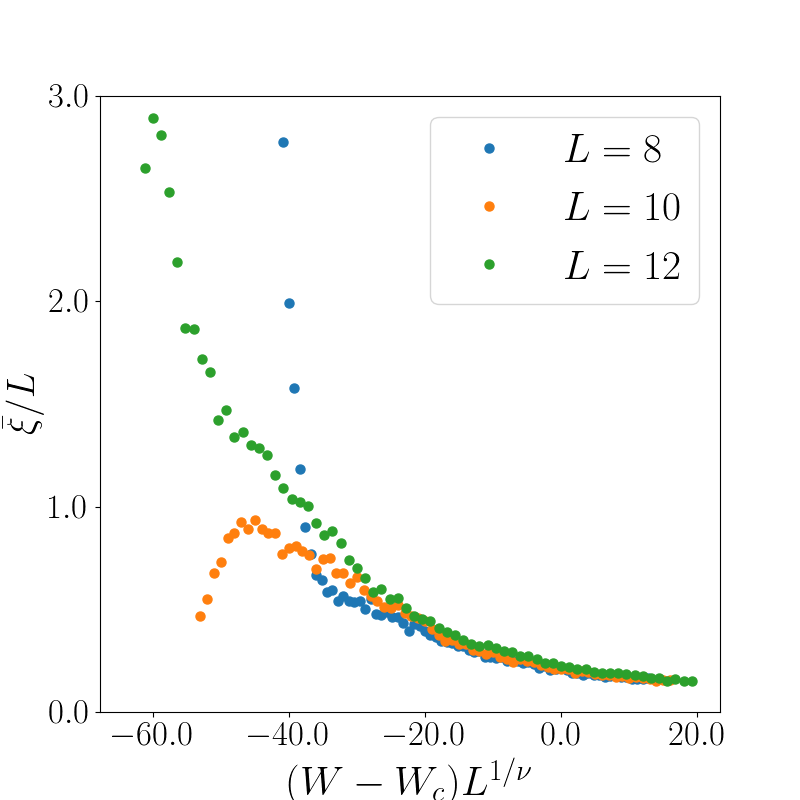}
\caption{}
\label{U=2}
\end{subfigure}
\\
\begin{subfigure}{.22\textwidth}
\includegraphics[scale=.225]{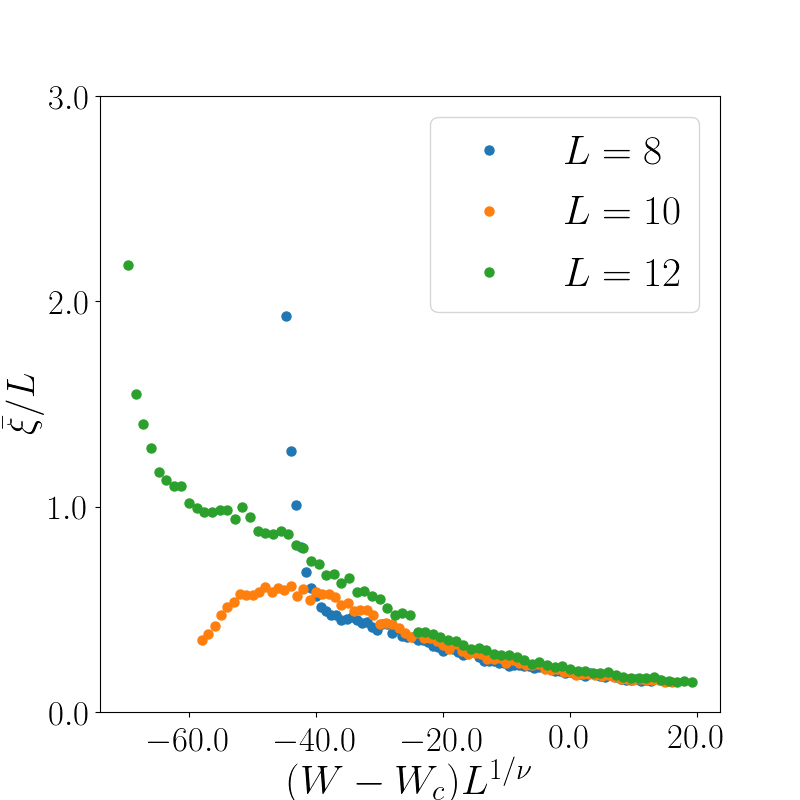}
\caption{}
\label{U=3}
\end{subfigure}
\hfill
\begin{subfigure}{.22\textwidth}
\includegraphics[scale=.225]{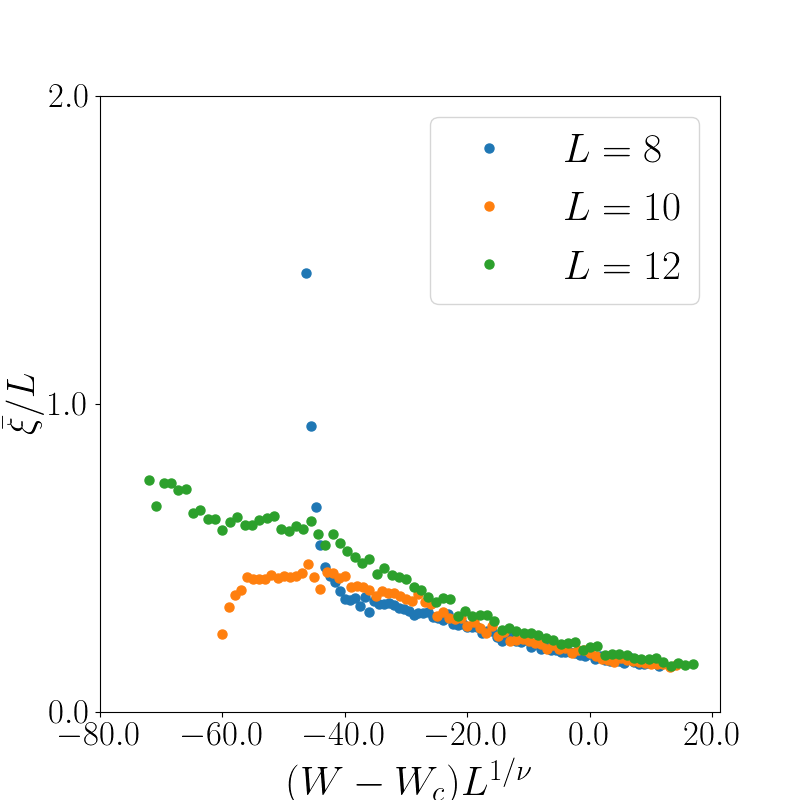}
\caption{}
\label{U=4}
\end{subfigure}
\\
\begin{subfigure}{.22\textwidth}
\includegraphics[scale=.225]{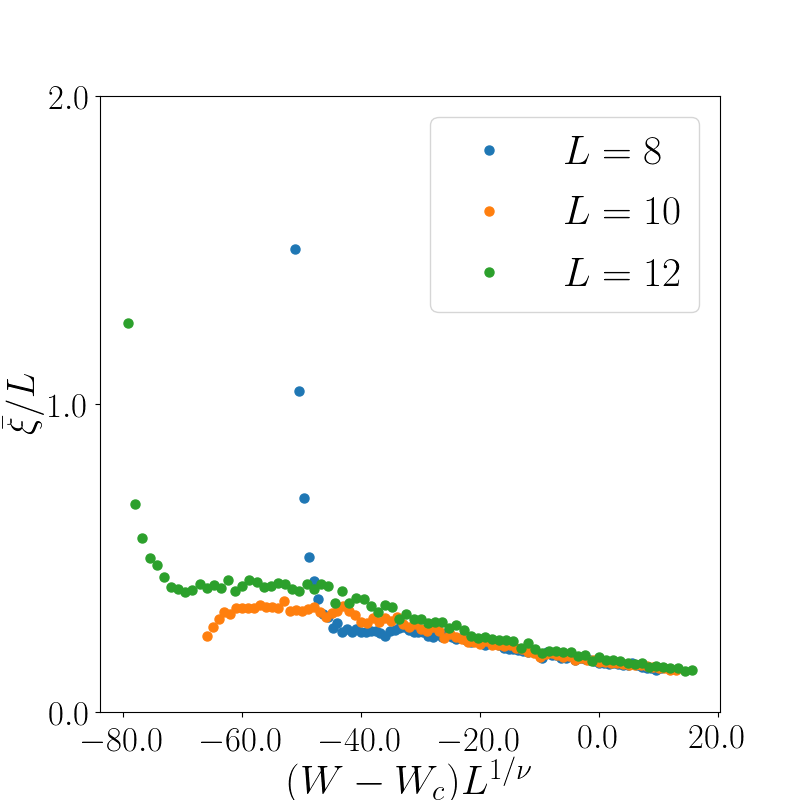}
\caption{}
\label{U=5}
\end{subfigure}

\caption{Finite-size scaling collapse for $\bar{\xi}/L = \avg{g_c}^{-1}/L$ at energy density $\epsilon = 0.5$ for a) $U=1$ , b) $U=2$, c)$U=3$, d) $U=4$, e) $U=5$. The critical exponent extracted in all cases is $\nu \approx 1$; the critical disorder strengths are $W_c \approx 4.8, 5.4, 5.9, 6.1, 6.7$ for $U=1,2,3,4,5$, respectively. All quantities are computed in the half-filling sector, and averaged over 200 disorder realizations.}
\label{FSS} 
\end{figure}

\begin{figure}
\centering 
\begin{subfigure}{.23\textwidth}
\includegraphics[scale=.15]{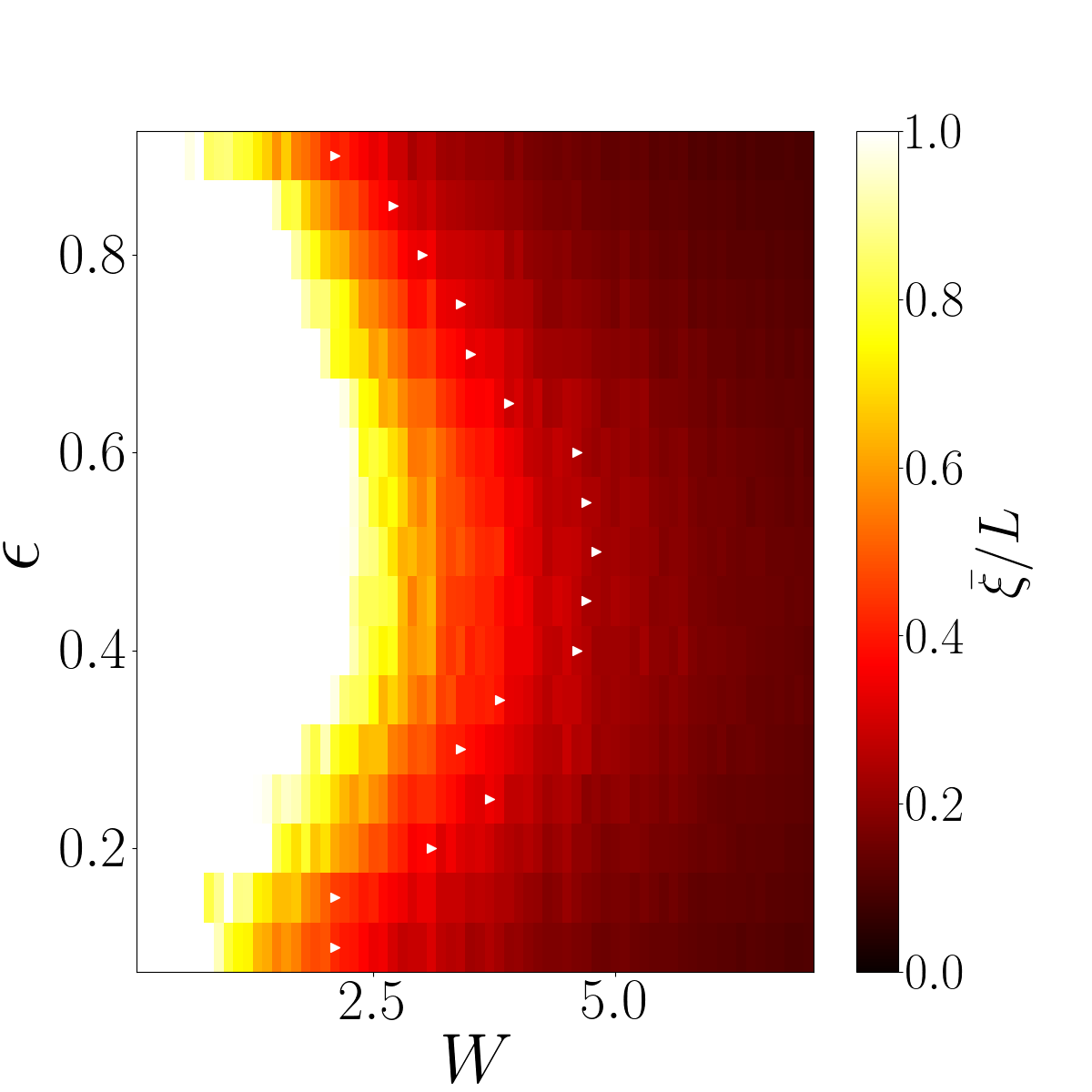}
\caption{}
\label{U=1}
\end{subfigure}
\hfill
\begin{subfigure}{.23\textwidth}
\includegraphics[scale=.15]{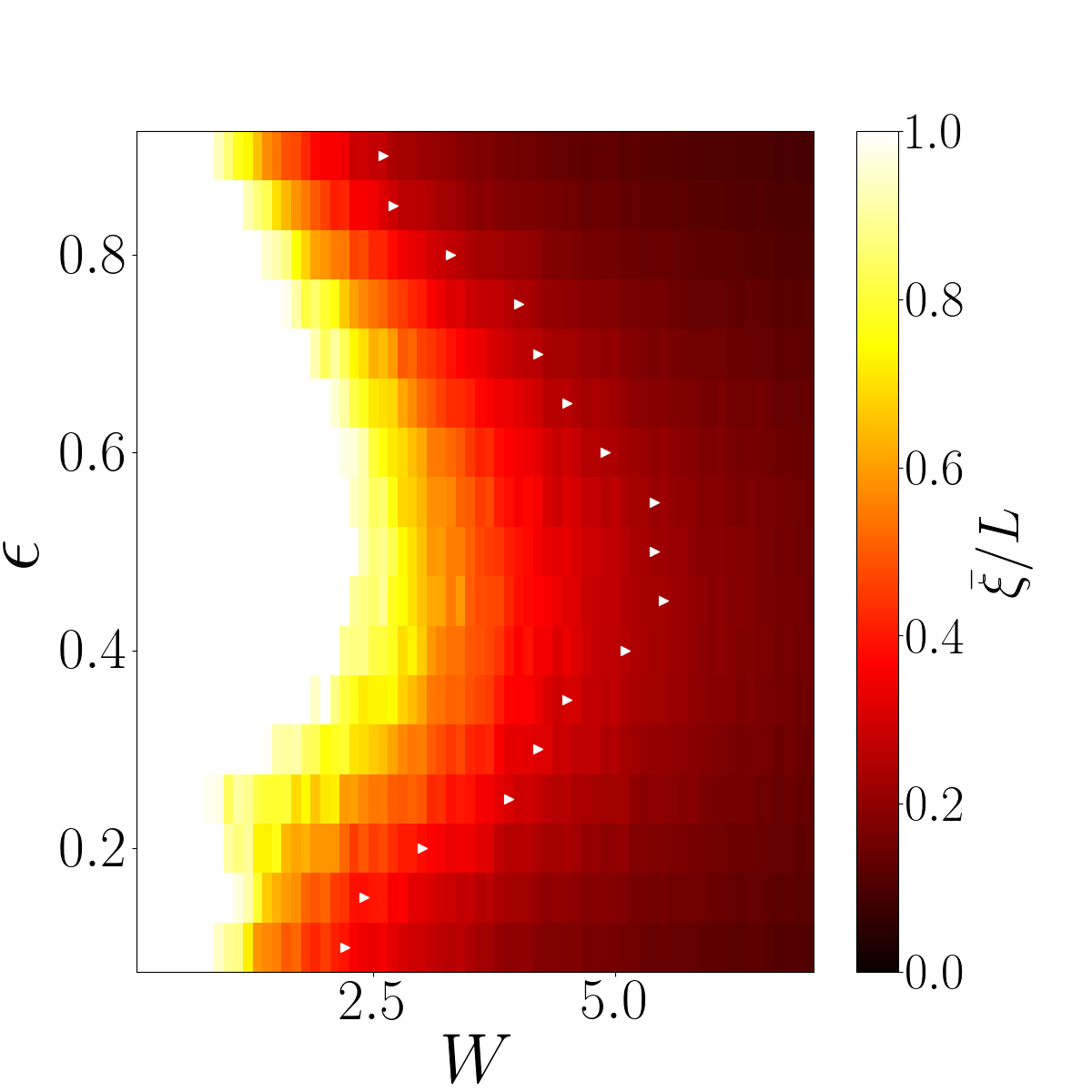}
\caption{}
\label{U=2}
\end{subfigure}
\\
\begin{subfigure}{.23\textwidth}
\includegraphics[scale=.15]{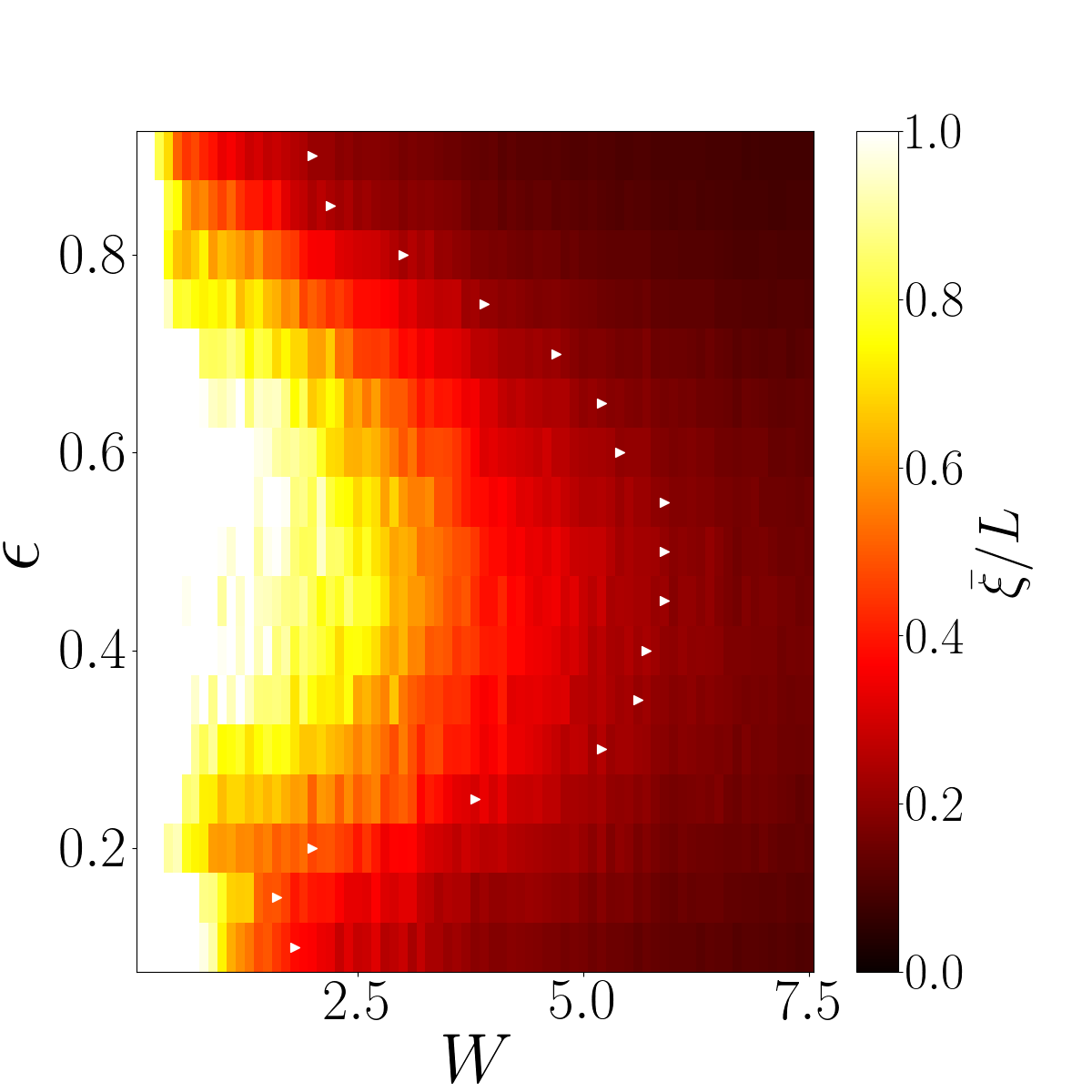}
\caption{}
\label{U=3}
\end{subfigure} 
\hfill
\begin{subfigure}{.23\textwidth}
\includegraphics[scale=.15]{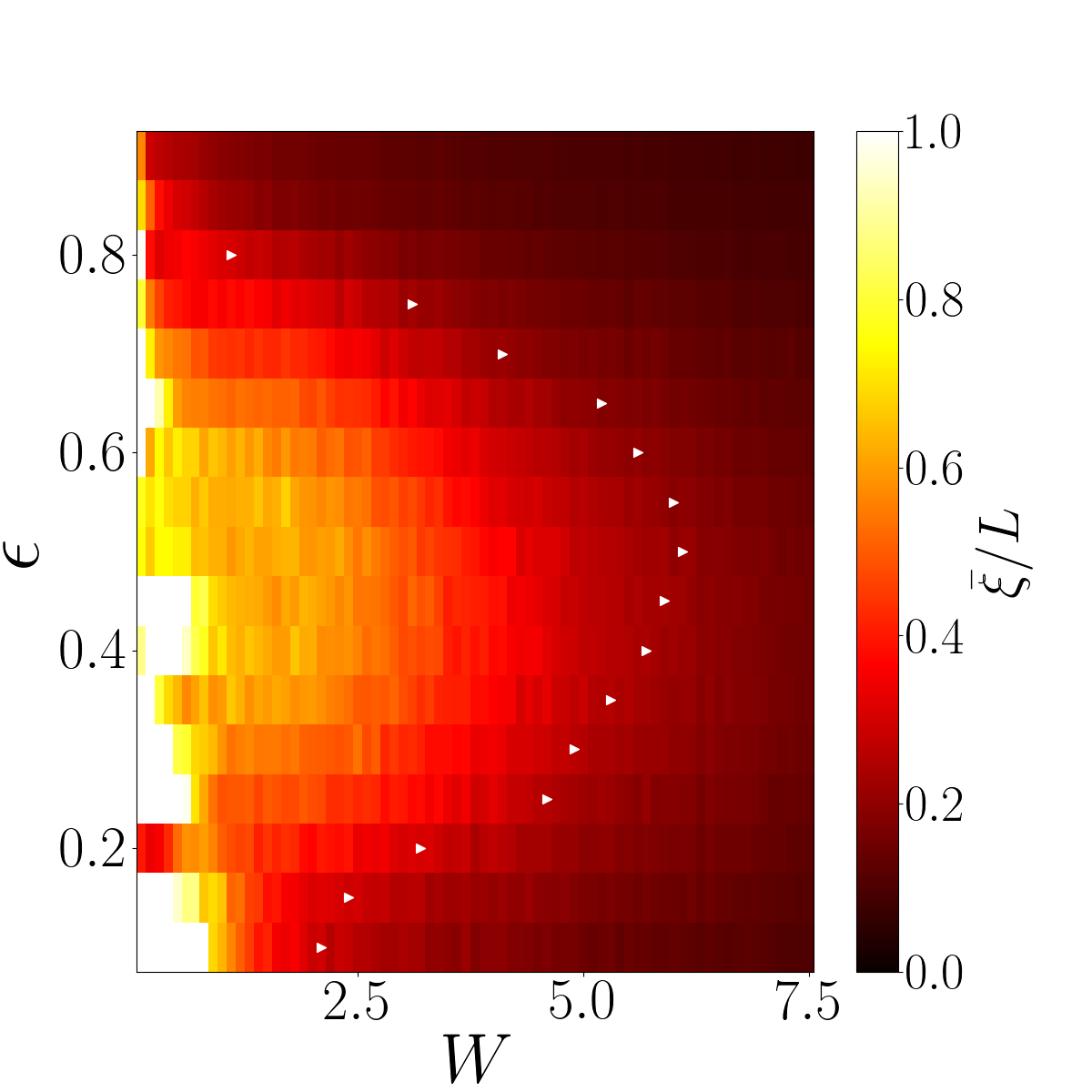}
\caption{}
\label{U=4}
\end{subfigure}
\\
\begin{subfigure}{.22\textwidth}
\includegraphics[scale=.15]{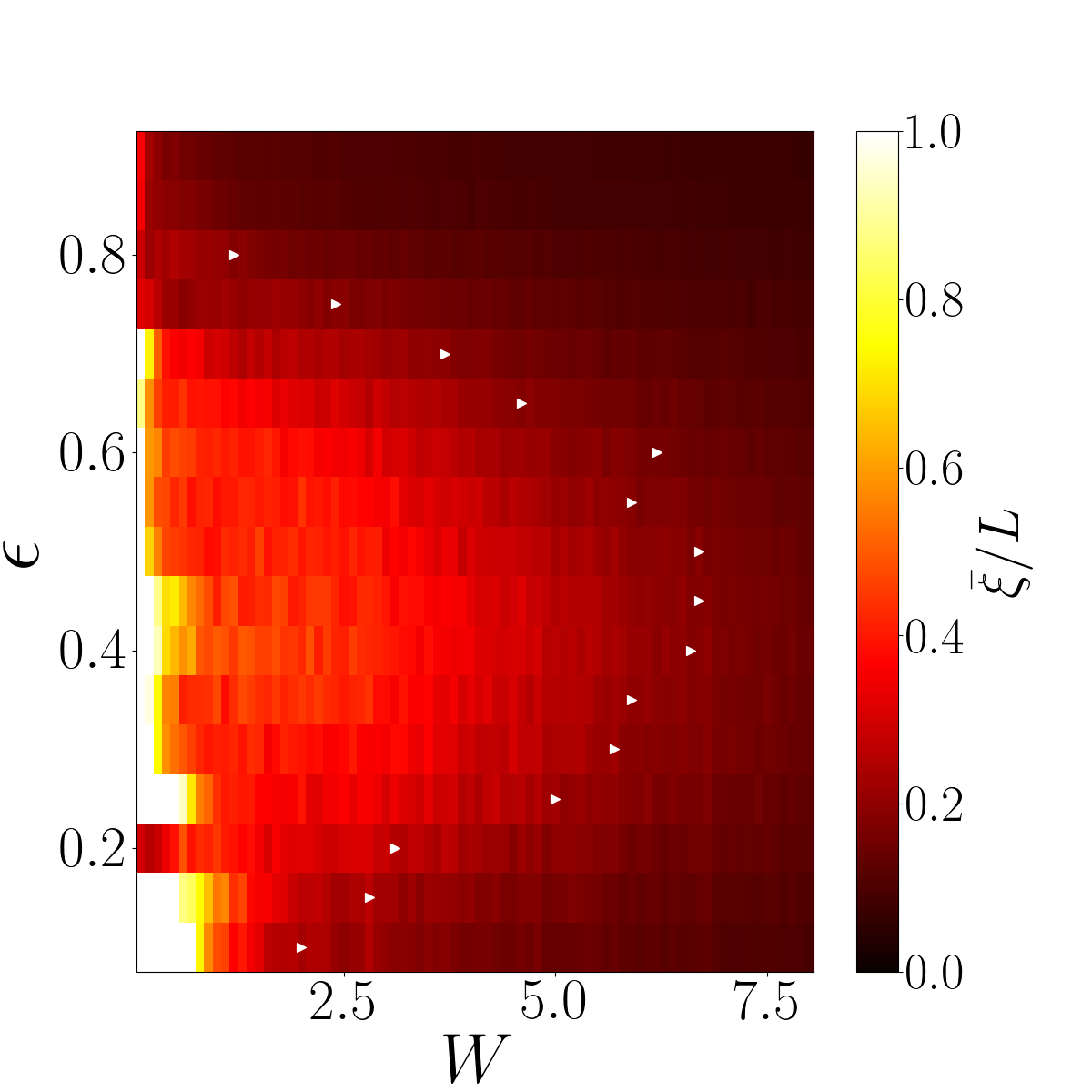}
\caption{}
\label{U=5}
\end{subfigure}

\caption{Phase portraits for $\bar{\xi}/L = \avg{g_c}^{-1}/L$ in the $\epsilon-W$ plane, for a) $U=1$ , b) $U=2$, c) $U=3$, d) $U=4$, e) $U=5$, for a chain of size $L=12$. White triangles mark the critical disorder strengths obtained from our finite size scaling collapse. All quantities are computed in the half-filling sector, and averaged over 200 disorder realizations.}
\label{phase-portraits} 
\end{figure}

To that end, we compute $g_c(\epsilon)$ (as described in Appendix \ref{appendix-numerical-procedure}) for the Hamiltonian \eqref{H}, with bare hopping $t=1$ (making it the energy scale of our system) and on-site disorders sampled from a uniform distribution $w_i \sim \text{Uni}[-W,W]$. Figure \ref{FSS} shows finite size-scaling of $\bar{\xi}/L = \avg{g_c}^{-1}/L$ (where $\avg{\cdot}$ denotes a disorder average) at the center of the spectrum ($\epsilon = 0.5$) in the half-filling sector. We examine a variety of interaction strengths $U = 1,\,2,\,3,\,4,\,5$, at system sizes $L = 8,\,10,\,12$, and average over 200 disorder realizations. We obtain good collapse of the data to the form
\begin{equation}
\frac{\bar{\xi}}{L} = f\left((W-W_c)L^{1/\nu}\right) ,
\label{FSS_ansatz}
\end{equation}
and in all cases find critical exponents $\nu \approx 1$. The resulting phase portraits in the $\epsilon$-$W$ plane are shown in Fig. \ref{phase-portraits}, with the transition points identified from the finite size scaling marked. We see the transition points capture the change from large (order of system size) to small localization length as we tune the disorder strength. The shape of the mobillity edge in the phase diagram also mirrors that of other finite size studies \cite{Luitz-MBL-trans}, and the critical disorder strengths agree well with \cite{NHMBL} .

\section{Additional Results Regarding Non-Hermitian Avalanches}
\label{appendix-NH}
\begin{figure}
\begin{center}
\begin{subfigure}{.22\textwidth}
\includegraphics[scale=.22]{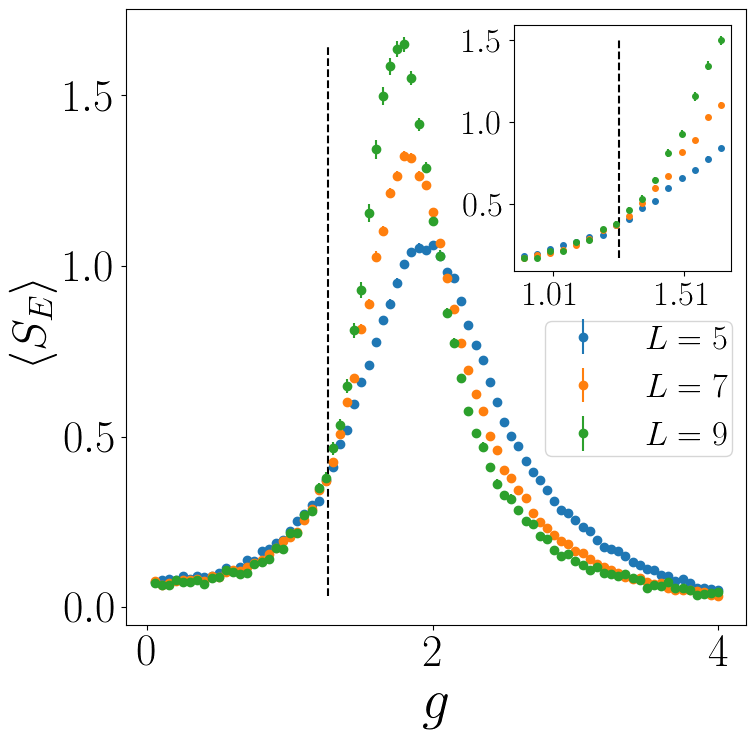}
\caption{}
\end{subfigure}
\hfill
\begin{subfigure}{.22\textwidth}
\includegraphics[scale=.22]{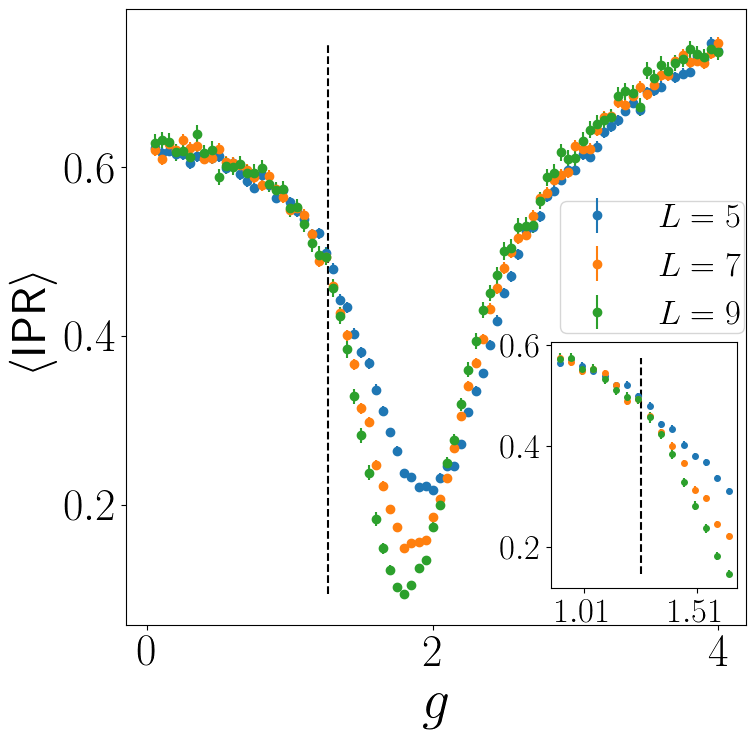}
\caption{}
\end{subfigure}\\
\begin{subfigure}{.24\textwidth}
\includegraphics[scale=.22]{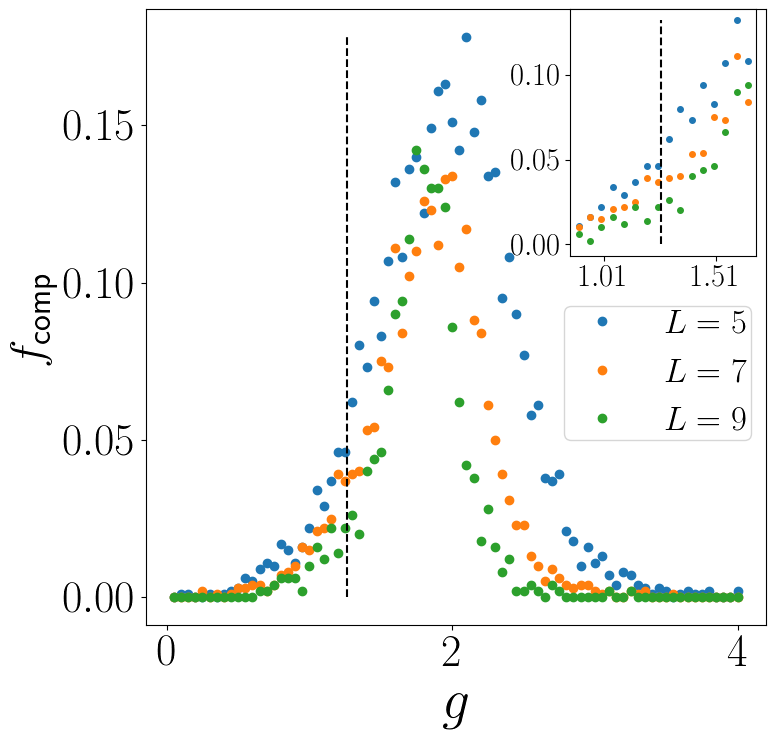}
\caption{}
\end{subfigure}
\end{center}
\caption{Disorder-averaged localization metrics of the eigenstates of the non-hermitian avalanche Hamiltonian \eqref{H_av_NH} for $e^{-1/\lambda} = 0.2$ and various $L$, as a function of $g$. Going clockwise: a) half-chain entanglement entropy $\langle S_E\rangle$, b) Inverse participation ratio $\langle IPR \rangle$, and c) Fraction $f_{\text{comp}}$ of disorder realizations whose central eigenvalue ($\epsilon = 0.5$) is non-real. Error bars on $\langle S_E\rangle$ and $\langle IPR \rangle$ indicate the standard error of the mean. As in Fig. \ref{NH_av_numerics}, we see a crossover from localized to delocalized behavior in panels a) and b) which occurrs roughly at the expected value of $g$ (dashed lines), along with a corresponding real-complex crossover in panel c).
All quantities were computed exactly as in Fig. \ref{NH_av_numerics}. }
\label{NH_av_numerics_0.2}
\end{figure}

\begin{figure}
\begin{center}
\begin{subfigure}{.22\textwidth}
\includegraphics[scale=.22]{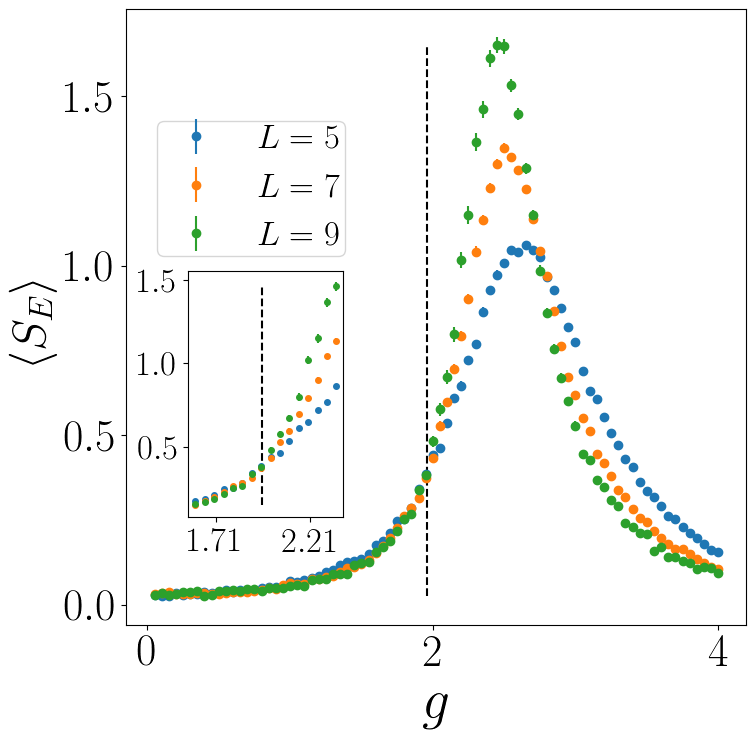}
\caption{}
\end{subfigure}
\hfill
\begin{subfigure}{.22\textwidth}
\includegraphics[scale=.22]{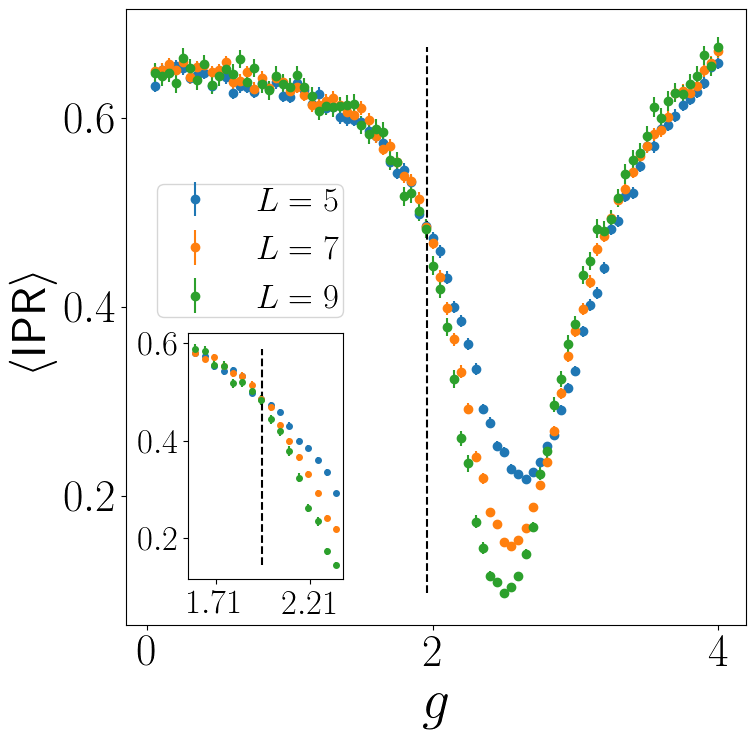}
\caption{}
\end{subfigure}\\
\begin{subfigure}{.24\textwidth}
\includegraphics[scale=.22]{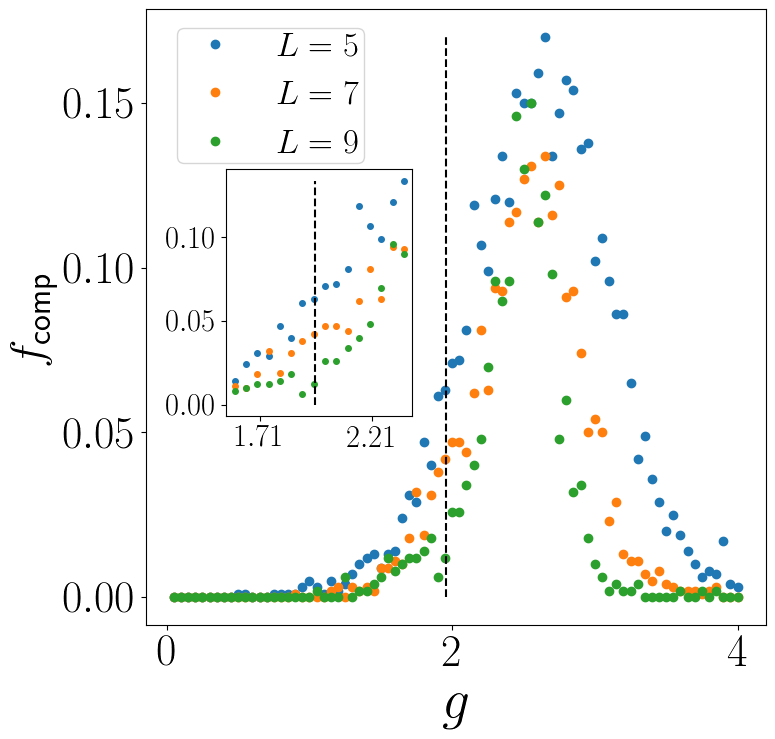}
\caption{}
\end{subfigure}
\end{center}
\caption{Disorder-averaged localization metrics of the eigenstates of the non-hermitian avalanche Hamiltonian \eqref{H_av_NH}, for $e^{-1/\lambda}=0.1$ and various $L$, as a function of $g$. Going clockwise: a) half-chain entanglement entropy $\langle S_E\rangle$, b) Inverse participation ratio $\langle IPR \rangle$, and c) Fraction $f_{\text{comp}}$ of disorder realizations whose central eigenvalue ($\epsilon = 0.5$) is non-real for various $L$ and $g$. Error bars on $\langle S_E\rangle$ and $\langle IPR \rangle$ indicate the standard error of the mean. We observe similar crossovers as seen in Fig. \ref{NH_av_numerics} and Fig. \ref{NH_av_numerics_0.2}. }
\label{NH_av_numerics_0.1}
\end{figure}

Here, we show additional numerical results supporting the existence of the non-hermitian avalanche mechanism, as in Fig. \ref{NH_av_numerics}. Results for a matrix element decay of $e^{-1/\lambda} = 0.2$ and $e^{-1/\lambda} = 0.1$ are shown in Fig. \ref{NH_av_numerics_0.2} and Fig. \ref{NH_av_numerics_0.1}, respectively. We see crossovers analogous to those observed in the main text, again occurring roughly at the expected value of $g = 1/\lambda - 1/\lambda_c$.

\section{Diagonalizing many-body hamiltonians via generalized displacement transformations}
\label{appendix-Mapping}

In this appendix, we detail a diagonalization algorithm that iteratively applies a series of transformations eliminating ``off-diagonal'' terms from a given fermionic Hamiltonian.

Consider fermions hopping on a 1-dimensional lattice with sites $S = \{i\}_{i=1}^L$. Let $c_i$, $c_i^{\dag}$, and $n_i$ be the fermionic annhiliation, creation, and number operators at a site $i$, respectively. A general (hermitian) Hamiltonian then has the form:
\begin{equation}
H = V + \sum_{A,B} t_{AB}I_{AB} ,
\label{H_gen}
\end{equation} 
where $V$ is a diagonal operator, $A,B$ are disjoint subsets of $S$, $t_{AB}$ is a diagonal operator acting only on sites in $S \setminus A\cup B$, and $I_{AB}$ is the operator for hopping between $A$ and $B$:
\begin{equation}
I_{AB} = \left(\prod_{a \in A}c_a^{\dag}\right)\left(\prod_{b \in B}c_b\right) + \text{ h.c. }
\label{I_AB}
\end{equation}
We neglect here the possibility of complex phases multiplying the hopping terms - our algorithm presented below is straightforwardly modified in this case, and the essential results do not change.

Suppose we now want to eliminate the hopping $I_{AB}$ via a unitary change of basis. We can do this by the following transformation
\begin{equation}
H \mapsto e^{\mathcal{O}_{AB} J_{AB}}\,H\,e^{-\mathcal{O}_{AB}J_{AB}} ,
\label{SW}
\end{equation}
where 
\begin{equation}
J_{AB} = \left(\prod_{a\in A} c_a^{\dag}\right)\left(\prod_{b \in B} c_b\right) - \text{h.c.} ,
\label{J_AB}
\end{equation}
and $\mathcal{O}_{AB}$ is a diagonal operator acting only on sites in $S \setminus A \cup B$ satisfying
\begin{equation}
\tan\left(2\mathcal{O}_{AB}\right) = \frac{2t_{AB}}{\Delta_{AB} V} ,
\label{tan_O}
\end{equation}
where 
\begin{equation}
\Delta_{AB} V := V\big{\rvert}_{\substack{A \text{ occupied},\\B \text{ unoccupied}}} - V\big{\rvert}_{\substack{B \text{ occupied},\\A \text{ unoccupied}}} .
\end{equation}
Application of the transformation \eqref{SW} to the Hamiltonian \eqref{H_gen} results in a Hamiltonian with no terms
of the form $\mathcal{D}\,I_{AB}$ present (where $\mathcal{D}$ is a diagonal operator). Generically, such a Hamiltonian will include new diagonal terms, as well as new hopping terms (created by the action of the transformation on the other hopping terms in \eqref{H_gen}).\\

Now, we can repeatedly apply transformations of the form \eqref{SW} to eliminate all hopping terms in the original Hamiltonian \eqref{H_gen}. As mentioned previously, each application of such a transformation generates new hopping terms; in the case of repeated transformations, we can actually re-introduce hoppings that were eliminated by previous transformations. This is not a problem, as the terms that re-introduced will have smaller and smaller ``magnitude'' over time. To be more precise, let us write out the unitary map in \eqref{SW} in closed form:
\begin{equation}
\resizebox{.38\textwidth}{!}{$e^{\pm\mathcal{O}_{AB}J_{AB}} = \mathbb{I} + [\cos(\mathcal{O}_{AB})-\mathbb{I}]P_{AB} \pm \sin(\mathcal{O}_{AB})J_{AB} ,$}
\label{SW_expanded} 
\end{equation}
where $P_{AB}$ is the projector onto the subspace in which all sites in either $A$ or $B$ (but not both) are filled:
\begin{equation}
\resizebox{.38\textwidth}{!}{$\displaystyle P_{AB} = \left(\prod_{a\in A}n_a\right)\left(\prod_{b\in B}(\mathbb{I} - n_b)\right) +  \left(\prod_{b\in B}n_b\right)\left(\prod_{a\in A}(\mathbb{I} - n_a)\right) .$}
\label{proj}
\end{equation}
In the form \eqref{SW_expanded}, we can see that every new hopping term introduced by a transformation of the form \eqref{SW} will include an operator multiplication by a trigonemtric function of $\mathcal{O}_{AB}$, which will have operator norm $< 1$. If we iterate such transformations, we can see that the re-introduced hopping terms will converge (in operator norm) to zero. Thus, in the limit of infinitely many transformations of the form \eqref{SW} applied, all the hopping (non-diagonal) terms will aproach zero, and we will be left with a diagonal Hamiltonian. Numerically, we must cut the procedure off when the largest (given by magnitude of largest matrix element) off-diagonal term has norm below some threshold - we choose this to be $10^{-8}$. 

Note also that by applying the same sequence of transformations to the creation/annhilation operators, we can obtain creation/annhilation operators for the local integrals of motion for the Hamiltonian \eqref{H_gen} (assuming it is in the MBL phase). Additionally, if we track the evolution of the hopping terms as we iterate these transformations - or leave certain hopping terms untouched by our transformations - we will see the amplitudes evolve as a function of iteration number. This is reminiscent of a discretized version of the Wegner flow employed by \cite{Quito-Titum-Pekker-Refael-Wegner-Flow}. Indeed, this intuition is how we obtain the ``avalanche-like'' hopping amplitudes described in the main text.

As a final note, this algorithm can be generalized to non-hermitian Hamiltonians. In such a case, the transformation to eliminate a term of the form
\begin{equation}
\scalebox{.93}{$\displaystyle I^{\alpha}_{AB} = \left(\prod_{a \in A}c_a^{\dag}\right)\left(\prod_{b \in B}c_b\right) + \alpha\left(\prod_{b \in B}c_b^{\dag}\right)\left(\prod_{a \in A}c_a\right) ,$}
\label{I_alpha_AB}
\end{equation}
where $\alpha$ is diagonal on $S \setminus A \cup B$ - is constructed analogously, but now with $\cos(\mathcal{O}_{AB}) \mapsto \cos(\sqrt{\alpha}\,\mathcal{O}_{AB})$ and $\sin(\mathcal{O}_{AB}) \mapsto \sin(\sqrt{\alpha}\,\mathcal{O}_{AB})/\sqrt{\alpha}$ in \eqref{tan_O} and \eqref{SW_expanded}. Note that, if any of the eigenvalues of $\alpha$ are negative, $\cos(\sqrt{\alpha}\,\mathcal{O}_{AB})$ and $\sin(\sqrt{\alpha}\,\mathcal{O}_{AB})$ are no longer necessarily bounded in operator norm, and convergence of the algorithm is no longer assured. This is why it is crucial to use the gauge freedom to eliminate the non-hermiticity from the part of the Hamiltonian we are diagonalizing, as described in the main text. 

These transformations \eqref{SW}, and the iterative procedure described here, are examples of the ``displacement transformations'' and diagonalization algorithm first proposed by Rademaker and Ortu\~{n}o \cite{Rademaker-Ortuno}.

\section{Derivation of Cumulative Density Function at the MBL transition}
\label{appendix-CDF}
In this appendix, we seek to derive an analytic expression for the distribution of $g_c$ at the MBL transition, based on the (generalized) avalanche condition \eqref{g_c_relation}. As it turns out, it is easier to derive the \textit{cumulative} density function (CDF) of $g_c$. To that end, let us consider the following more general problem.

Let $\{x_i\}_{i=1}^N$ be a collection of random variables, and $\alpha$ a real number. Consider the following extremization problem:
\begin{equation}
M_L = \max_{1 \leq j \leq L}\left\{ \sum_{i=1}^jx_i  - (j-1)\alpha\right\} := \max_{1 \leq j \leq L} m_i .
\label{M_L}
\end{equation}
We wish to find the CDF of $M_L$, given the probability densities $\rho_i(x_i)$ of the $x_i$'s. 

To write down the generic form of this CDF, we first make the following observation:
\begin{equation}
P(M_L \leq x) = P(m_1 \leq x, m_2 \leq x,\ldots, m_L \leq x)
\label{CDF_joint}
\end{equation}
That is, $M_L \leq x$ iff $m_j \leq x$ for every value of $j$. Futhermore, we can see that $m_j \leq x$ iff $x_j \leq x + (j-1)\alpha -  \sum_{i=1}^{j-1}x_i$. Hence,
\begin{equation}
P(m_j \leq x) = P\left( x_j \leq x + (j-1)\alpha - \sum_{i=1}^{j-1}x_i \right) .
\label{CDF_mj}
\end{equation}
The right hand side is the CDF of $x_j$, evaluated at a point that depends on the other $x_i$. Using this in the right side of \eqref{CDF_joint}, we obtain an expression for $P(M_L \leq x)$ in terms of the PDFs of the $x_j$'s (i.e., the $\rho_j$'s):
\begin{equation}
P(M_L \leq x) = \prod_{j=1}^L \int_{-\infty}^{x + (j-1)\alpha - \sum_{i=1}^{j-1}x_i}\intd x_j\, \rho_j(x_j) .
\label{CDF_generic}
\end{equation}
At this point, we can compute the integral over $x_L$ to obtain an integral expression involving the CDF of $x_L$, but we cannot go any farther without knowledge of the $\rho_i$'s.

To that end, let us now take each of the $x_i$'s to be exponentially distributed - that is, $\rho_i(x) = e^{-x}$ for $x \geq 0$, and $0$ for $x < 0$. We can now perform the integrals in \eqref{CDF_generic} for any finite value of $L$, and obtain
\begin{equation}
\scalebox{.9}{$\displaystyle P(M_L \leq x) = 1 - e^{-x}\left[ 1 + x\sum_{j=1}^{L-1}\frac{e^{-j\alpha}}{j!}(x + j\alpha)^{j-1}\right] $} .
\label{CDF_exp}
\end{equation}
This is the desired CDF of $M_L$, assuming $x_i \sim \text{Exp}[1]$ for each $i$. The CDF for a more general exponential distribution $x_i \sim  \text{Exp}[\lambda]$ can be obtained by making the replacements $x,\, \mapsto \lambda x$, $\alpha \mapsto \lambda \alpha$.  

Let us now connect back to the exceptional points. We have the generalized ``avalanche criterion'' for $g_c$:
\begin{equation}
g_c \approx \frac{1}{F}\max_{1\leq \ell \leq N_{\ell}} \left[ -\log\mathcal{A}_{\ell}(\epsilon) - (\ell-1)\alpha\right] + g_0.
\label{g_c-av-criterion}
\end{equation}
Comparing \eqref{M_L} with this  generalized criterion, and assuming $\mathcal{A}_{\ell}(\epsilon) = \prod_{i=1}^{\ell} \chi_{i}(\epsilon)$, we see that $g_c = \frac{1}{F}M_{N_{\ell}}$, with $x_i = -\log \chi_{i}(\epsilon)$. Assuming $\chi_{i} \sim \text{Uni}[0,1]$ at the transition (as observed in Section  \ref{mapping-results}), it follows that $-\log \chi_{i}(\epsilon) \sim \text{Exp}[1]$. Therefore, the result \eqref{CDF_exp} with the replacements $x \mapsto F(x-g_0)$, $L \mapsto N_{\ell}$ and $\alpha = \log 2/2$ should describe the distribution of $g_c(\epsilon)$ at the MBL crossover.

Note that the general formula \eqref{CDF_generic}, with appropriate replacements, gives the distribution for $g_c$ whenever an avalanche criterion of the form \eqref{g_c-av-criterion} holds. In particular, if the avalanche criteria holds deep in the MBL regime, this formula should still give the distribution of $g_c$. Of course, the $\chi_i(\epsilon)$ are not necessarily uniformly distributed away from the crossover region; nonetheless, replacing $\rho_i$ by the appropriate distribution of $-\log \chi_i(\epsilon)$ should yield the correct distribution for $g_c(\epsilon)$.

\section{Additional results from Numerical Fits of Probability distributions}
\label{appendix-Distribution-Fits}

In this appendix, we show some additional results from the numerical fits of the CDFs of $g_c(\epsilon)$ described in section \ref{distributions}.

\begin{figure}
\begin{subfigure}{.22\textwidth}
\includegraphics[scale=.35]{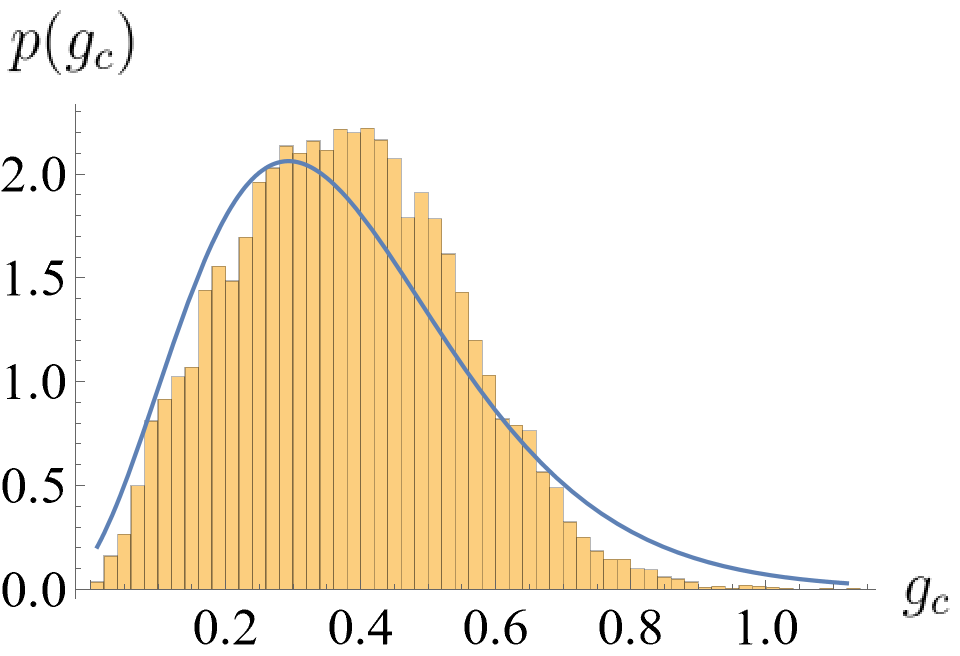}
\caption{}
\label{CDF_fits_PDF_U=1}
\end{subfigure}
\hfill
\begin{subfigure}{.22\textwidth}
\includegraphics[scale=.35]{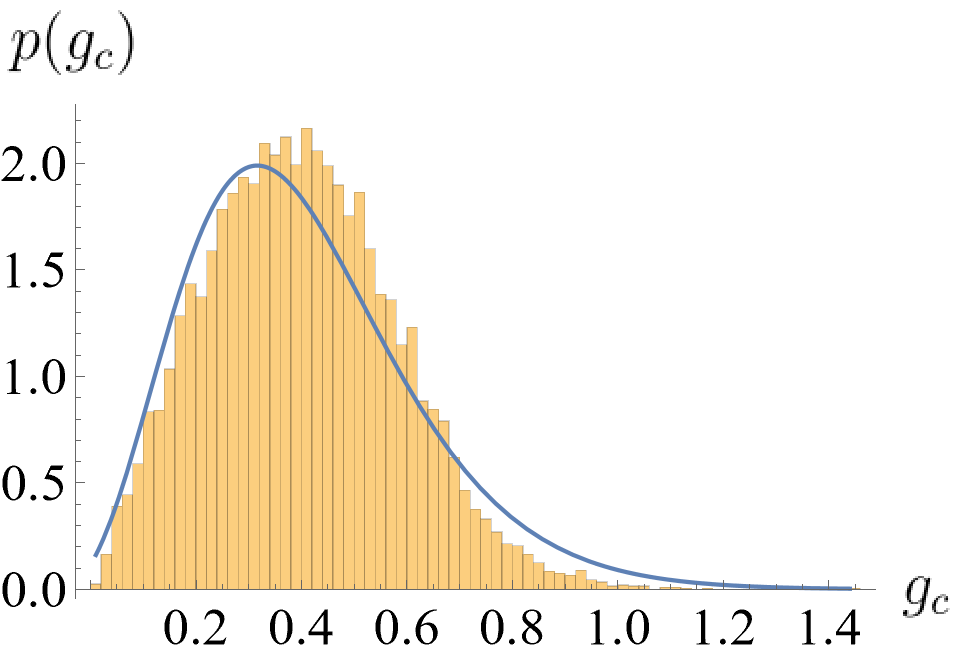}
\caption{}
\label{CDF_fits_PDF_U=2}
\end{subfigure}
\\
\begin{subfigure}{.22\textwidth}
\includegraphics[scale=.35]{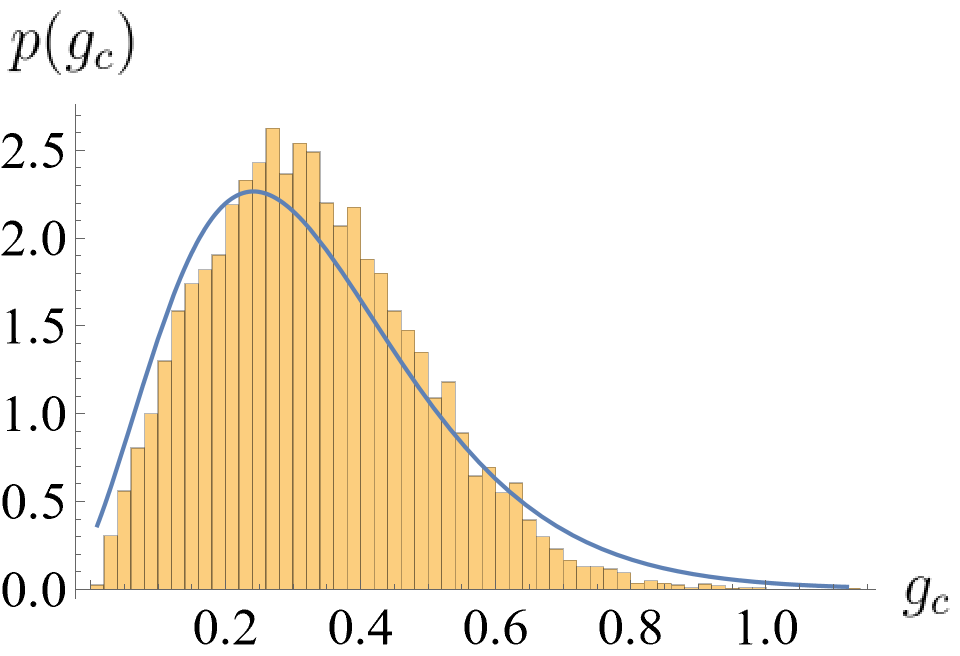}
\caption{}
\label{CDF_fits_PDF_U=3}
\end{subfigure}
\hfill
\begin{subfigure}{.22\textwidth}
\includegraphics[scale=.35]{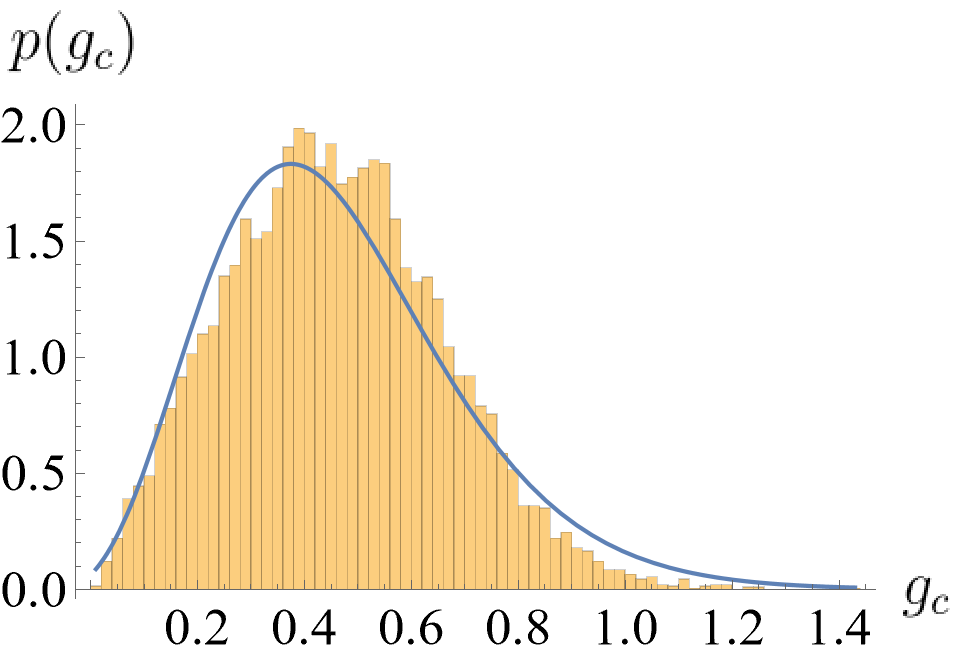}
\caption{}
\label{CDF_fits_PDF_U=4}
\end{subfigure}
\\
\begin{subfigure}{.22\textwidth}
\includegraphics[scale=.35]{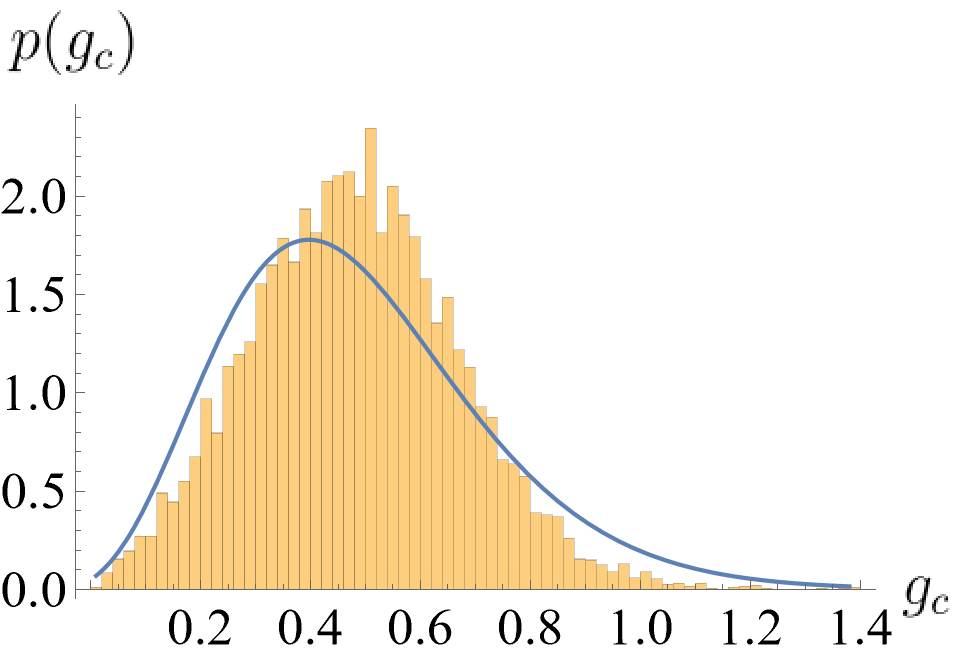}
\caption{}
\label{CDF_fits_PDF_U=5}
\end{subfigure}

\caption{Numerically observed probability density functions of $g_c$ for $\epsilon = 0.5$ at the critical point, for interaction strengths a) $U=1$ ($W=4.8$) , b) $U=2$ ($W = 5.4$), c) $U=3$ ($W = 5.9$), d) $U=4$ ($W = 6.1$), e) $U=5$ ($W = 6.7$). The blue solid lines overlaid are the pdfs obtained from differentiating the best fit mixture distributions in Fig. \ref{CDF_fits}.}
\label{CDF_fits_PDFs}
\end{figure}

Figure \ref{CDF_fits_PDFs} shows the numerically observed probability density function for $g_c(\epsilon)$, along with the analytic PDF obtained by differentiating the best fit mixture distribution of section \ref{distributions}, for the data in Fig \ref{CDF_fits}. We can see that the PDFs generated from our fit have thicker tails (i.e., approach zero more slowly) than the numerically observed PDFs, and that the peak of the numerically observed PDF is larger than the peak of the fitted distribution. Despite these discrepancies, the PDFs from the fits still capture the observed behavior qualitatively well, especially at small $g_c$ and for $U=3$.

\begin{figure}
\begin{subfigure}{.22\textwidth}
\includegraphics[scale=.35]{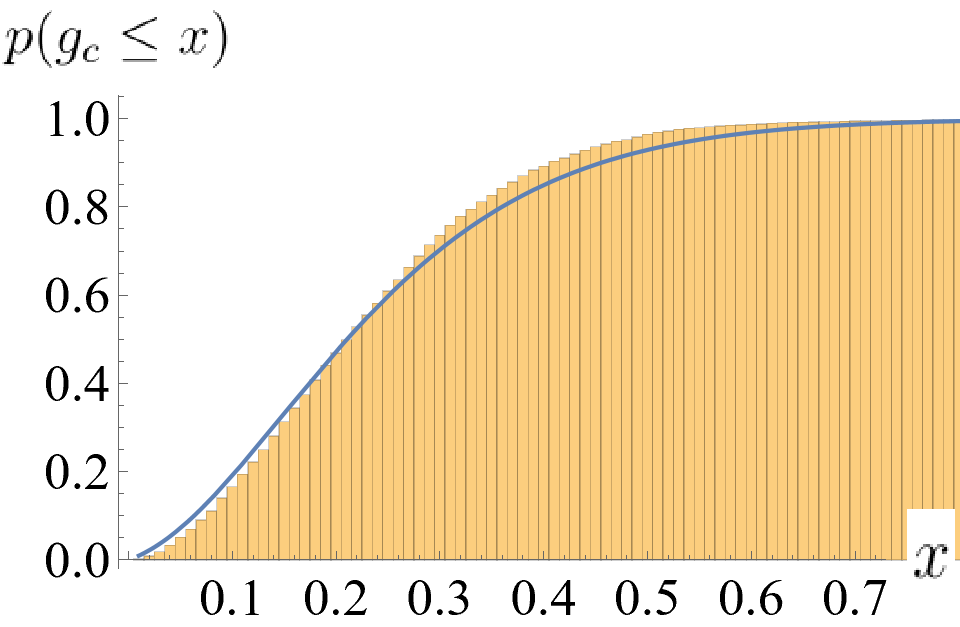}
\caption{}
\label{CDF_fits_min_U=1}
\end{subfigure}
\hfill
\begin{subfigure}{.22\textwidth}
\includegraphics[scale=.35]{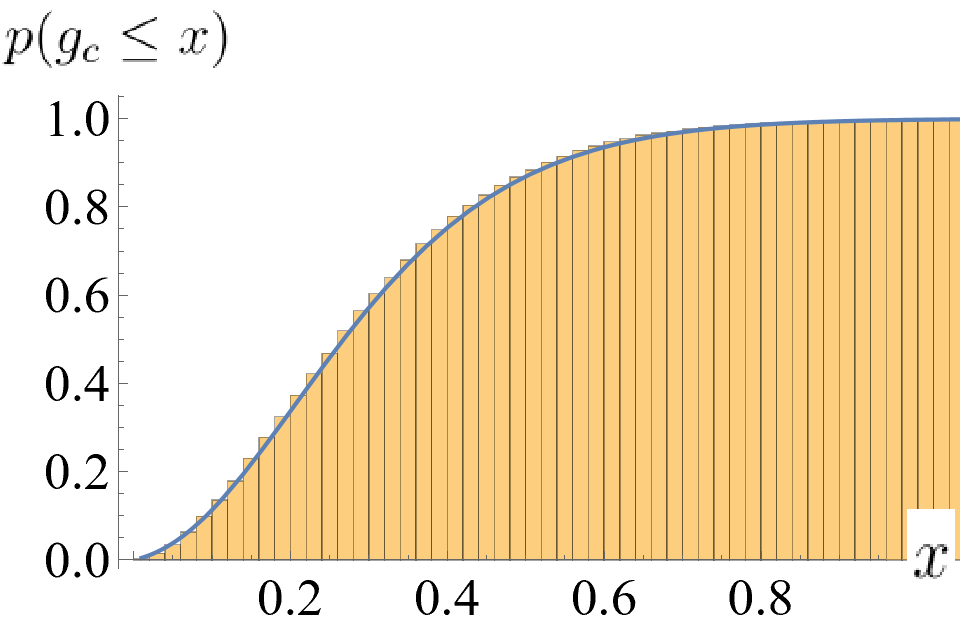}
\caption{}
\label{CDF_fits_min_U=2}
\end{subfigure}
\\
\begin{subfigure}{.22\textwidth}
\includegraphics[scale=.35]{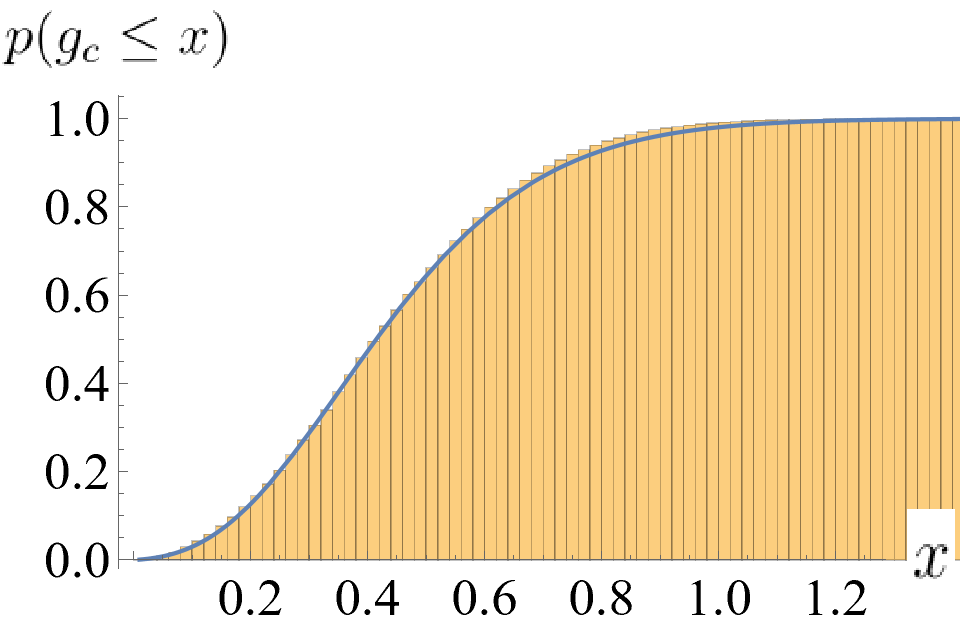}
\caption{}
\label{CDF_fits_min_U=3}
\end{subfigure}
\begin{subfigure}{.22\textwidth}
\includegraphics[scale=.35]{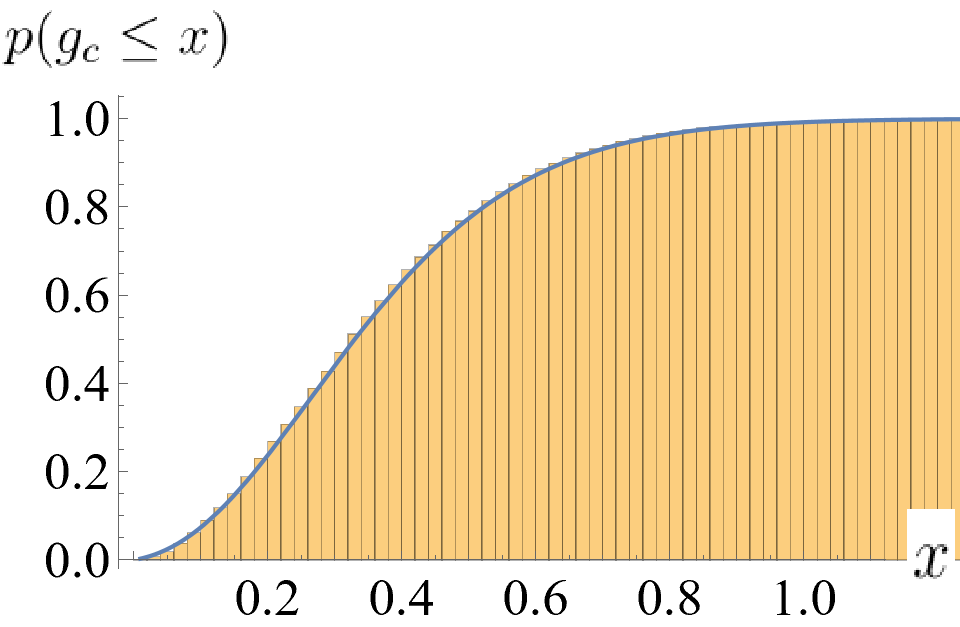}
\caption{}
\label{CDF_fits_min_U=4}
\end{subfigure}
\\
\begin{subfigure}{.22\textwidth}
\includegraphics[scale=.35]{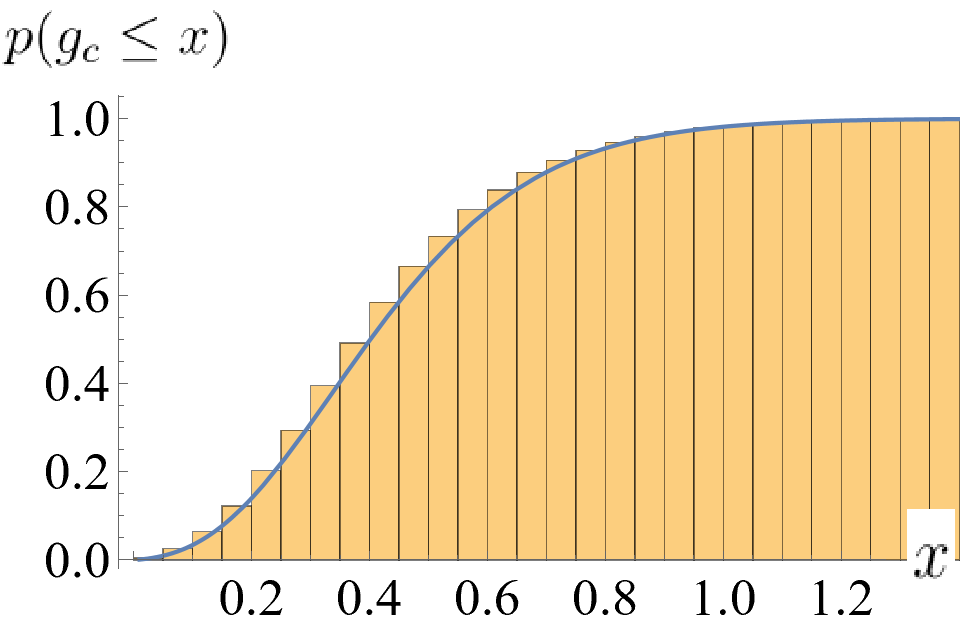}
\caption{}
\label{CDF_fits_min_U=5}
\end{subfigure}

\caption{Numerically observed CDFs at the critical point, with the best fit mixture distribution overlaid, for a) $U=1$, $\epsilon = 0.3$ ($W = 3.4$), b) $U=2$, $\epsilon=0.25$ ($W = 3.9$), c) $U=3$, $\epsilon = 0.8$ ($W = 3.0$), d) $U=4$, $\epsilon = 0.3$ ($W = 4.9$), e) $U = 5$, $\epsilon = 0.3$ ($W = 5.7$). The energy densities shown are the energy whose fitted distribution minimized the squared error \eqref{squared_error}, for each interaction strength. We can see excellent agreement between observation and fit, suggesting that the non-hermitian avalanche mechanism developed in the main text is an even better description of delocalization at the band edges. The parameters extracted from the fit are a) $\beta_8 = 0.756$, $\beta_9 = 0.244$ b) $\beta_7 = 0.794$, $\beta_8 = 0.206$,  c) $\beta_4 = 0.481$, $\beta_5 = 0.518$, d) $\beta_5 = 0.313$, $\beta_6 = 0.527$, $\beta_8 = 0.125$, $\beta_9 = .035$, e) $\beta_1 = 0.036$, $\beta_4 = 0.124$, $\beta_5 = 0.840$, and $g_0 = 0$ in all cases. All quantities are computed exactly as in Fig. \ref{CDF_fits}.}
\label{CDF_fits_min_error}
\end{figure}

\begin{figure}
\begin{subfigure}{.22\textwidth}
\includegraphics[scale=.35]{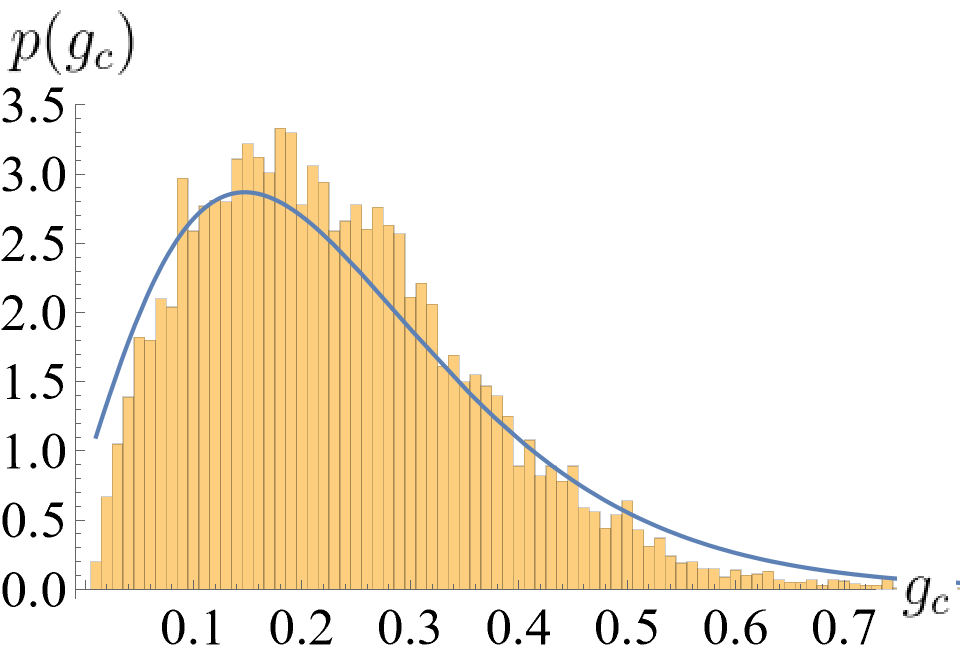}
\caption{}
\label{CDF_fits_min_U=1_PDF}
\end{subfigure}
\hfill
\begin{subfigure}{.22\textwidth}
\includegraphics[scale=.35]{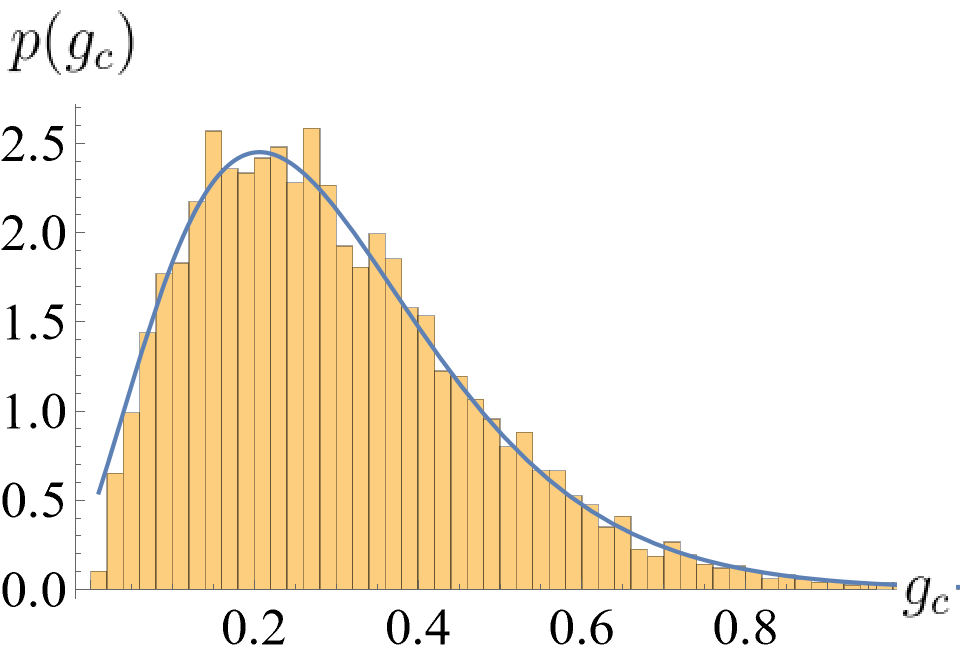}
\caption{}
\label{CDF_fits_min_U=2_PDF}
\end{subfigure}
\\
\begin{subfigure}{.22\textwidth}
\includegraphics[scale=.35]{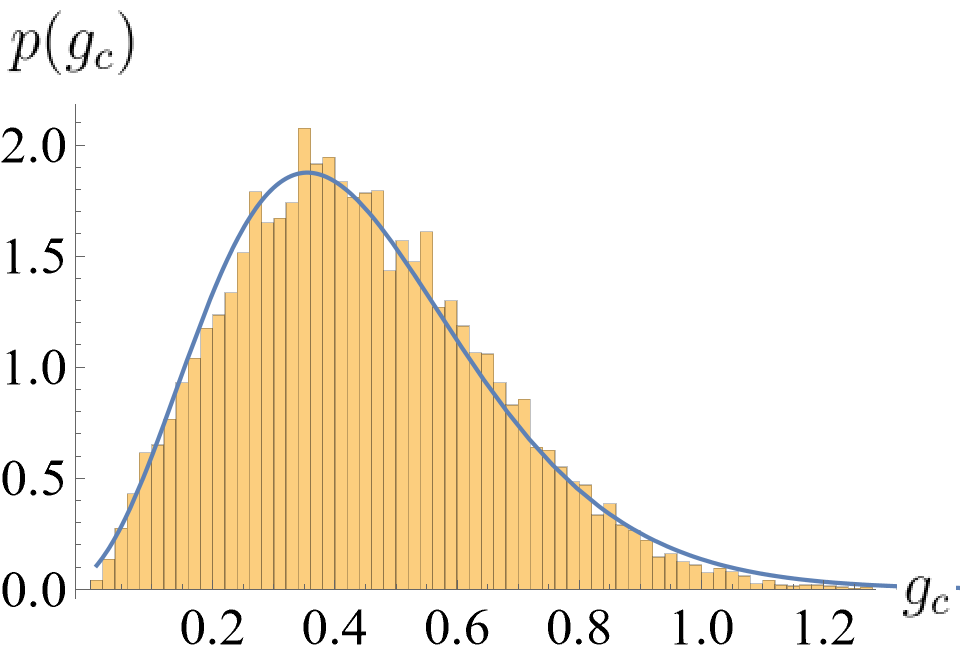}
\caption{}
\label{CDF_fits_min_U=3_PDF}
\end{subfigure}
\begin{subfigure}{.22\textwidth}
\includegraphics[scale=.35]{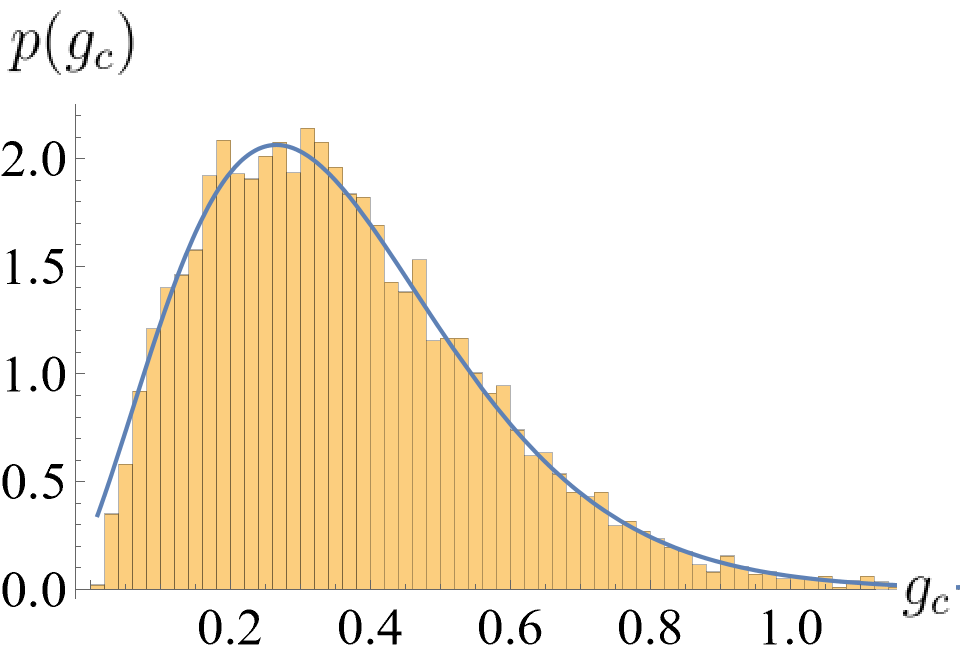}
\caption{}
\label{CDF_fits_min_U=4_PDF}
\end{subfigure}
\\
\begin{subfigure}{.22\textwidth}
\includegraphics[scale=.35]{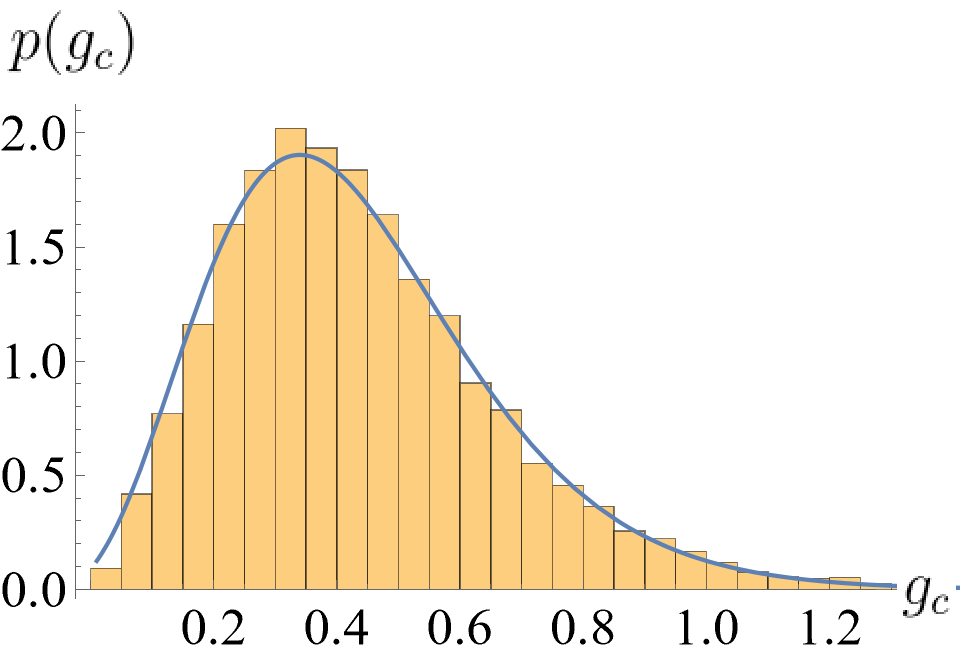}
\caption{}
\label{CDF_fits_min_U=5_PDF}
\end{subfigure}

\caption{Numerically observed PDFs at the MBL transition, with the PDFs from the best fit mixture distribution overlaid for, a) $U=1$, $\epsilon = 0.3$ ($W = 3.4$), b) $U=2$, $\epsilon=0.25$ ($W = 3.9$), c) $U=3$, $\epsilon = 0.8$ ($W = 3.0$), d) $U=4$, $\epsilon = 0.3$ ($W = 4.9$), e) $U = 5$, $\epsilon = 0.3$ ($W = 5.7$). We see excellent agreement again, confirming what we observed in Fig \ref{CDF_fits_min_error}. In contrast with Fig. \ref{CDF_fits_PDFs}, we also see that the tails of the observed distribution are well-described by the PDFs from the fits. All quantities are computed exactly as in Fig. \ref{CDF_fits} and \ref{CDF_fits_min_error}.} 
\label{CDF_fits_min_error_PDFs}
\end{figure}

As mentioned in the main text, we observed the numerical fits to be better towards the edges of the spectrum. To exemplify this, Fig. \ref{CDF_fits_min_error} shows the fits for the energy densities whose fitted mixture distribution had the lowest squared error (defined by \eqref{squared_error}), and Fig. \ref{CDF_fits_min_error_PDFs} shows the corresponding PDFs. We see excellent agreement between the observed and fitted distributions, especially for $U\geq 2$.

\section{Average Avalanche Condition}
\label{appendix-avg}

\begin{figure}
\begin{subfigure}{.22\textwidth}
\includegraphics[scale=.16]{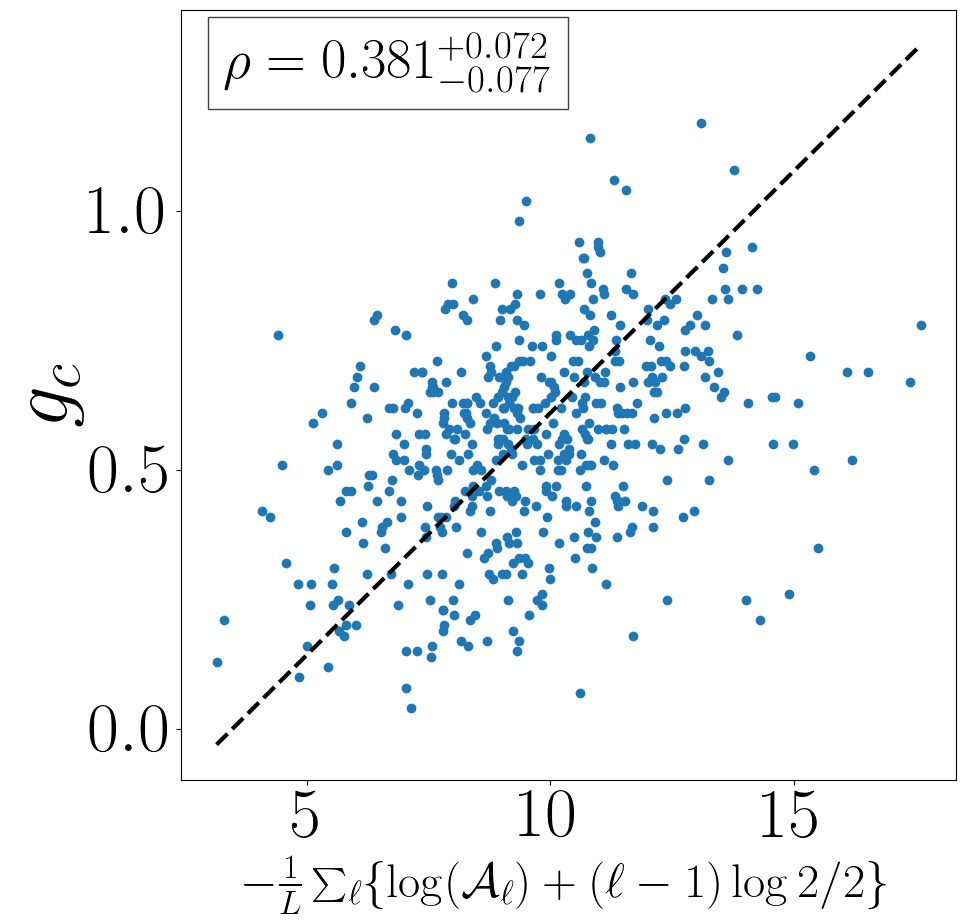}
\caption{}
\label{realizations_gs_amps_U=1}
\end{subfigure}
\hfill
\begin{subfigure}{.22\textwidth}
\includegraphics[scale=.16]{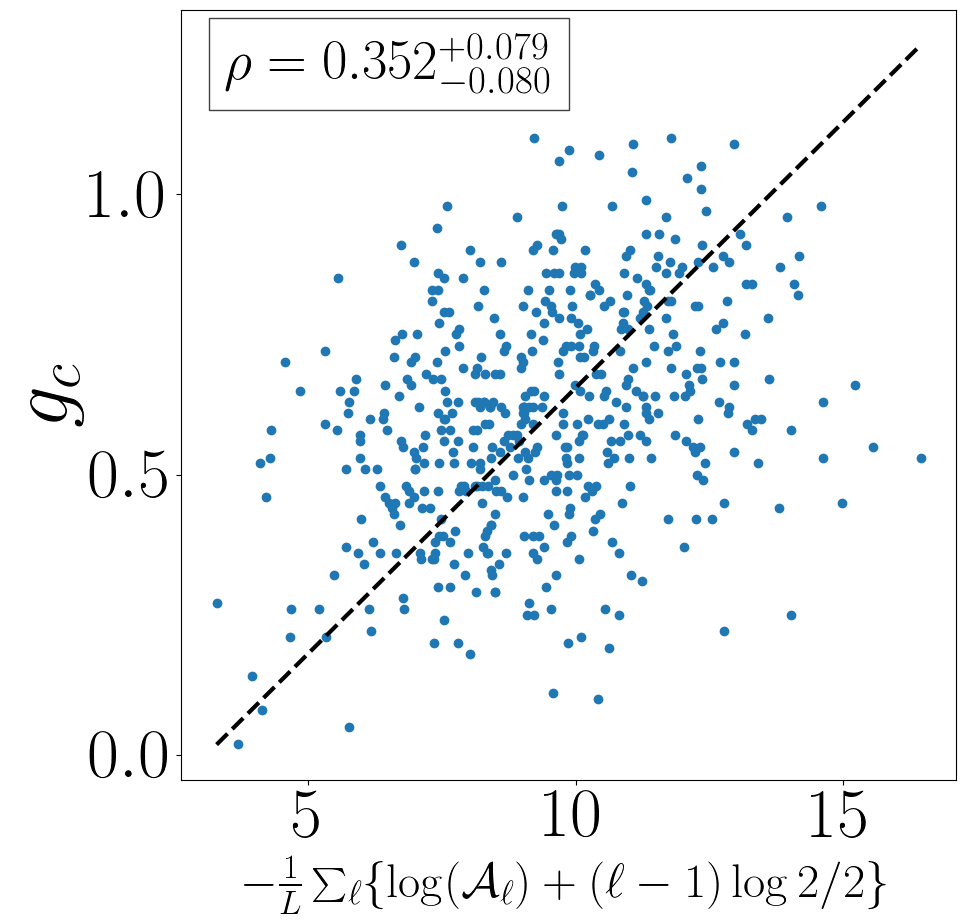}
\caption{}
\label{realizations_gs_amps_U=2}
\end{subfigure}
\\
\begin{subfigure}{.22\textwidth}
\includegraphics[scale=.16]{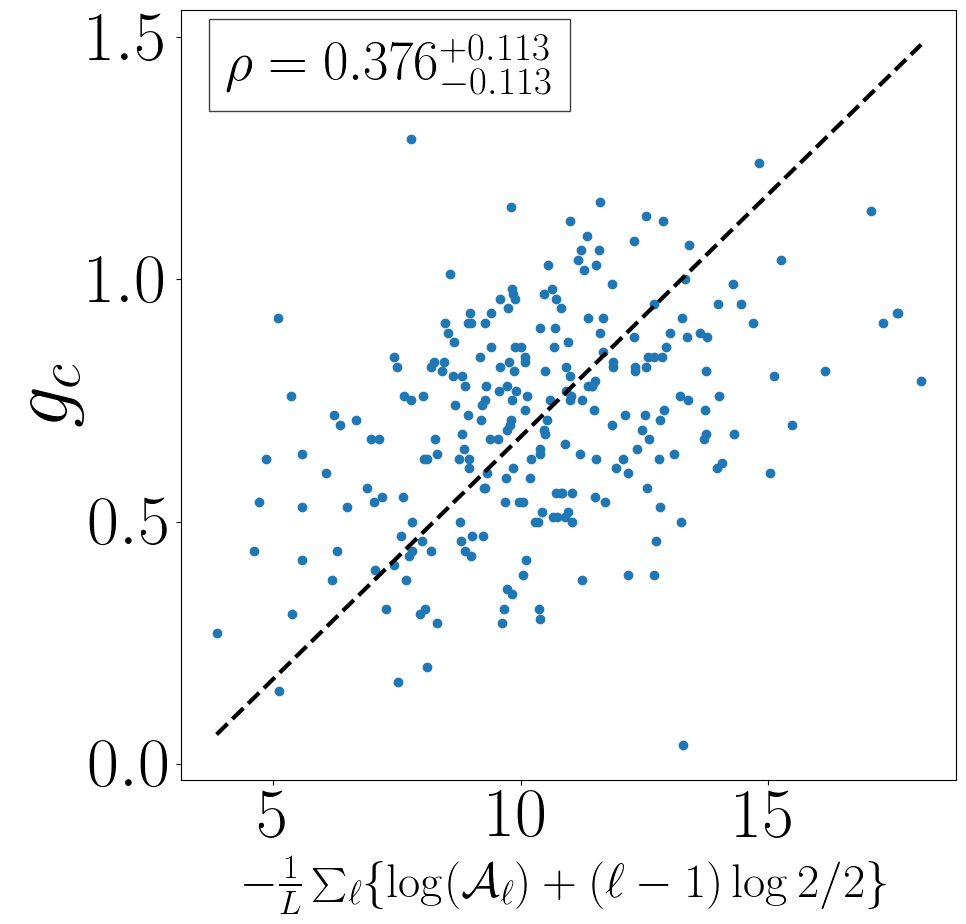}
\caption{}
\label{realizations_gs_amps_U=3}
\end{subfigure}
\hfill
\begin{subfigure}{.22\textwidth}
\includegraphics[scale=.16]{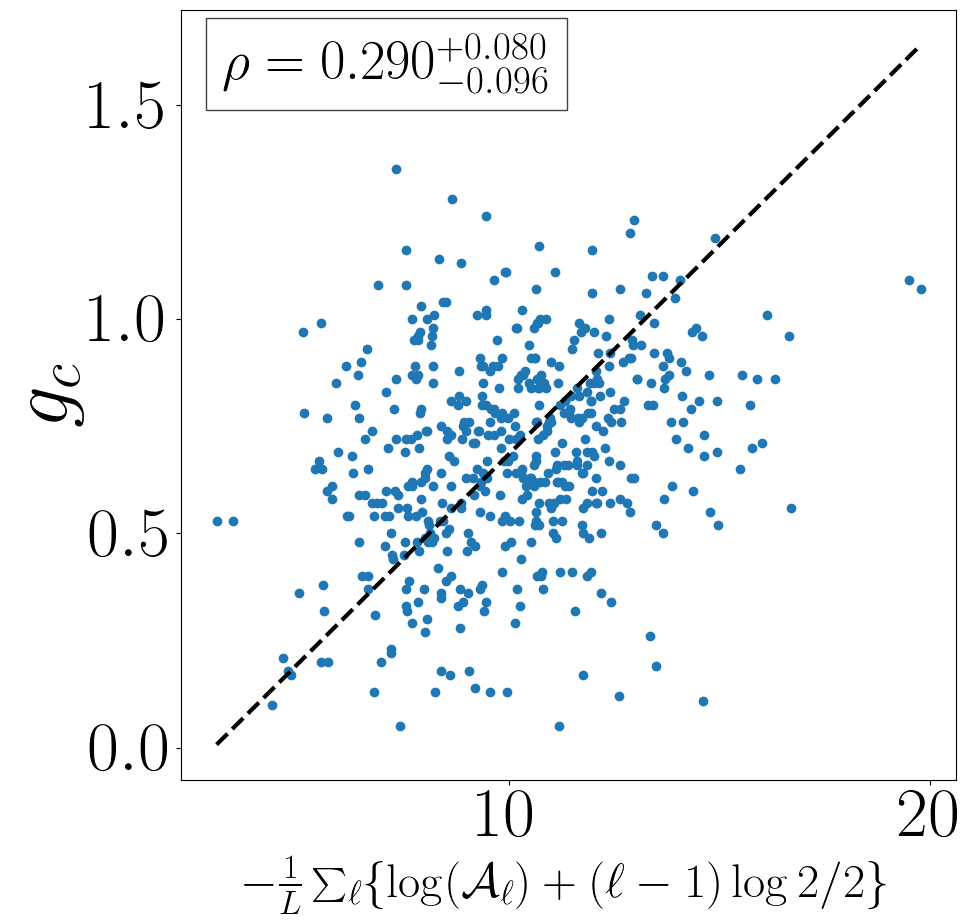}
\caption{}
\label{realizations_gs_amps_U=4}
\end{subfigure}
\\
\begin{subfigure}{.22\textwidth}
\includegraphics[scale=.16]{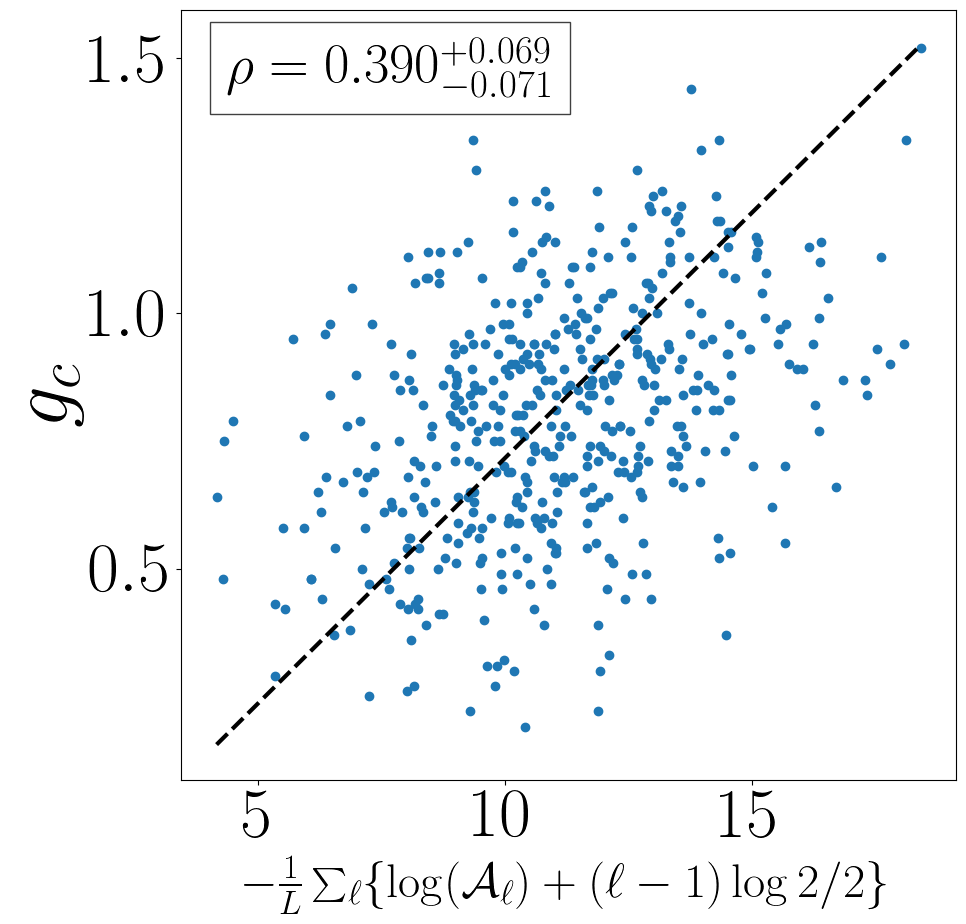}
\caption{}
\label{realizations_gs_amps_U=5}
\end{subfigure}

\caption{Example scatter plots of exceptional points $g_c(\epsilon)$ and ``avalanche parameter'' $\bar{\mathscr{A}}(\epsilon)$ (given by \eqref{A_avg}), as in Fig. \ref{gs_amps_noavg}. Each point represents a single disorder realization (for a fixed disorder strength), with the ordinate and abscissa $g_c$ and $\bar{\mathscr{A}}$ at the center of the spectrum ($\epsilon = 0.5$), respectively. The various panels show interaction/disorder strengths a) $U=1, W=5.5$, b) $U=2, W=6.0$, c) $U=3, W=7.0$, d) $U=4, W=7.0$, e) $U=5, W=8.0$. Similar to Fig. \ref{gs_amps_noavg}, we see no clear visual relationship between $g_c(\epsilon)$ and $\mathscr{A}(\epsilon)$, but weak positive correlation $\rho \sim 0.3$. For reference, we also overlay the disorder averaged fits of Fig. \ref{gs_amps_avgs} (dashed lines). All data and quantities were computed exactly as in Fig. \ref{gs_amps_noavg}.}
\label{gs_amps_noavg_avg}
\end{figure}

\begin{figure}
\begin{subfigure}{.22\textwidth}
\includegraphics[scale=.175]{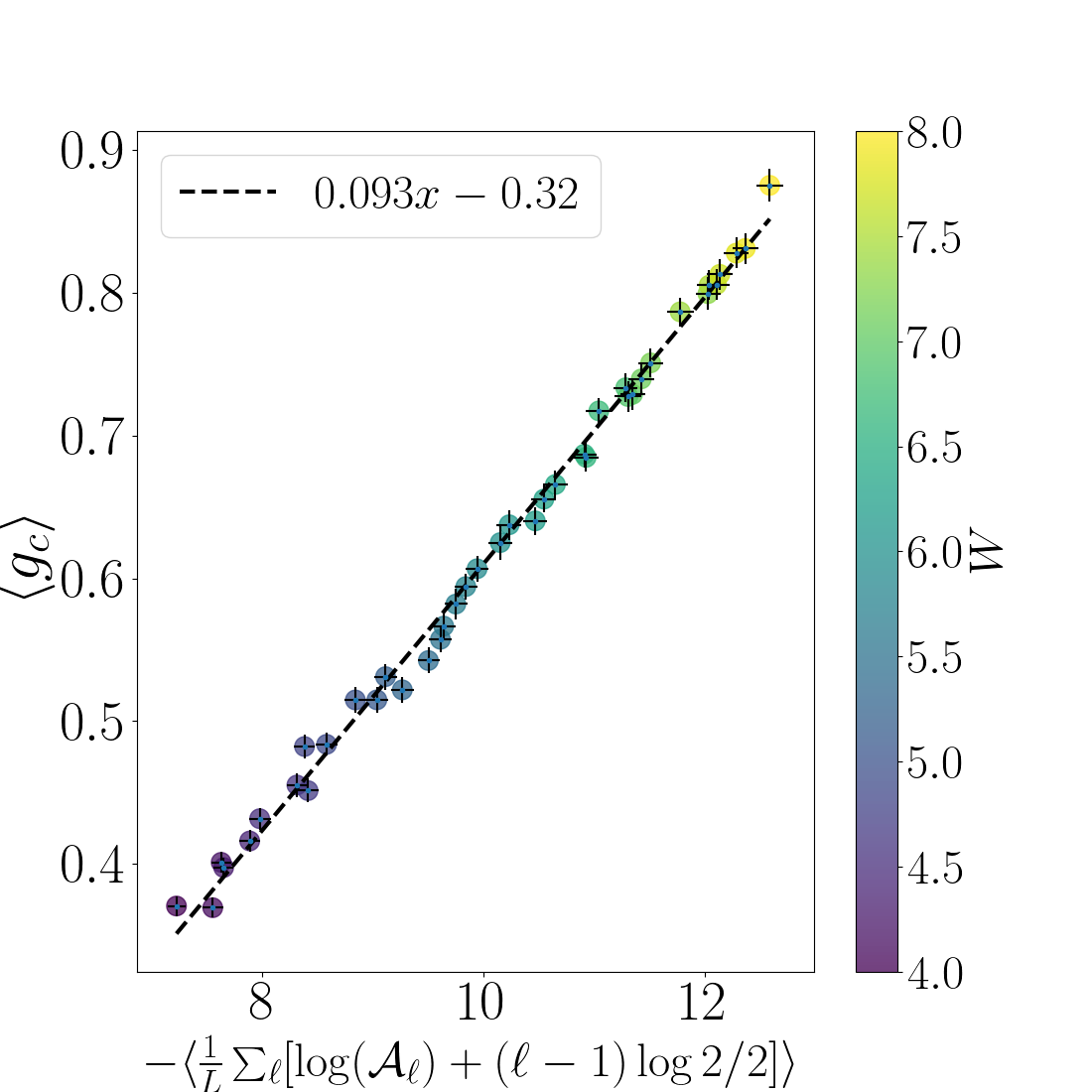}
\caption{}
\label{gs_amps_avgs_U=1}
\end{subfigure}
\hfill
\begin{subfigure}{.22\textwidth}
\includegraphics[scale=.175]{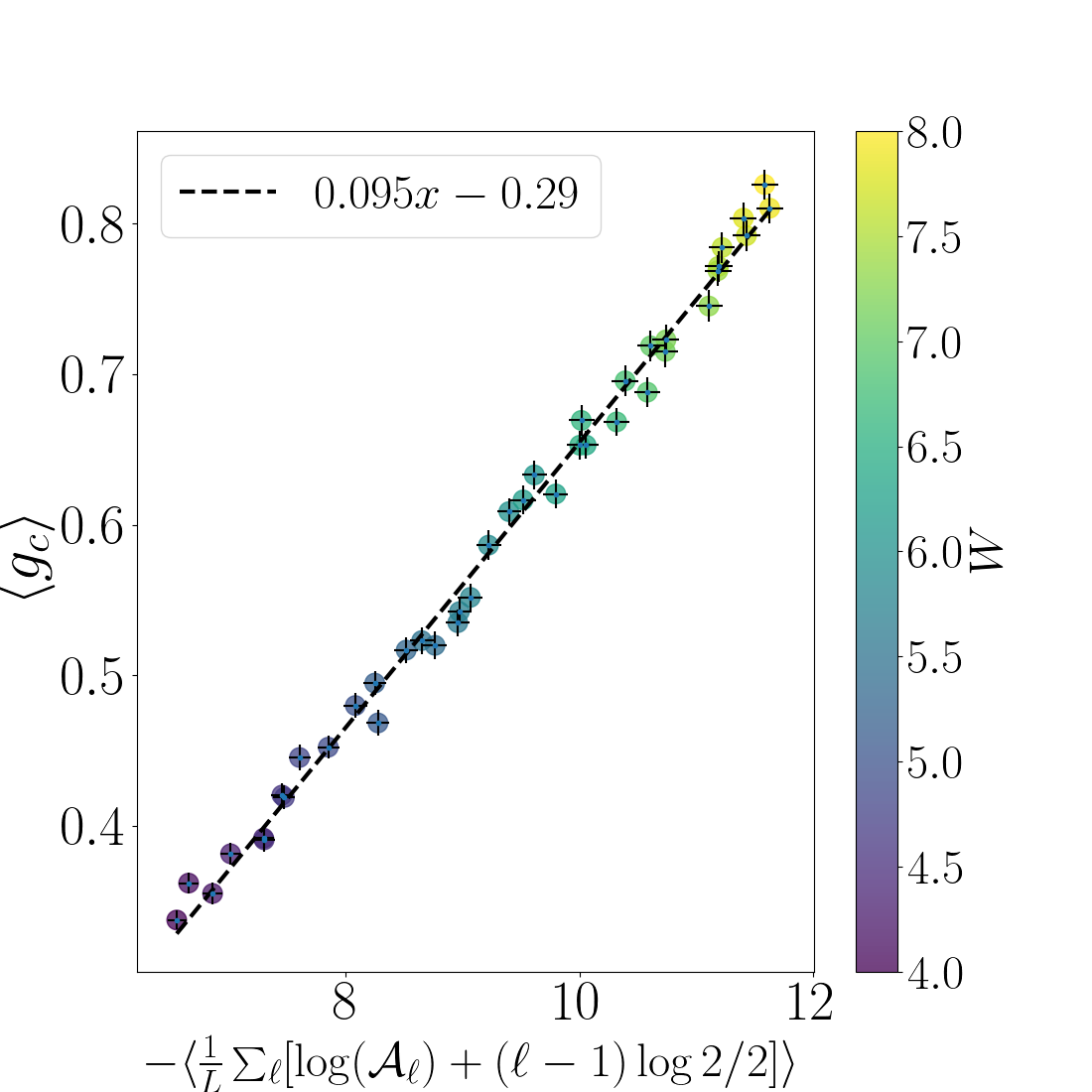}
\caption{}
\label{gs_amps_avgs_U=2}
\end{subfigure}
\\
\begin{subfigure}{.22\textwidth}
\includegraphics[scale=.175]{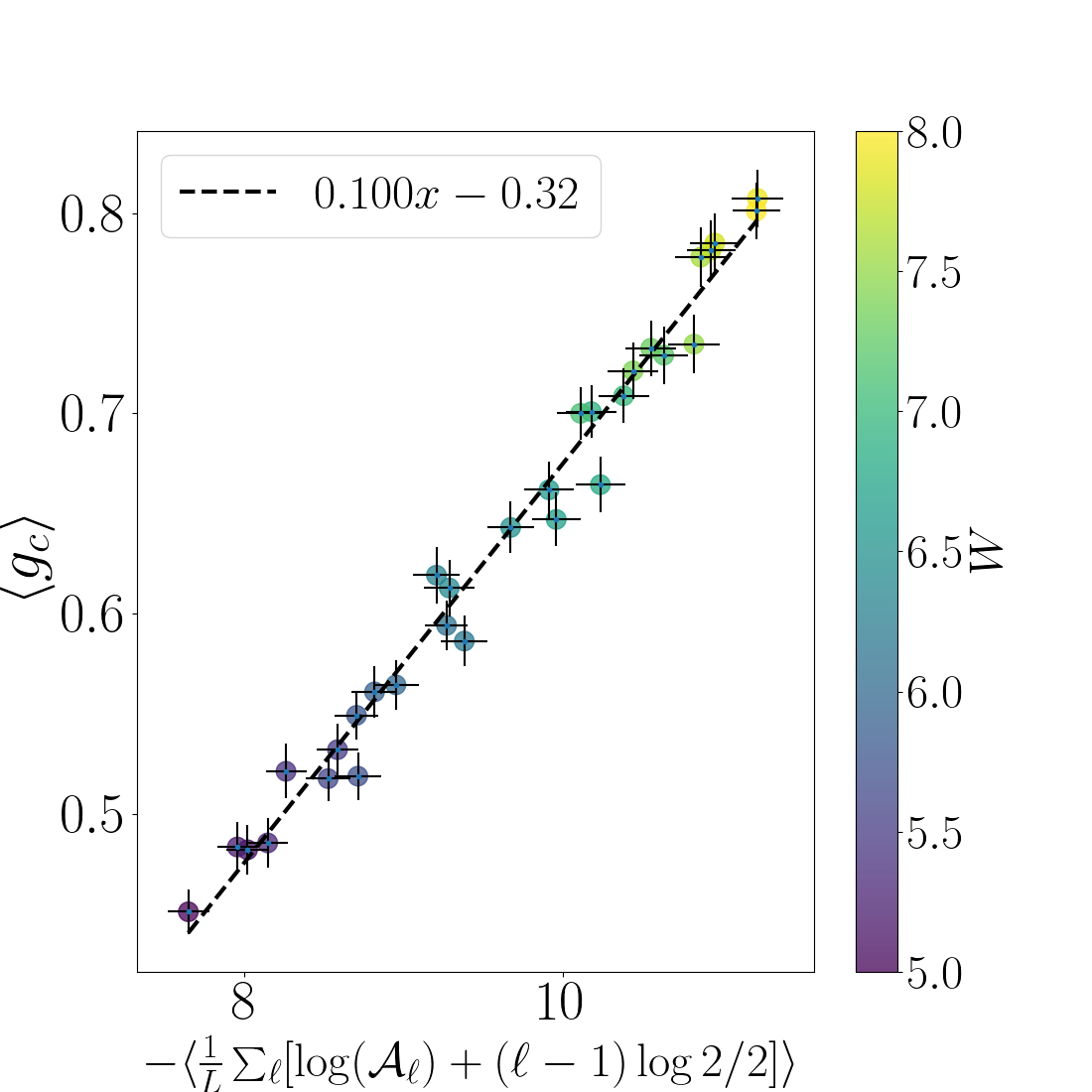}
\caption{}
\label{gs_amps_avgs_U=3}
\end{subfigure}
\hfill
\begin{subfigure}{.22\textwidth}
\includegraphics[scale=.175]{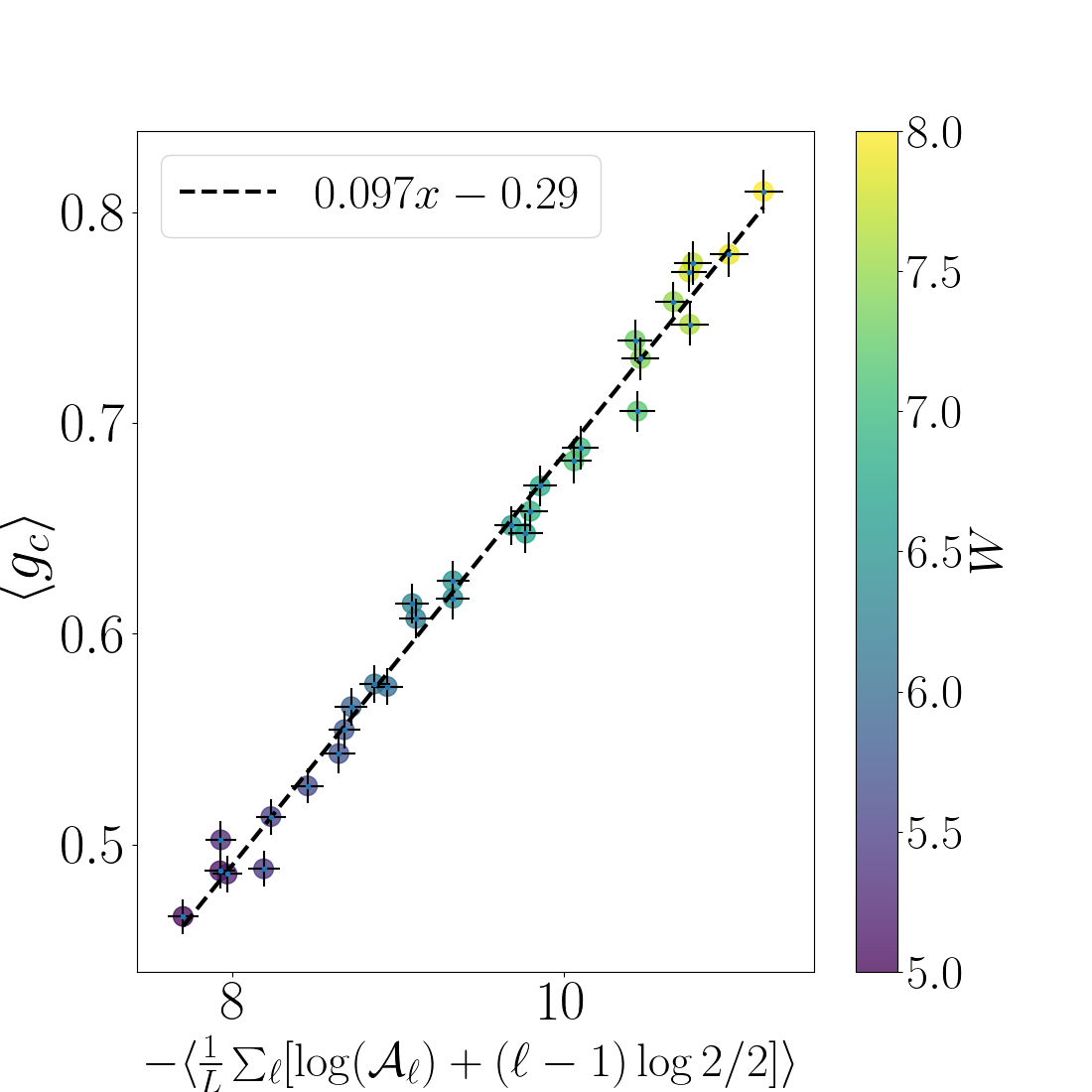}
\caption{}
\label{gs_amps_avgs_U=4}
\end{subfigure}
\\
\begin{subfigure}{.22\textwidth}
\includegraphics[scale=.175]{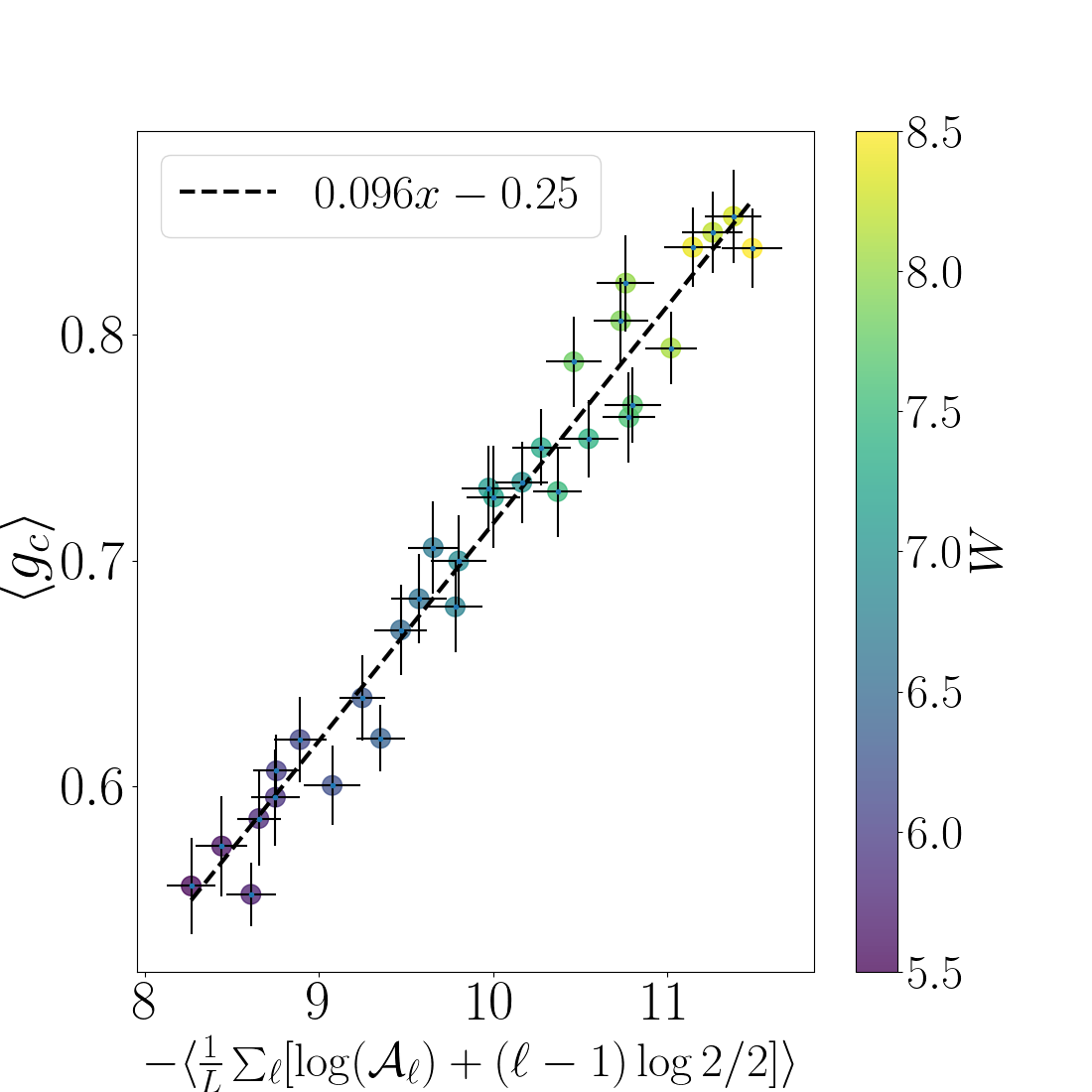}
\caption{}
\label{gs_amps_avgs_U=5}
\end{subfigure}

\caption{Scatter plot of disorder-averaged $g_c(\epsilon)$ and $\bar{\mathscr{A}}(\epsilon)$ for a variety of disorder strengths. The scatter plot is made exactly as in in Fig. \ref{gs_amps} with $\bar{\mathscr{A}}$ replacing $\mathscr{A}$. Shown are results at the center of the spectrum ($\epsilon = 0.5$) for interaction strengths a) $U=1$, b) $U=2$, c) $U=3$, d) $U=4$, e) $U=5$. Comparing with the fits in Fig. \ref{gs_amps}, we see comparable fit quality but different fit parameters.}
\label{gs_amps_avgs}
\end{figure}

In the process of relating $g_c(\epsilon)$ to the amplitudes $\mathcal{A}_{\ell}(\epsilon)$, we found empirically that $\avg{g_c(\epsilon)}$ was also well-described by a linear relationship with
\begin{equation}
\bar{\mathscr{A}}(\epsilon) := -\frac{1}{L}\sum_{\ell=1}^L\left[\log \mathcal{A}_{\ell}(\epsilon) + (\ell-1)\frac{\log 2}{2}\right] .
\label{A_avg}
\end{equation} 
This quantity is an \textit{average} over l-bits of the difference between hopping and level spacing, instead of the \textit{extremum} present in \eqref{A_min}. As discussed in Section \ref{g_c-and-avalanche}, such a relationship may be better suited to describing a delocalization mechanism in which the number of l-bits that need to hybridize with the bath for the system to undergo a real-complex transition varies between disorder realizations. Figure \ref{gs_amps_noavg_avg} shows a scatterplot of $g_c(\epsilon)$ versus $\bar{\mathscr{A}}(\epsilon)$ for individual disorder realizations, akin to Fig. \ref{gs_amps_noavg}. As in Fig. \ref{gs_amps_noavg}, we see a noisy, approximately  linear relationship. We perform linear fits of $\avg{g_c(\epsilon)}(W)$ vs $\avg{\bar{\mathscr{A}}(\epsilon)}(W)$, in analogy with Fig. \ref{gs_amps}; the results of these fits are shown in Fig. \ref{gs_amps_avgs}, and the extracted fit parameters are shown in Table \ref{linear_fit_avgs}. The quality of fit is comparable to those shown in the main text for $\avg{g_c(\epsilon)}$ vs $\avg{\mathscr{A}(\epsilon)}$, though the fit parameters are different.

\begin{table*}
\centering
\begin{tabular}{|c|c|c|c|c|c|c|c|c|c|c|}
\hline 
 & \multicolumn{2}{|c|}{ $U=1$} & \multicolumn{2}{|c|}{ $U=2$}  & \multicolumn{2}{|c|}{$U=3$} & \multicolumn{2}{|c|}{ $U=4$} & \multicolumn{2}{|c|}{$U=5$} \\ 
 \hline
 $\epsilon$ & Slope & Intercept & Slope & Intercept & Slope & Intercept & Slope & Intercept & Slope & Intercept \\
 \hline 
 $0.2$ & $.099 \pm .002$ & $-.332 \pm .018$ & $.093 \pm .002$ & $-.224 \pm .016$ & $.096 \pm .004 $ & $-.232 \pm  .041$ & $.094 \pm .002$ & $-.198 \pm .024$ & $.099 \pm .004$ & $-.199 \pm .041$\\
 \hline
$0.3$ & $.094 \pm .001$ & $-.315 \pm .014$ & $.092 \pm .001$ & $-.259 \pm .012$ & $.101 \pm .003$ & $-.326 \pm .033$ & $.095 \pm .002$ & $-.267 \pm .020$ & $.103 \pm .003$ & $-.318 \pm .030$\\
 \hline
$0.4$ & $.095 \pm .001$ & $-.340 \pm .012 $ & $.095 \pm .001$ & $-.296 \pm .011$ & $.101 \pm .003$ & $-.335 \pm .026$ & $.100 \pm .002$ & $-.315 \pm .019$ & $.107 \pm .004$ & $-.361 \pm .041$\\
 \hline
$0.5$ & $.093 \pm .001$ & $-.324 \pm  .011$ & $.095 \pm .001$ & $-.294 \pm .011$ & $.100 \pm .003$ & $-.323 \pm .026$ & $.097 \pm .002$ & $-.289 \pm .019$ & $.096 \pm .004$ & $-.246 \pm .042$\\
 \hline
$0.6$ & $.092 \pm .001$ & $-.301 \pm .012$ & $.097 \pm .001$ & $-.298 \pm .012$ & $.100 \pm .003$ & $-.311 \pm .029$ & $.097 \pm .002$ & $-.263 \pm .020$ & $.111 \pm .006$ & $-.339 \pm .058$\\
 \hline
$0.7$ & $.097 \pm .001$ & $-.323 \pm .014$ & $.095 \pm .001$ & $-.252 \pm .014$ & $.098 \pm .003$ & $-.254 \pm .035$ & $.100 \pm .003$ & $-.231 \pm .029$ & $.111 \pm .004$ & $-.323 \pm .042$\\
 \hline
$0.8$ & $.094 \pm .002$ & $-.258 \pm .019$ & $.097 \pm .002$ & $-.222 \pm .019$ & $.090 \pm .005$ & $-.099 \pm .054$ & $.095 \pm .004$ & $-.112 \pm .042$ & $.095 \pm .005$  & $-.564 \pm .063$\\
 \hline
\end{tabular}
\caption{Best fit parameters for a linear fit of $\avg{g_c(\epsilon)}$ vs $\avg{\bar{\mathscr{A}}(\epsilon)}$, across the spectrum for various interaction strengths. The linear fits, and the uncertainties on the fit parameters, are computed analogously as in Table \ref{linear_fit}. Comparing with Table \ref{linear_fit}, we see the fit parameters deviate significantly - in particular, the slope here no longer agrees with 1/12, and the intercepts are closer to zero.}
\label{linear_fit_avgs}
\end{table*}

\begin{figure}
\begin{subfigure}{.22\textwidth}
\includegraphics[scale=.35]{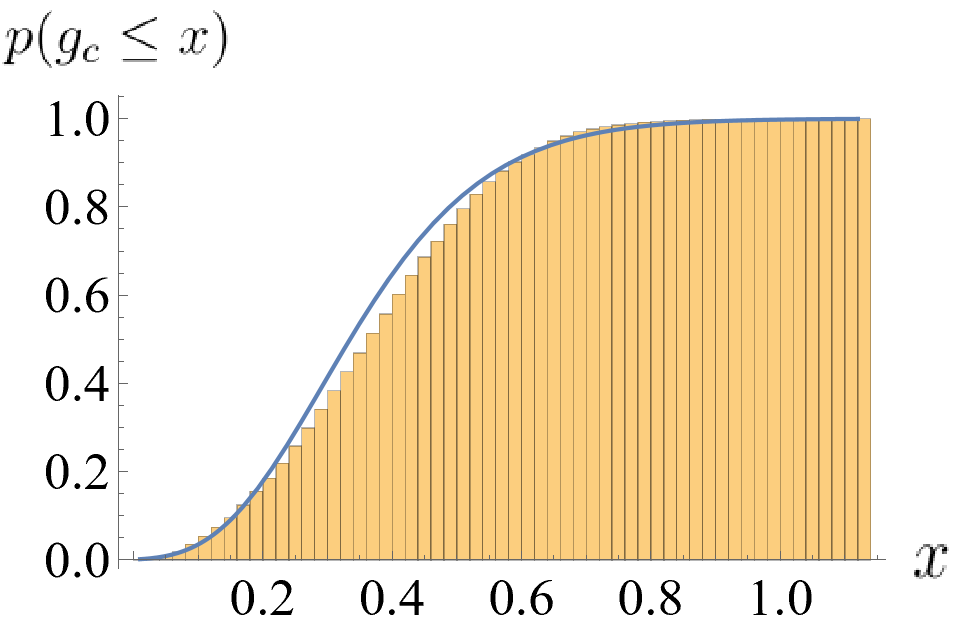}
\caption{}
\label{CDF_fits_avg_U=1}
\end{subfigure}
\hfill
\begin{subfigure}{.22\textwidth}
\includegraphics[scale=.35]{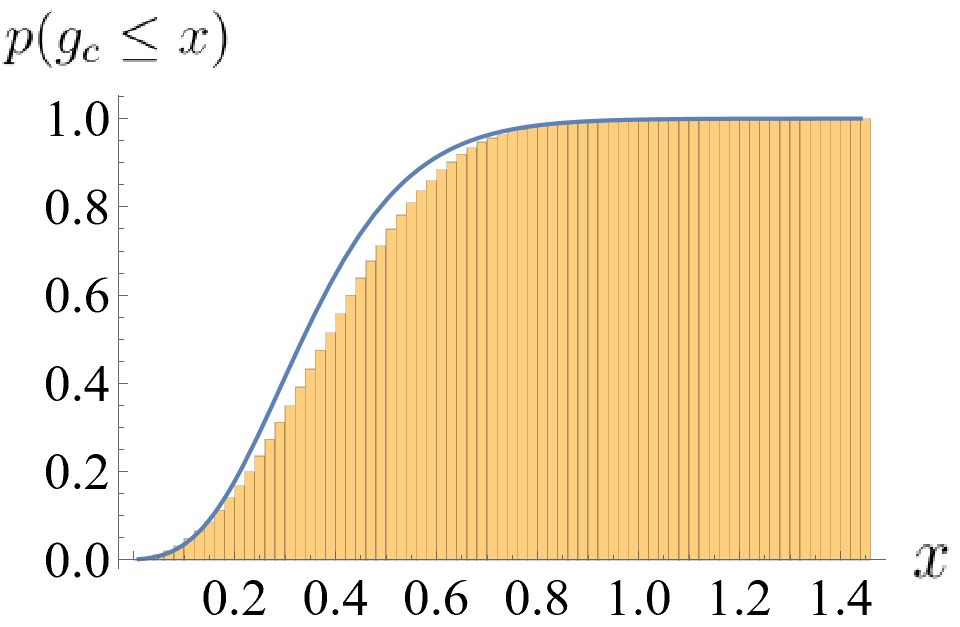}
\caption{}
\label{CDF_fits_avg_U=2}
\end{subfigure}
\\
\begin{subfigure}{.22\textwidth}
\includegraphics[scale=.35]{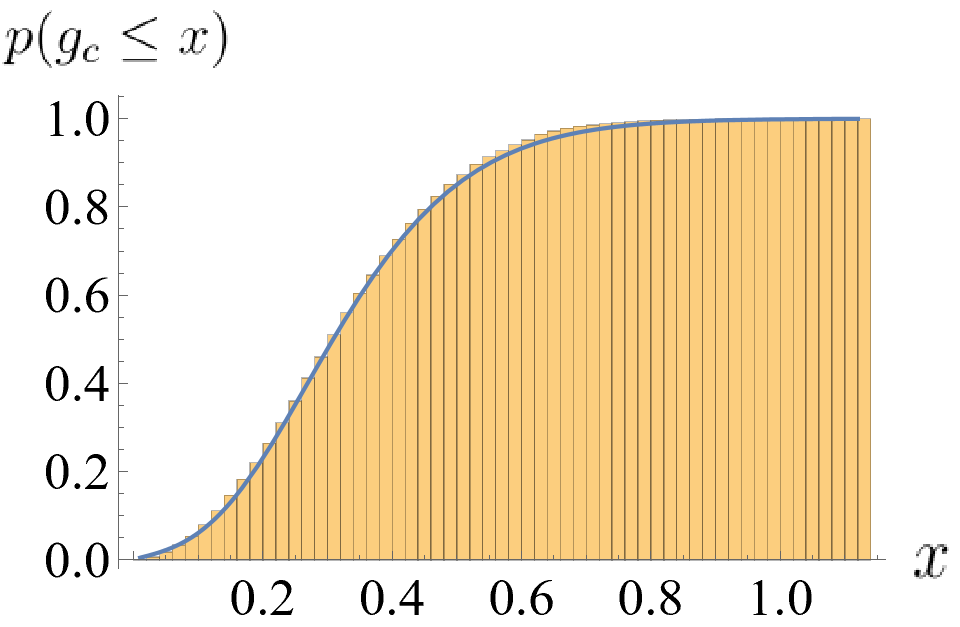}
\caption{}
\label{CDF_fits_avg_U=3}
\end{subfigure}
\hfill
\begin{subfigure}{.22\textwidth}
\includegraphics[scale=.35]{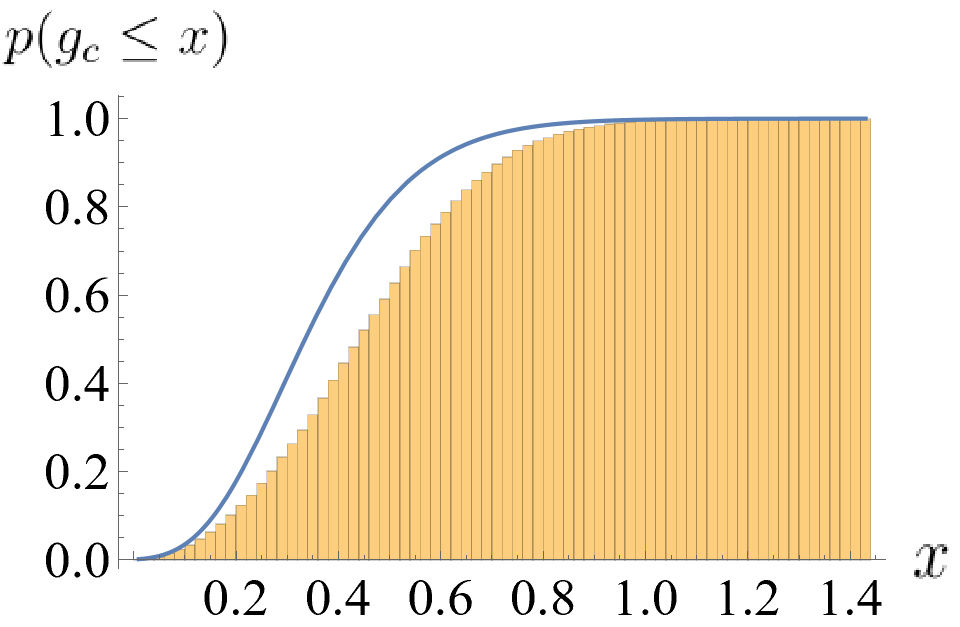}
\caption{}
\label{CDF_fits_avg_U=4}
\end{subfigure}
\\
\begin{subfigure}{.22\textwidth}
\includegraphics[scale=.35]{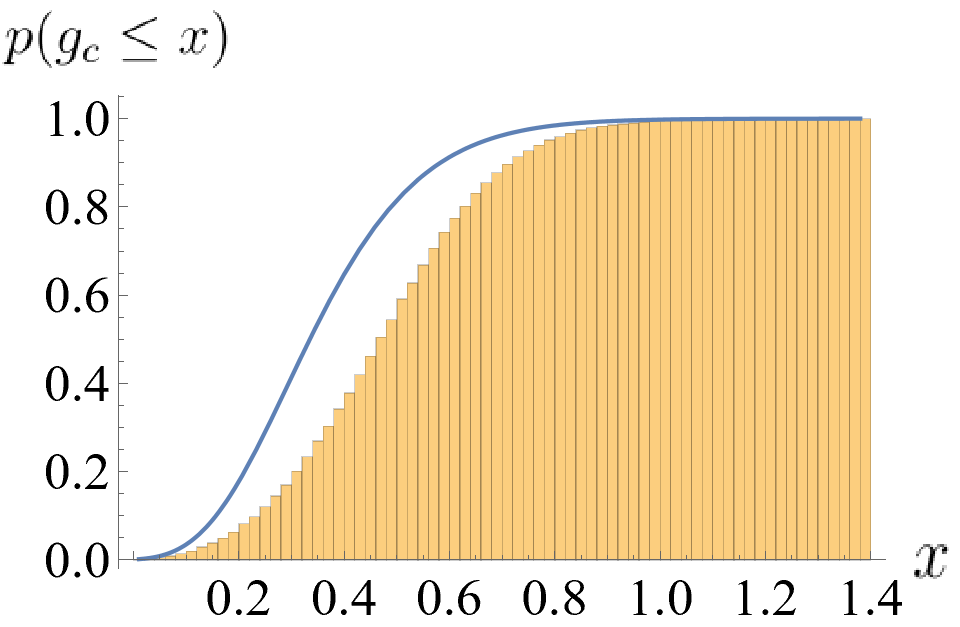}
\caption{}
\label{CDF_fits_avg_U=5}
\end{subfigure}

\caption{Numerically observed cumulative density functions of $g_c$ for $\epsilon = 0.5$ at the critical point, for interaction strengths a) $U=1$ ($W=4.8$) , b) $U=2$ ($W = 5.4$), c) $U=3$ ($W = 5.9$), d) $U=4$ ($W = 6.1$), e) $U=5$ ($W = 6.7$). The blue solid lines overlaid are best fit mixture distributions of the form \eqref{CDF_mixture} (now with $P_{F}^{N_{\ell}}$ now given by \eqref{CDF_gen_avg}). We see reasonably good agreement, roughly on par or better than of Fig. \ref{CDF_fits}, for $U \leq 3$, and significant deviation for $U > 3$. This suggests the mixture distribution derived from \eqref{CDF_gen_avg} is a good descriptor at low interaction strengths, whereas that derived from \eqref{CDF_gen} in the main text is far better at larger interaction strengths. The parameters extracted from the fit are a) $\beta_1 = 1.0$ b) $\beta_1 = 1.0$,  c) $\beta_1 = 0.211$, $\beta_2 = 0.765$, $\beta_{11} = 0.023$ d) $\beta_1 = 1.0$, e) $\beta_1 = 1.0$, and $g_0 = 0$ in all cases. All quantities are computed in chains of size $L=12$ in the half-filling sector, and from 10000 disorder realizations. When computing the squared error \eqref{squared_error}, we sample $n = 100$ points uniformly spaced between the smallest and largest observed values of $g_c$.}
\label{CDF_fits_avg}
\end{figure}

\begin{figure}
\begin{subfigure}{.22\textwidth}
\includegraphics[scale=.35]{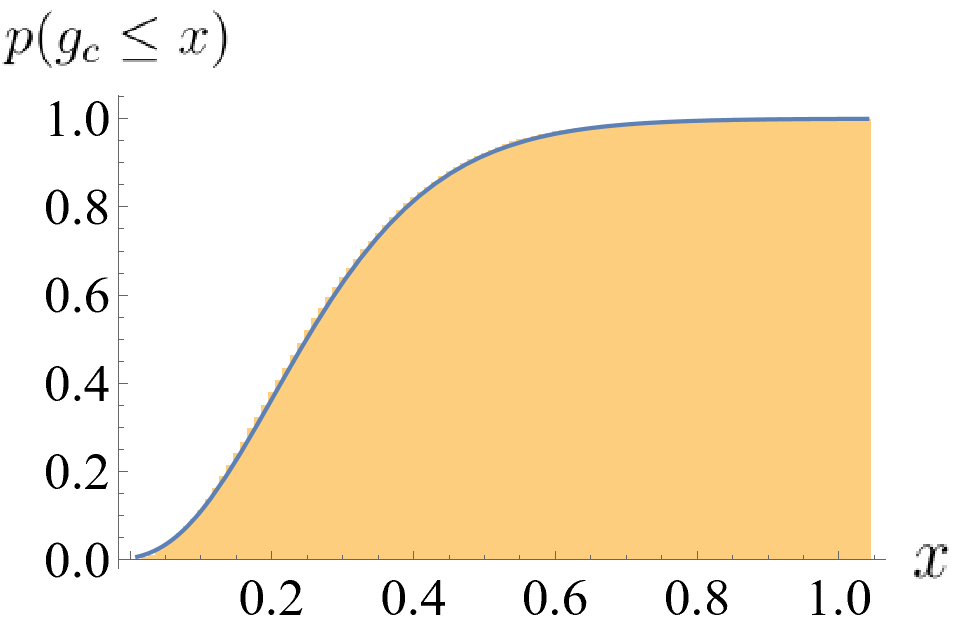}
\caption{}
\label{CDF_fits_min_U=1}
\end{subfigure}
\hfill
\begin{subfigure}{.22\textwidth}
\includegraphics[scale=.35]{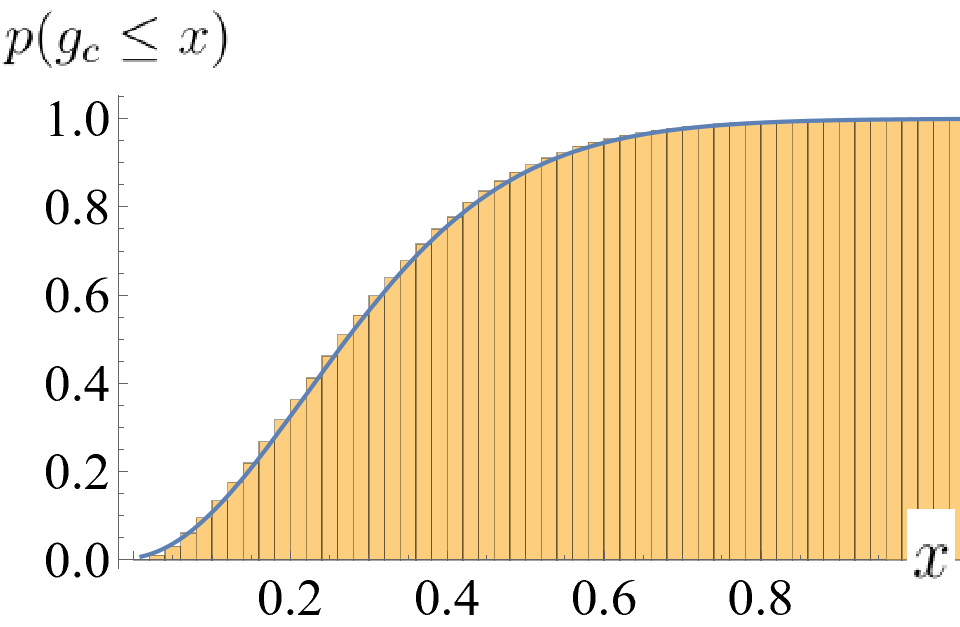}
\caption{}
\label{CDF_fits_min_U=2}
\end{subfigure}
\\
\begin{subfigure}{.22\textwidth}
\includegraphics[scale=.35]{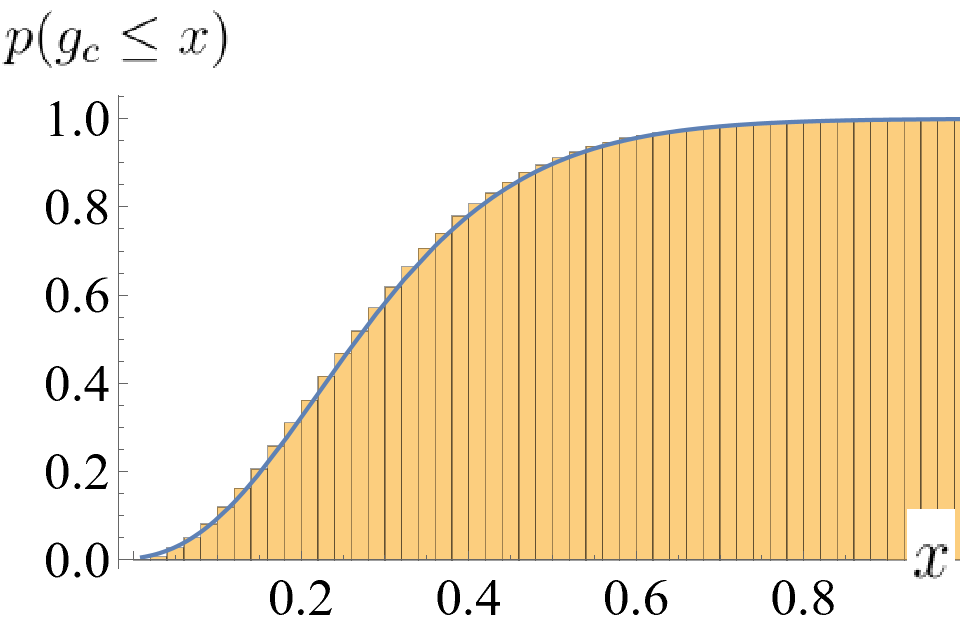}
\caption{}
\label{CDF_fits_min_U=3}
\end{subfigure}
\hfill
\begin{subfigure}{.22\textwidth}
\includegraphics[scale=.35]{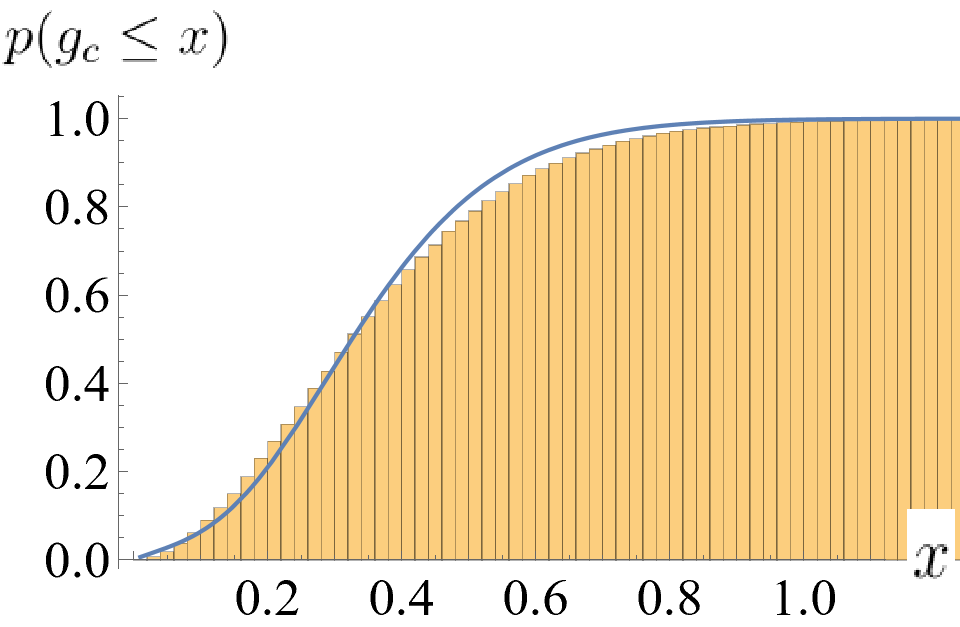}
\caption{}
\label{CDF_fits_min_U=4}
\end{subfigure}
\\
\begin{subfigure}{.22\textwidth}
\includegraphics[scale=.35]{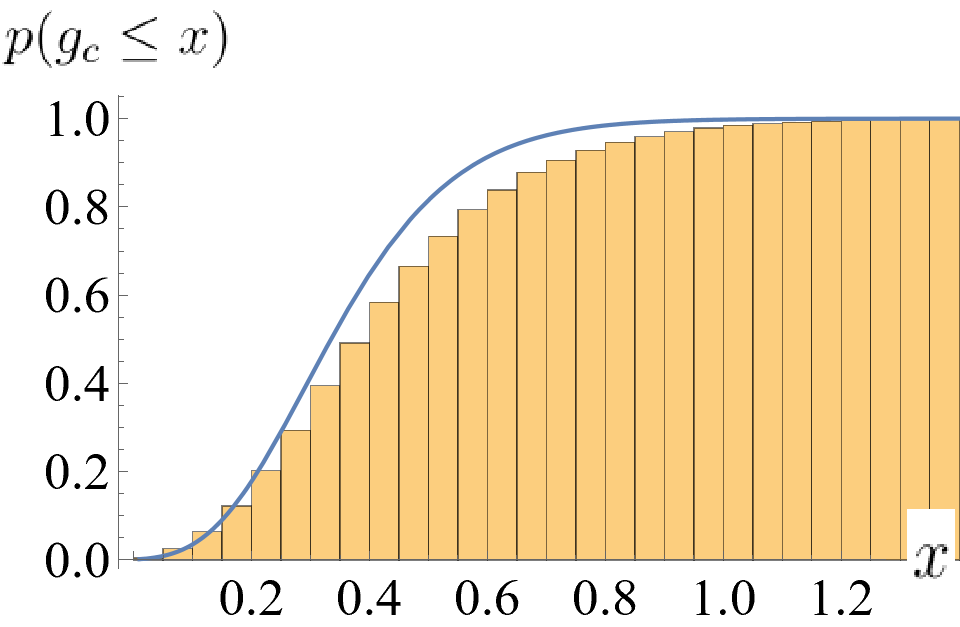}
\caption{}
\label{CDF_fits_min_U=5}
\end{subfigure}

\caption{Numerically observed cumulative density functions of $g_c$ for $\epsilon = 0.5$ at the critical point, for interaction strengths a) $U=1$, $\epsilon = 0.8$ ($W = 3.0$), b) $U=2$, $\epsilon = 0.3$ ($W = 4.2$), c) $U=3$, $\epsilon = 0.35$ ($W = 5.6$), d) $U=4$, $\epsilon = 0.3$ ($W = 4.9$), e) $U = 5$, $\epsilon = 0.3$ ($W = 5.7$). As in Fig. \ref{CDF_fits_avg}, the mixture distributions are constructed with the $P_F^{N_{\ell}}$ given by \eqref{CDF_gen_avg}. The energy densities shown are the energy whose fitted distribution minimized the squared error \eqref{squared_error}, for each interaction strength. We can see excellent agreement between observation and fit at weaker interactions strengths, suggesting that averaging over the differences $\mathcal{A}_{\ell}(\epsilon) + (\ell-1)\log 2/2$ is an effective descriptor of the physics at the band edges in the weaker interacting regime. The parameters extracted from the fit are a) $\beta_5 = 0.603$, $\beta_6 = 0.396$, $\beta_{11} = .001$, b) $\beta_1 = 0.308$, $\beta_2 = 0.405$, $\beta_7 = 0.189$, $\beta_8 = 0.097$,  c) $\beta_1 = 0.095$, $\beta_2 = 0.056$, $\beta_3 = 0.676$, $\beta_7 = 0.173$, d) $\beta_1 = 0.957$, $\beta_{11} = 0.043$,  e) $\beta_1 = 1.0$, and $g_0 = 0$ in all cases. All quantities are computed exactly as in Fig. \ref{CDF_fits_avg}}
\label{CDF_fits_min_error_avg}
\end{figure}

\begin{figure}
\begin{subfigure}{.22\textwidth}
\includegraphics[scale=.35]{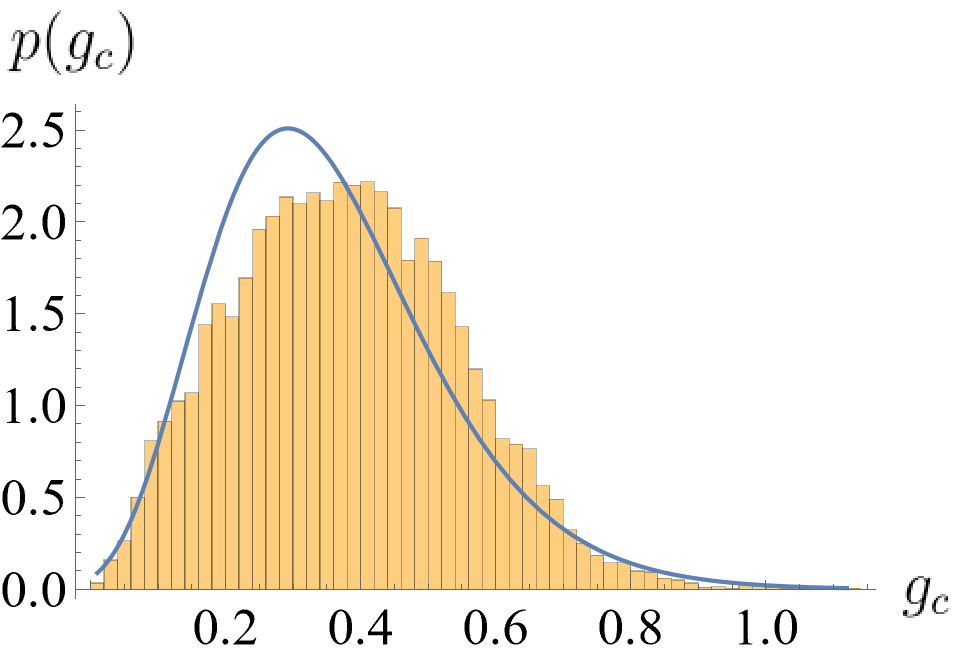}
\caption{}
\label{CDF_fits_PDF_U=1_avg}
\end{subfigure}
\hfill
\begin{subfigure}{.22\textwidth}
\includegraphics[scale=.35]{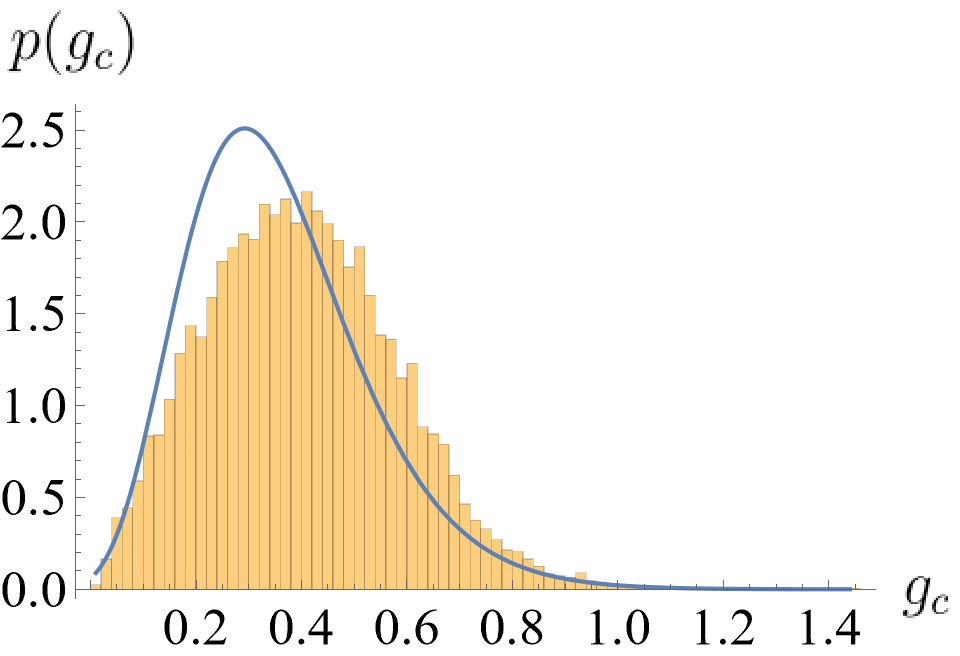}
\caption{}
\label{CDF_fits_PDF_U=2_avg}
\end{subfigure}
\begin{subfigure}{.22\textwidth}
\includegraphics[scale=.35]{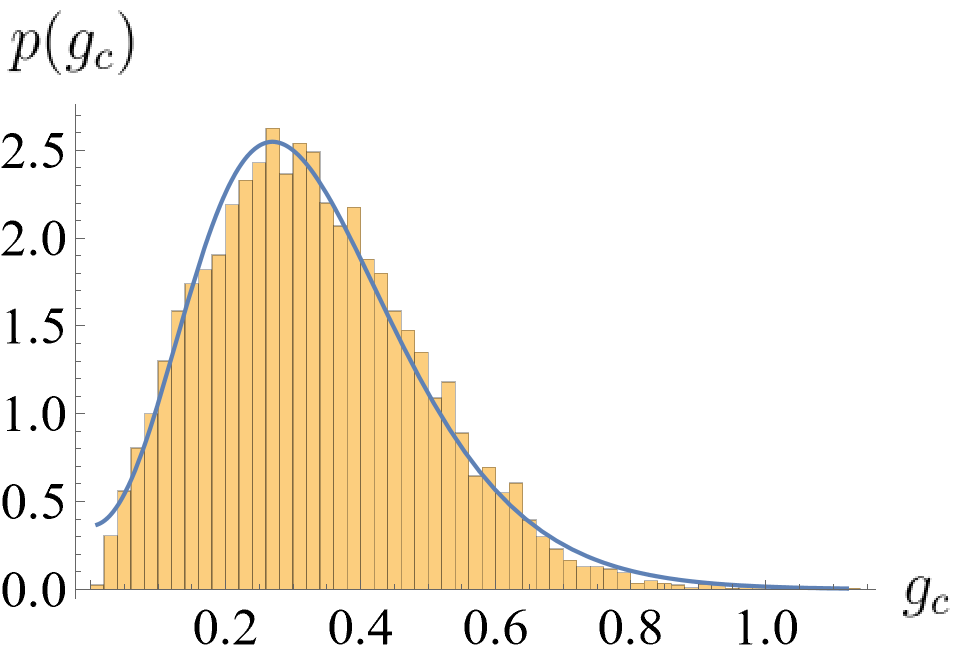}
\caption{}
\label{CDF_fits_PDF_U=3_avg}
\end{subfigure}
\hfill
\begin{subfigure}{.22\textwidth}
\includegraphics[scale=.35]{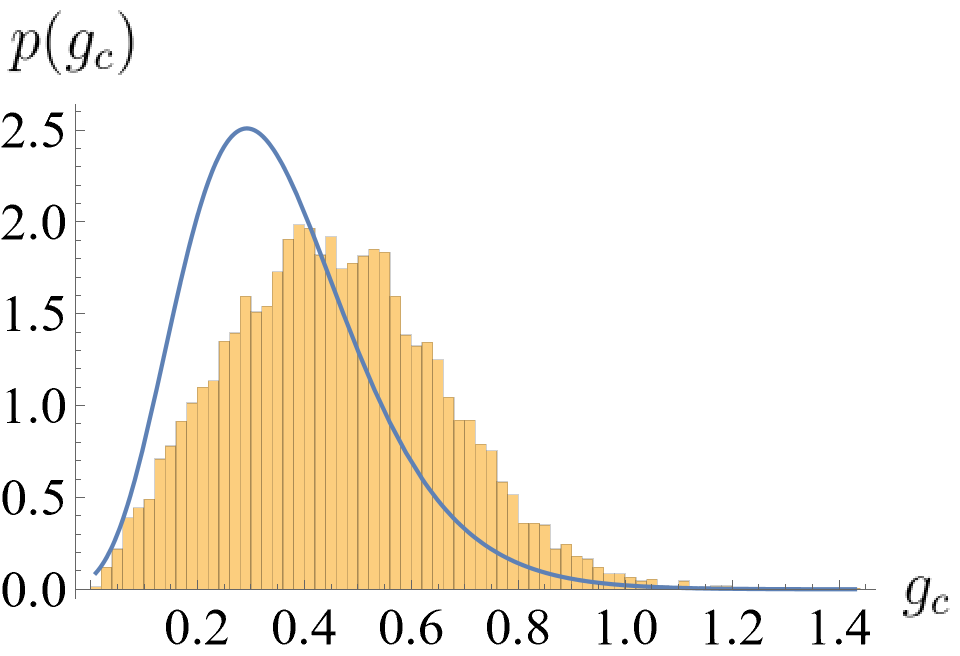}
\caption{}
\label{CDF_fits_PDF_U=4_avg}
\end{subfigure}
\\
\begin{subfigure}{.22\textwidth}
\includegraphics[scale=.35]{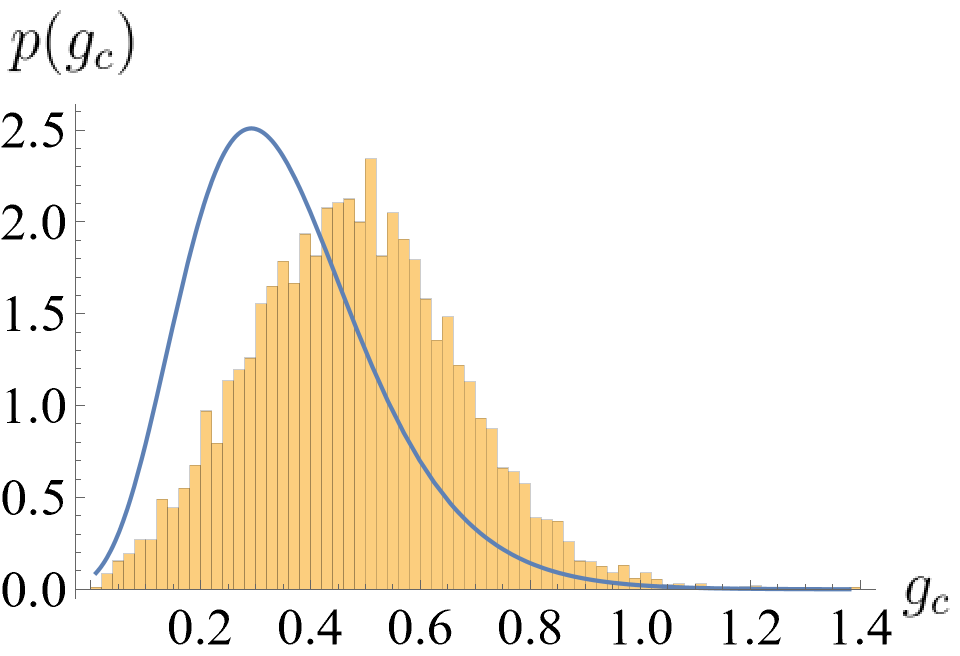}
\caption{}
\label{CDF_fits_PDF_U=5_avg}
\end{subfigure}

\caption{Numerically observed probability density functions of $g_c$ for $\epsilon = 0.5$ at the critical disorder strength, for interaction strengths a) $U=1$ ($W=4.8$) , b) $U=2$ ($W = 5.4$), c) $U=3$ ($W = 5.9$), d) $U=4$ ($W = 6.1$), e) $U=5$ ($W = 6.7$). The blue solid lines overlaid are the pdfs obtained from differentiating the best fit mixture distributions shown in Fig. \ref{CDF_fits_avg}.}
\label{CDF_fits_PDFs_avg}
\end{figure}

\begin{figure}
\begin{subfigure}{.22\textwidth}
\includegraphics[scale=.35]{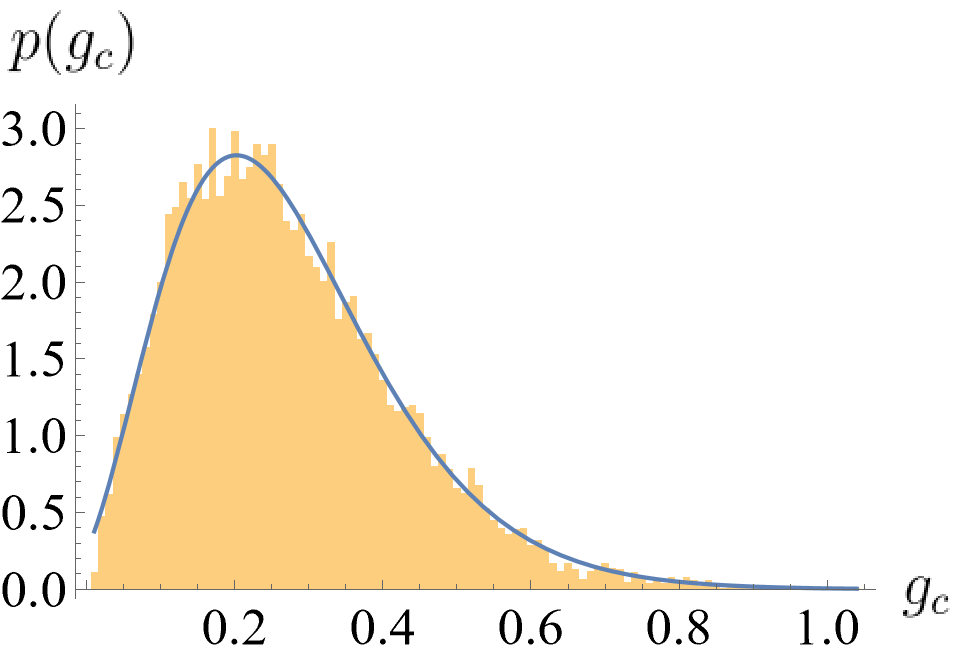}
\caption{}
\label{CDF_fits_min_PDFs_U=1_avg}
\end{subfigure}
\hfill
\begin{subfigure}{.22\textwidth}
\includegraphics[scale=.35]{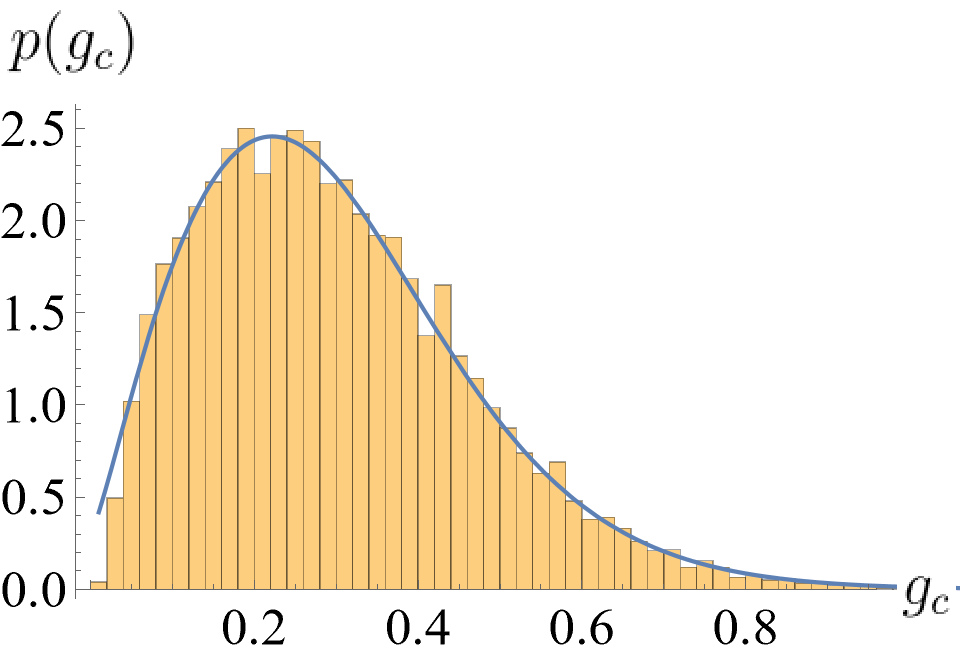}
\caption{}
\label{CDF_fits_min_PDFs_U=2_avg}
\end{subfigure}
\\
\begin{subfigure}{.22\textwidth}
\includegraphics[scale=.35]{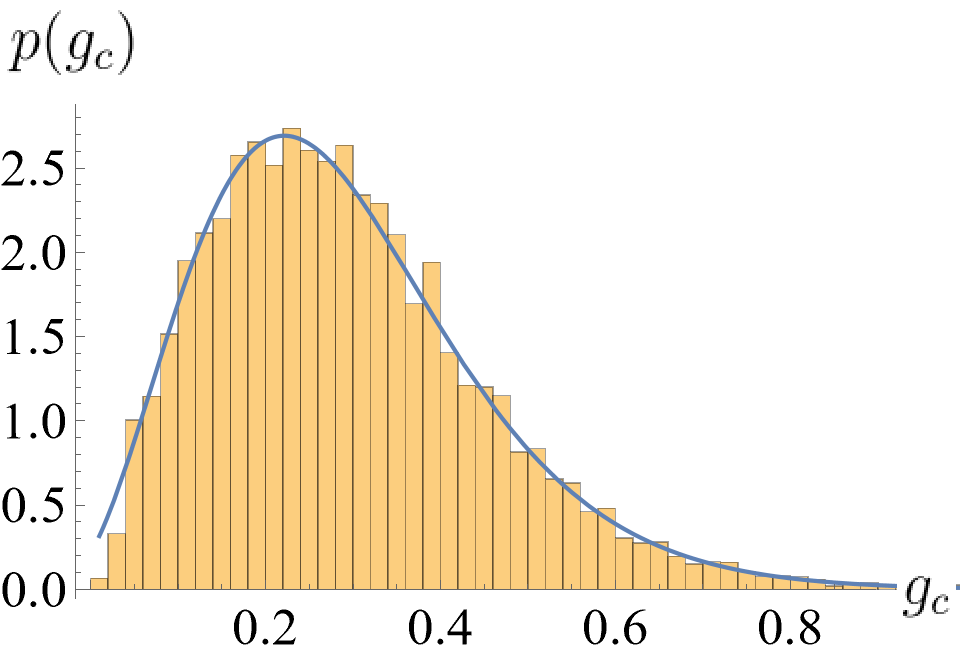}
\caption{}
\label{CDF_fits_min_PDFs_U=3_avg}
\end{subfigure}
\hfill
\begin{subfigure}{.22\textwidth}
\includegraphics[scale=.35]{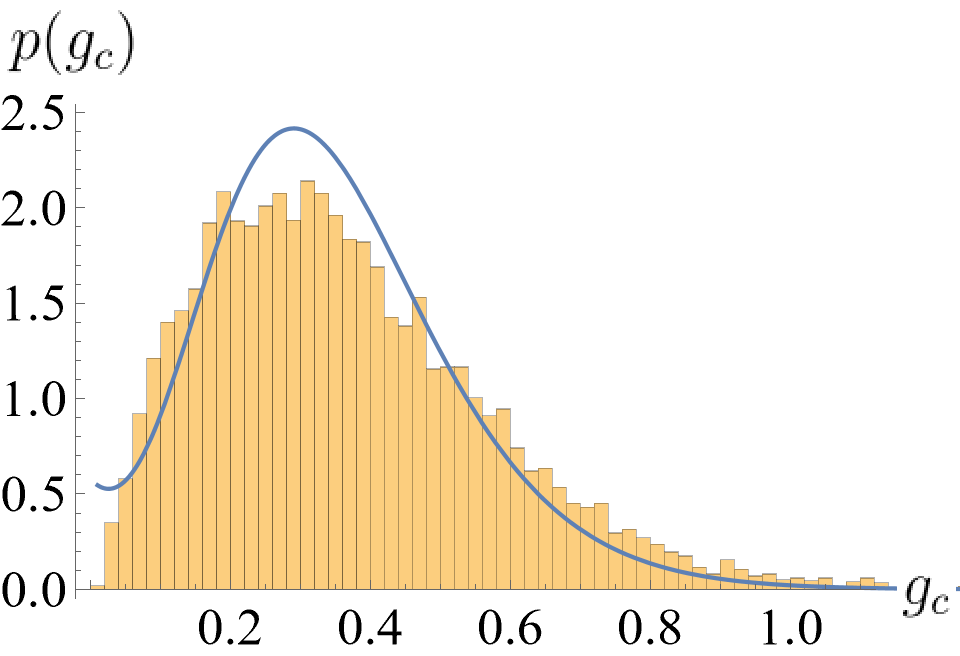}
\caption{}
\label{CDF_fits_min_PDFs_U=4_avg}
\end{subfigure}
\\
\begin{subfigure}{.22\textwidth}
\includegraphics[scale=.35]{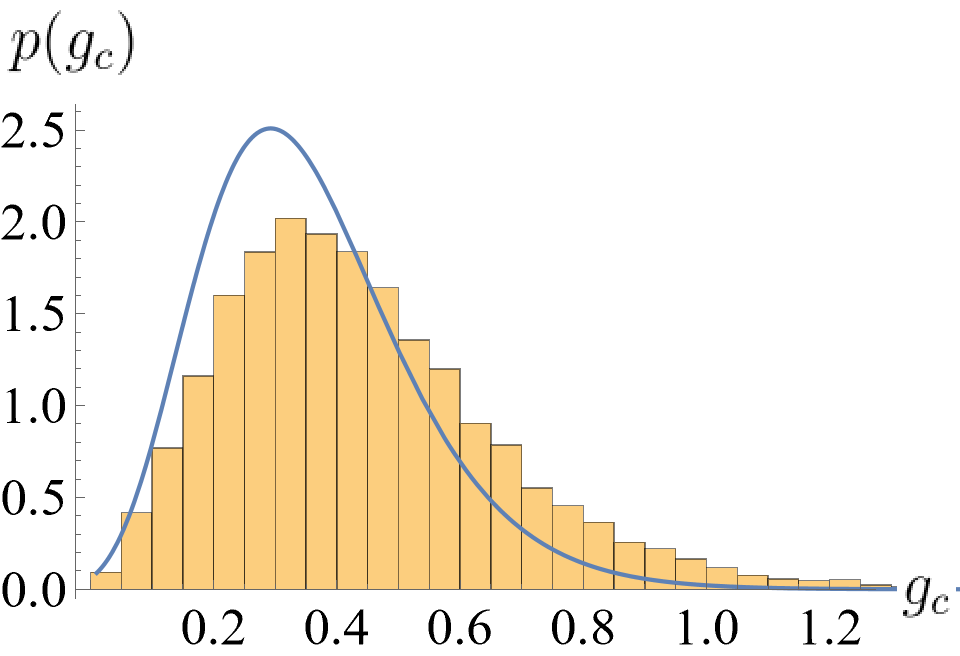}
\caption{}
\label{CDF_fits_min_PDFs_U=5_avg}
\end{subfigure}

\caption{Numerically observed probability density functions of $g_c$ at the critical disorder strength for a) $U=1$, $\epsilon = 0.8$ ($W = 3.0$), b) $U=2$, $\epsilon = 0.3$ ($W = 4.2$), c) $U=3$, $\epsilon = 0.35$ ($W = 5.6$), d) $U=4$, $\epsilon = 0.3$ ($W = 4.9$), e) $U = 5$, $\epsilon = 0.3$ ($W = 5.7$). The blue solid lines overlaid are the pdfs obtained from differentiating the best fit mixture distributions shown in Fig. \ref{CDF_fits_min_error_avg}. } 
\label{CDF_fits_min_error_PDFs_avg}
\end{figure}

We can derive distributions for $g_c(\epsilon)$ at the transition, in analogy with Section \ref{distributions}. To that end, assume that a generalized relation 
\begin{equation}
g_c(\epsilon) \approx \frac{1}{F}\frac{1}{N_{\ell}}\sum_{i=1}^{N_{\ell}}\left[-\log \mathcal{A}_{\ell}(\epsilon) - (\ell-1)\alpha \right]
\label{g_c_relation_avg}
\end{equation} 
holds. As before, let us write $\mathcal{A}_{\ell}(\epsilon) = \prod_{i=1}^{\ell}\chi_i(\epsilon)$, so that
$$g_c(\epsilon) \approx \frac{1}{F}\left[-\sum_{i=1}^{N_{\ell}} \frac{N_{\ell}-i+1}{N_{\ell}}\log\left(\chi_i(\epsilon)\right) - \frac{N_{\ell}-1}{2}\alpha\right] .$$
Assuming again $\chi_{i}(\epsilon) \sim \text{Uni}[0,1]$ at the MBL crossover, $-k\log(\chi_i(\epsilon))$ is exponentially distributed with rate parameter $1/k$, hence $g_c$ is (up to the shift $(N_{\ell}-1)\alpha/2F := \Delta$) a sum of exponential random variables with rate parameters $\lambda_i = F N_{\ell}/(N_{\ell}-i+1)$ ($i = 1,\ldots,N_{\ell}$). Such a sum is hypoexponentially distributed; the cumulative density for $g_c$ is then \cite{Cox-Renewal-Theory}
\begin{equation}
P_F^{N_{\ell}}(g_c \leq x) = 1 - \sum_{i=1}^{N_{\ell}}\left(\prod_{\substack{j=1 \\ j \neq i}}^{N_{\ell}} \frac{\lambda_j}{\lambda_j-\lambda_i}\right)e^{-\lambda_i(x+\Delta)} .
\label{CDF_gen_avg}
\end{equation}
As before, we construct a mixture distribution of the form \eqref{CDF_mixture}, and fit the $\beta_i$'s by minimizing the squared error \eqref{squared_error}. Examples at the center of the spectrum are shown in Fig. \ref{CDF_fits_avg} and at the band edges in Fig. \ref{CDF_fits_min_error_avg}. The resulting probability densities are shown in Figs. \ref{CDF_fits_PDFs_avg} and \ref{CDF_fits_min_error_PDFs_avg}. We note that, for lower interaction strengths, we find these fits yield comparable or slightly better (depending on the energy density) squared errors than those of section \ref{distributions}, whereas for stronger interactions the fits described in the main text are clearly superior.

\bibliography{Paper}% Produces the bibliography via BibTeX.

\end{document}